\documentclass[conference,onecolumn,10pt]{IEEEtran}

\usepackage[dvips]{graphicx}
\usepackage{subfigure}
\usepackage{algorithm}
\usepackage{algorithmic}

\begin{document}
\renewcommand{\baselinestretch}{1}
\title{Scale-Free Overlay Topologies with Hard Cutoffs for
Unstructured Peer-to-Peer Networks}

\author{\authorblockN{Hasan Guclu$^\dagger$ and Murat Yuksel$^\ddagger$}
\authorblockA{$^\dagger$Center for Nonlinear Studies, Theoretical Division\\
Los Alamos National Laboratory, Los Alamos, NM 87545}
\authorblockA{$^\ddagger$Computer Science and Engineering Department\\
University of Nevada - Reno, Reno, NV 89557\\
{\ttfamily guclu@lanl.gov, yuksem@cse.unr.edu} }}

\maketitle

\begin{abstract}
In unstructured peer-to-peer (P2P) networks, the overlay topology
(or connectivity graph) among peers is a crucial component in
addition to the peer/data organization and search. Topological
characteristics have profound impact on the efficiency of search on
such unstructured P2P networks as well as other networks. It has
been well-known that search on small-world topologies of $N$ nodes
can be as efficient as $O(\ln N)$, while scale-free (power-law)
topologies offer even better search efficiencies like as good as
$O(\ln \ln N)$ for a range of degree distribution exponents.
However, generation and maintenance of such scale-free topologies
are hard to realize in a distributed and potentially uncooperative
environments as in the P2P networks.

A key limitation of scale-free topologies is the high load (i.e.
high degree) on very few number of hub nodes. In a typical
unstructured P2P network, peers are not willing to maintain high
degrees/loads as they may not want to store large number of entries
for construction of the overlay topology. So, to achieve fairness
and practicality among all peers, hard cutoffs on the number of
entries are imposed by the individual peers, which limits
scale-freeness of the overall topology. Thus, efficiency of the
flooding search reduces as the size of the hard cutoff does. We
investigate construction of scale-free topologies with hard cutoffs
(i.e. there are not any major hubs) and effect of these hard cutoffs
on the search efficiency. Interestingly, we observe that the
efficiency of normalized flooding and random walk search algorithms
increases as the hard cutoff decreases.
\end{abstract}

\begin{keywords}
unstructured peer-to-peer networks, scale-free, power-law, search
efficiency, cutoff
\end{keywords}

\renewcommand{\baselinestretch}{1.75}

\section{Introduction}
In decentralized P2P networks, the overlay topology (or connectivity
graph) among peers is a crucial component in addition to the
peer/data organization and search. Topological characteristics have
profound impact on the efficiency of search on P2P networks as well
as other networks. It has been well-known that search on small-world
topologies can be as efficient as $O(\ln N)$ \cite{kleinberg-2000},
and this phenomenon has recently been studied on P2P networks
\cite{ZGG02,MSZ-2005,HLY-2006,IRF04}.

The best search efficiency in realistic networks can be achieved
when the topology is scale-free (power-law), which offers search
efficiencies like $O(\ln \ln N)$. However, generation and
maintenance of such scale-free topologies are hard to realize in a
distributed and potentially uncooperative environments as in the P2P
networks. Another key limitation of scale-free topologies is the
high load (i.e. high degree) on very few number of hub nodes. In a
typical unstructured P2P network, peers are not willing to maintain
high degrees/loads as they may not want to store large number of
entries for construction of the overlay topology. So, to achieve
fairness and practicality among all peers, \emph{hard} cutoffs on
the number of entries are imposed by the individual peers. These
hard cutoffs might limit \emph{scale-freeness} of the overall
topology, by which we mean having a network with a power-law degree
distribution from which an exponent can be obtained properly. Thus,
it is expected that the search efficiency reduces as the size of the
hard cutoff does.

The primary focus of this paper is to (i) investigate construction
of scale-free topologies with hard cutoffs (i.e. there are not any
major hubs) in a distributed manner without requiring global
topology information at the time when nodes join and (ii) to
investigate the effect of these hard cutoffs on the search
efficiency.

The rest of the paper is organized as follows: First, in the rest of
this section, we provide motivation for this work, outline key
dimensions to be considered, and briefly indicate major
contributions and findings of the work. Then, we survey previous
work on P2P networks in Section~\ref{sec:survey}. In
Section~\ref{sec:scale-free-topologies}, we survey the previous work
on scale-free topology generation and cover two specific models that
we use: Preferential Attachment (PA), Configuration Model (CM). We
introduce our practical topology generation methodologies,
Hop-and-Attempt Preferential Attachment (HAPA) and
Discover-and-Attempt Preferential Attachment (DAPA), in
Section~\ref{sec:local-heuristics}. In
Section~\ref{sec:simulations}, we present our simulations of three
different search algorithms (i.e., Flooding (FL), Normalized
Flooding (NF), and Random Walk (RW)) on topologies generated by the
models PA, CM, HAPA, and DAPA. We conclude by summarizing the work
and outlining future directions in Section~\ref{sec:summary}.

\subsection{Motivation and Key Considerations}
The search efficiency on small-world and scale-free topologies is
well-known. Although scale-free topologies are superior in search
efficiency, their super-hub-based structure makes them vulnerable to
threats and impractical due to unfair assignment of network load on
a very small subset of all nodes. As peers in a P2P network are
typically not fully cooperative, protocols cannot rely on methods
working with full cooperation of peers. For example, peers may not
want to store large number of entries for construction of the
overlay topology, i.e. connectivity graph. Even though
characteristics of the overlay topology is crucial in determining
the efficiency of the network, peers typically do not want to take
the burden of storing excessive amount of control information for
others in the network. Effect of this on the overlay topology
maintenance is that peers impose \emph{hard cutoffs} on the amount
of control information to be stored. Since P2P overlay topology
generation and maintenance are very important for realizing a
scalable unstructured P2P network, the main focus of this paper is
\emph{to investigate the effect of the hard cutoffs on the overall
search efficiency}.

A key issue is the construction of scale-free overlay topologies
without global information. There are several techniques to generate
a scale-free topology \cite{bara99,albert00}, which rely on
\emph{global} information about the current network when a new node
joins. Such global methods are not practical in P2P networks, and
\emph{local} heuristics in generating such scale-free overlay
topologies with hard cutoffs is the key issue, which we investigate
in this paper. In other words, each peer has to figure out the
optimal way of joining the P2P overlay by only using the locally
available (i.e. immediate/close neighbors) information, and also
causing a minimal inefficiency to the search mechanisms to be run on
the network.

%- We also explore, for each peer, the possibility of maintaining
%global statistics (e.g. max and min degree) about the overlay
%topology so that the peers can optimize their local decisions.

\subsection{Contributions and Major Results}
This paper touches an uncovered set of research problems relating to
tradeoffs between maximum number of links a peer can (or is willing
to) store and the efficiency of search on an overlay topology
composed of such peers. We defined the maximum number of links to be
stored by peers as the \emph{hard cutoff} for the degree of a peer
in the network as compared to \emph{natural cutoff} which
occurs due to finite-size effects. Our contributions include:
\begin{itemize}
\item{\emph{Scale-free topology generation methods:}} We studied two
well-known scale-free topology generation mechanisms (i.e.
Preferential Attachment (PA) and Configuration Model (CM)) which use
global information about the overlay topology within the context of
unstructured P2P networks. We introduced two novel mechanisms (i.e.
HAPA and DAPA) that use local topology information solely or
partially.

\item{\emph{Search efficiency on scale-free topologies with hard
cutoffs:}} Through extensive simulations, we studied efficiency of
FL, NF, and RW on the topologies generated by the four mechanisms
PA, CM, HAPA, and DAPA.

\item{\emph{Guidelines for designing peer join algorithms for
unstructured P2P networks:}} Our study yielded several guidelines
for peers to join to a Gnutella-like unstructured P2P network, so
that the search performance of the overall overlay topology remains
high.
\end{itemize}

Our study of hard cutoffs enlightened several interesting issues. We
found that hard cutoffs may not always affect the search performance
adversely, unlike the expected wisdom that the value of the
power-law exponent in the degree distribution of the topology is
directly related to search performance. We showed that search
performance depends on the particular search algorithm being used as
well as on the topological characteristics including the exponent of
the degree distribution, connectedness (minimum degree is a measure
for it in scale-free networks), hard cutoff, and locality.

We also showed that there is an interplay between the connectedness
and the degree distribution exponent for a fixed cutoff; more
specifically, if connectedness is too low in the topology, then one
can improve search performance by applying smaller hard cutoffs. Our
simulation experiments clearly showed that practical search
algorithms like NF or repeated RWs can perform better on scale-free
topologies with smaller hard cutoffs as long as peers join
carefully, e.g. as in HAPA and DAPA mechanisms. As a particular
guideline for optimizing joining techniques of peers, we showed
that, as long as every peer is required to maintain a minimum of 2-3
links to the rest of the network rather than just one link, it is
possible to diminish negative effects of hard cutoffs on search
performance.

\section{Related Work}
\label{sec:survey}

Our work is related to peer-to-peer (P2P) network protocol designs
and topological analysis of complex networks. Previous work on P2P
network protocols can be classified into \emph{centralized} and
\emph{decentralized} ones. As centralized P2P protocols (e.g.
Napster \cite{napster}) proved to be unscalable, the majority of the
P2P research has focused on decentralized schemes. The decentralized
P2P schemes can be further classified into sub-categories:
\emph{structured}, \emph{unstructured}, and \emph{hybrid}.

In the structured P2P networks, data/file content of peers is
organized based on a keying mechanism that can work in a distributed
manner, e.g. Distributed Hash Tables (DHTs) \cite{RSS01}, CAN
\cite{can01}, Chord \cite{chord01}, Kademlia \cite{MaMa-2002},
Oceanstore \cite{oceanstore-2000}. The keying mechanism typically
maps the peers (or their content) to a logical search space, which
is then leveraged for performing efficient searches. Another
positive side of the structured schemes is the guarantees of finding
rare items in a timely manner. However, the cost comes from
complexity of maintaining the consistency of mapping the peers to
the logical search space, which typically causes considerable amount
of control traffic (e.g. join/leave messaging) for highly dynamic
P2P environments. Due to their capability of locating rare items,
structured approaches have been very well suited to a wide-range of
various applications, e.g.
\cite{CDKR-2003,RoDr-2001,CMM-2002,Kato-2002,DBKKMSB-2001}.

In contrast to the structured schemes, unstructured P2P networks do
not include a strict organization of peers or their content. Since
there is no particular keying or organization of the content, the
search techniques are typically based on flooding. Thus, the
searches may take very long time for rare items, though popular
items can be found very fast due to possible leveraging of locality
of reference \cite{BCAA-2004,SMZ03,CLL-2002} and caching/replication
\cite{CoSh-2002,LCCLS-2002}.

To balance the tradeoffs between structured and unstructured
schemes, hybrid approaches (e.g. \cite{SuGaMo03,ZhHu-2005,SHK-2003})
have attempted attaining a middle-ground between the costly
maintenance of global peer/data keying of structured schemes and the
high cost searches of unstructured schemes. Typically, hybrid
schemes include a localized (or reduced in size) data/peer keying
(e.g. DHT among neighbors instead of among all nodes) to achieve
faster discovery of rare items, and a probabilistic search among the
partially organized peers.

Since our work is more applicable to unstructured P2P networks, we
focus our survey in this section to that category of the previous
work. Also, since we propose to use scale-free topologies in
constructing the overlay P2P topology, we survey the literature on
scale-free network topologies in the next section.

\subsection{Unstructured P2P Networks}
The main focus of the research on unstructured P2P networks has been
the tradeoff between state complexity of peers (i.e. number of
records needed to be stored at each peer) and flooding-based search
efficiency. The minimal state each peer has to maintain is the
\emph{list of neighbor peers}, which construct the overlay topology.
Optionally, peers can maintain \emph{forwarding table}s (also
referred as routing tables in the literature) for data items in
addition to the list of neighbor peers. Thus, we can classify
unstructured P2P networks into two based on the type(s) of state
peers maintain: (i) \emph{per-data} unstructured P2P networks (i.e.
peers maintain both the list of neighbor peers and the per-data
forwarding table), and (ii) \emph{non-per-data} unstructured P2P
networks (i.e. peers maintain only the list of neighbor peers).

Non-per-data schemes are mainly Gnutella-like schemes
\cite{gnutella}, where search is performed by means of flooding
query packets. Search performance over such P2P networks has been
studied in various contexts, which includes pure random walks
\cite{GMS04}, probabilistic flooding techniques \cite{KXZ05,GMS05},
and systematic filtering techniques \cite{Xu03}.

Per-data schemes (e.g. Freenet \cite{freenet}) can achieve better
search performances than non-per-data schemes, though they impose
additional storage requirements to peers. By making the peers
maintain a number of $<$key,pointer$>$ entries peers direct the search
queries to more appropriate neighbors, where ``key'' is an
identifier for the data item being searched and the ``pointer'' is
the next-best neighbor to reach that data item. This capability
allows peers to leverage associativity characteristics of search
queries \cite{CFK03}. Studies ranged from grouping peers of similar
interests (i.e. peer associativity) \cite{IRF04,CFK03} to exploiting
locality in search queries (i.e. query associativity)
\cite{CRBLS03,SMZ03}.

Our work is applicable to both per-data and non-per-data
unstructured P2P networks, since we focus on the interactions
between search efficiency and topological characteristics.

\section{Scale-Free (Power-Law) Network Topologies}
\label{sec:scale-free-topologies}

Recent research shows that many natural and artificial systems such
as the Internet \cite{faloutsos99}, World Wide Web
\cite{albert00_2}, scientific collaboration network \cite{bara02},
and e-mail network \cite{ebel02} have power-law degree
(connectivity) distributions. These systems are commonly known as
power-law or scale-free networks since their degree distributions
are free of scale (i.e. not a function of the number of network
nodes $N$) and follow power-law distributions over many orders of
magnitude. This phenomenon has been represented by the probability
of having nodes with $k$ degrees as $P(k) \sim k^{-\gamma}$ where
$\gamma$ is usually between $2$ and $3$ \cite{bara99}. Scale-free
networks have many interesting properties such as high tolerance to
random errors and attacks (yet low tolerance to attacks targeted to
hubs) \cite{alb00}, high synchronizability
\cite{guclu05,toro03,korniss06}, and resistance to congestion
\cite{toroczkai04_1,toroczkai04_2}.

The origin of the scale-free behavior can be traced back to two
mechanisms that are present in many systems, and have a strong
impact on the final topology \cite{bara99}. First, networks are
developed by the addition of new nodes that are connected to those
already present in the system. This mechanism signifies continuous
expansion in real networks. Second, there is a higher probability
that a new node is linked to a node that already has a large number
of connections. These two features led to the formulation of
a growing network model first proposed by Barab\'asi and Albert that
generates a scale-free network for which $P(k)$ follows a power law with
$\gamma$$=$$3$. This model is known as \textit{preferential
attachment} (PA or rich get richer mechanism) and the resulting network
is called \textit{Barab\'asi-Albert} network \cite{bara99,albert00}.

\begin{figure*}
\begin{center}
\begin{tabular}{ccc}
\includegraphics[keepaspectratio=true,angle=0,width=60mm]
{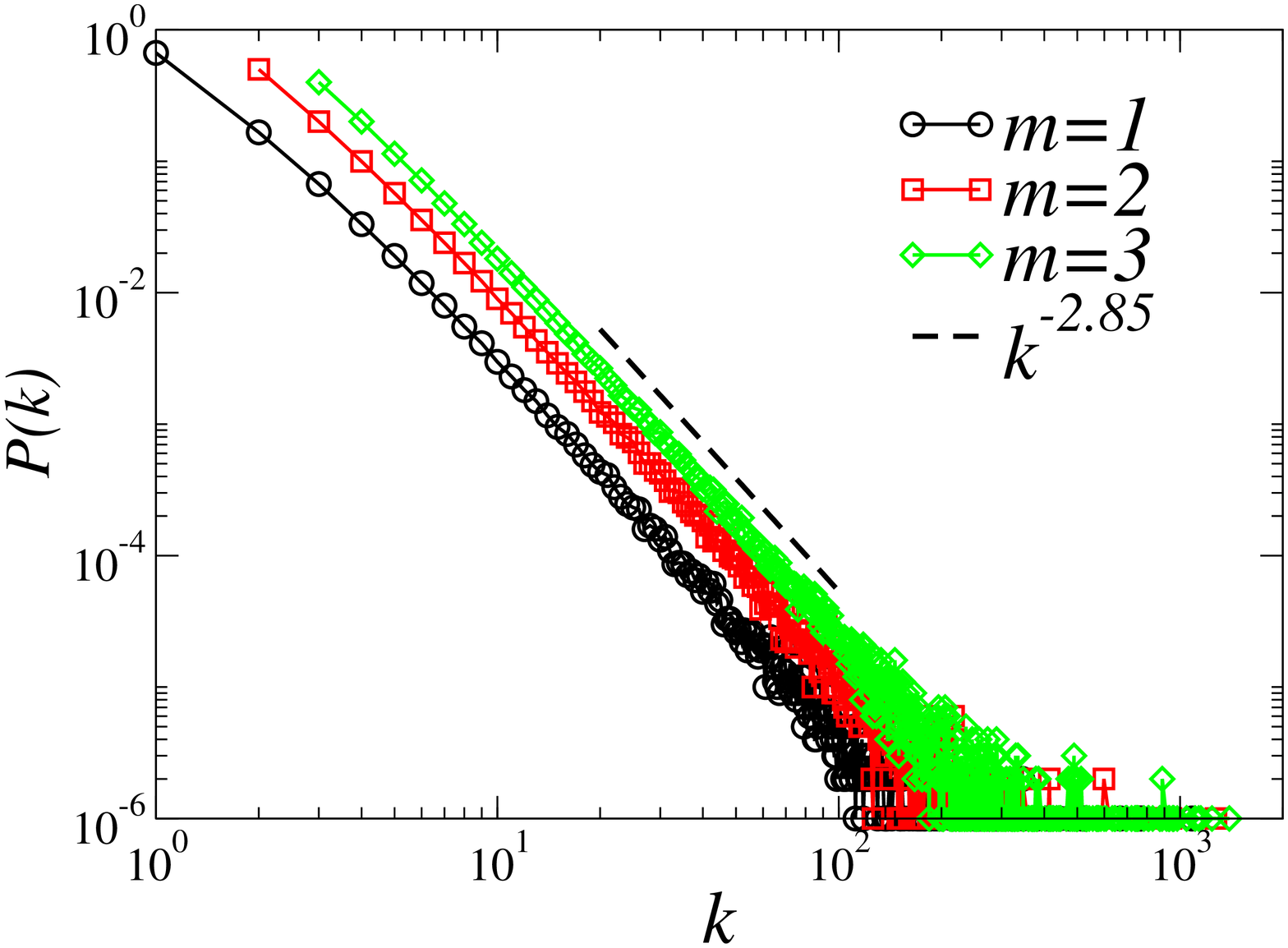} & \hspace{-2mm}
\includegraphics[keepaspectratio=true,angle=0,width=60mm]
{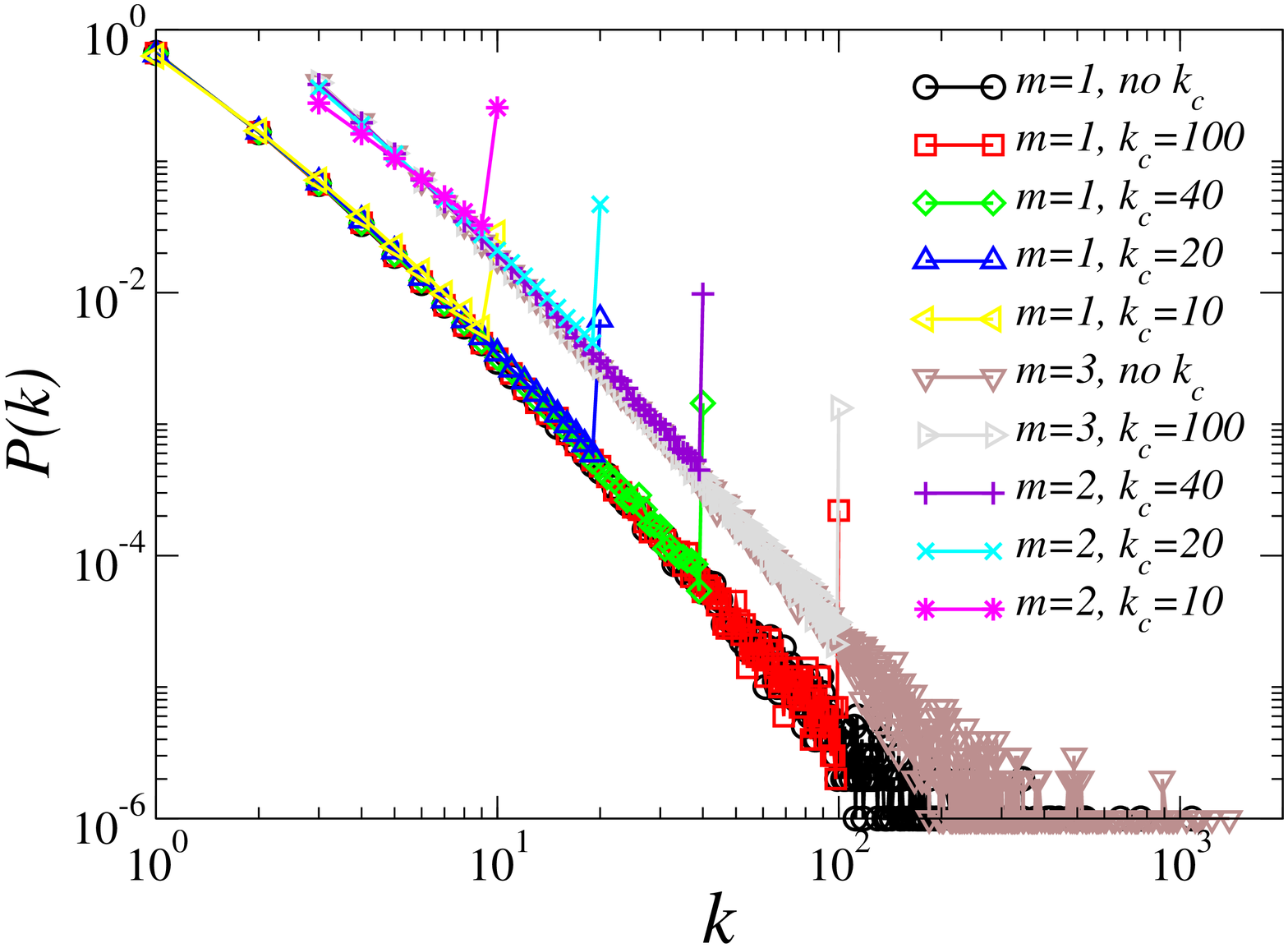} & \hspace{-2mm}
\includegraphics[keepaspectratio=true,angle=0,width=60mm]
{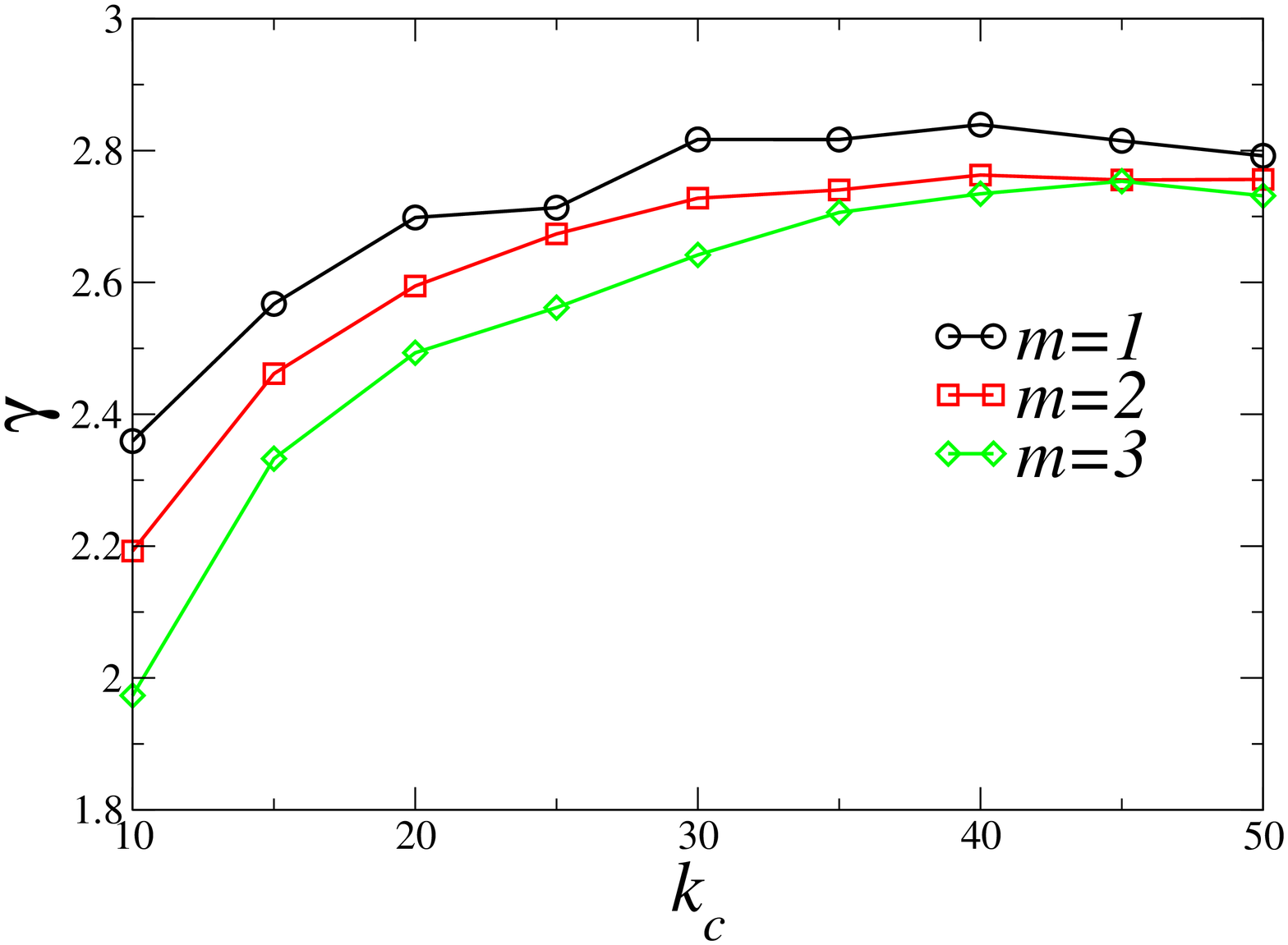} \vspace{-5mm} \\
\small{(a) $P(k)$ without hard cutoff.} & \small{(b) $P(k)$ for
different hard cutoffs.} & \small{(c) $P(k)$ exponents vs hard
cutoffs.}
\end{tabular}
\end{center}
\caption{Degree distributions, $P(k)$ for various networks
generated by the PA model. The number of nodes is $N$$=$$10^5$ and
for every data point 10 different realizations of the network have
been used. (a) The power-law fits to the distributions have
exponents between $(-2.9$$,$$-2.8)$. The dashed line is a power-law
function with exponent $\gamma=2.85$. Simulations show that the exponent
$\gamma=3$ is attainable only for very large networks.}
\label{fig_pk-pa}
\end{figure*}

In this study, we use a simple version of the PA model
\cite{bara99}. The model evolves by one node at a time and this new
node is connected to $m$ (number of stubs) different existing nodes
with probability proportional to their degrees, i.e.,
$P_i$$=$$\frac{k_i}{\sum_j k_j}$ where $k_i$ is the degree of the
node $i$. The average degree per node in the resulting network is
$2m$. Fig.~\ref{fig_pk-pa}(a) shows the degree distributions of
scale-free networks generated by the PA model with different $m$
values. The links are regarded as bidirectional links; however, the
results can easily be generalized to directed networks as well
\cite{albert00}. The special case of the PA procedure is when the
number of stubs is one (i.e. $m$$=$$1$) in which a scale-free tree
without clustering (loops) is generated. The algorithm for the PA
model is presented in Appendix~\ref{app_pa}.

Scale-free networks are very robust against random failures and
attacks since the probability to hit the hub nodes (few nodes with
very large degree) is very small and attacking the satellite nodes
with just a few degree does not harm the network. On the other hand,
deliberate attacks targeted to hubs through which most of the
traffic go can easily shatter the network and severely damage the
overall communication in the network. For the same reason the
Internet is called ``robust yet fragile'' \cite{doyle05} or
``Achilles heel'' \cite{alb00, alb00_2}.

Scale-free networks also have \textit{small-world} \cite{watts98}
properties. In small-world networks the diameter, or the mean hop
distance between the nodes scales with the system size (or the
number of network nodes) $N$ logarithmically, i.e., $d\sim \ln N$.
The scale-free networks with $2<\gamma<3$ have a much
smaller diameter and can be named \textit{ultra-small} networks
\cite{cohen03}, behaving as $d\sim\ln{\ln N}$. When
$\gamma=3$ and $m\geq 2$, $d$ behaves as in
$d\sim\ln N/\ln\ln N$. However, when $m=1$ for
$\gamma=3$ the Barab\'asi-Albert model turns into a tree and
$d \sim \ln N$ is obtained. Also when $\gamma > 3$, the diameter
still behaves logarithmically $d \sim \ln N$. These relationships
are summarized in Table~\ref{tab:diameter-behavior}. Since the
speed/efficiency of search algorithms strongly depend on the average
shortest path, scale-free networks have much better performance in
search than other random networks.

\renewcommand{\baselinestretch}{1}
\begin{table}
    \caption{Scale-free network diameter behavior}
    \label{tab:diameter-behavior}
    \begin{center}
\vspace{-5mm}
    \begin{tabular}{|c|c|c|}
    \hline
        Diameter & Exponent & \# of stubs \\
        $d$ & $\gamma$ & $m$ \\
    \hline
        $\ln \ln N$ & (2,3) & $\geq 1$ \\
        $\ln N / \ln \ln N$ & 3 & $\geq 2$ \\
        $\ln N$ & 3 & 1 \\
        $\ln N$ & $> 3$ & $\geq 1$ \\
    \hline
    \end{tabular}
\vspace{-5mm}
    \end{center}
\end{table}
\renewcommand{\baselinestretch}{1.75}

\subsection{The Cutoff}
One of the important characteristics of scale-free networks is the
degree cutoff (or maximum degree) due to the finite-size effects.
Aiello et al. \cite{aiello01} defined a \emph{natural} cutoff
$k_{nc} = k_{max}$ value of the degree, above which the expected
number of nodes is $1$, that is
\begin{equation}
N Pr(k=k_{nc}) \sim 1. \label{eq1}
\end{equation}
For a scale-free network, when one substitutes $P(k)$$\propto
$$k^{-\gamma}$
\begin{equation}
k_{nc}(N) \sim N^{1/\gamma}.
\label{eq2}
\end{equation}
This definition of natural cutoff degree used in many P2P network
studies lacks some mathematical rigor since it considers the
probability of a single point in a probability distribution, which
is not completely well-defined in the continuous $k$ limit for large
$N$ \cite{boguna04}.

A more physical definition of cutoff was given by Dorogovtsev et
al. \cite{doro02_2}, defining it as the value of the degree
above which one expects to find at most one vertex. This meant the
satisfaction of the condition:
\begin{equation}
N \int_{k_{nc}}^{\infty} P(k)dk \sim 1.
\label{eq3}
\end{equation}
Again by using the degree distribution for the scale-free network
and the exact form of probability distribution (i.e.,
$P(k)=(\gamma-1)m^{\gamma-1}/k^{\gamma}$), one obtains
\begin{equation}
k_{nc}(N) \sim mN^{1/(\gamma-1)},
\label{eq4}
\end{equation}
which is known as the \textit{natural} cutoff of the network. The
scaling of the natural cutoff can also be calculated by using the
extreme-value theory \cite{boguna04}. For the scale-free networks
generated by PA model ($\gamma$$=$$3$) the
natural cutoff becomes
\begin{equation}
k_{nc}(N) \sim m\sqrt{N}.
\label{eq5}
\end{equation}

\subsection{Preferential Attachment with Hard Cutoffs}

The natural cutoff may not be always attainable for most of the
scale-free networks due to technical reasons. One main reason is
that the network might have limitations on the number of links the
nodes can have. This is especially important for P2P networks in
which nodes can not possibly connect many other nodes. This requires
putting an artificial or \textit{hard} cutoff $k_c$ to the number of
links one node might have.

In order to see the effect of hard cutoff in PA, we simply did not
allow nodes to have links more than a fixed hard cutoff value during
the attachment process. This modified method generates a scale-free
network in which there are many nodes with degree fixed to hard
cutoff instead of a few very high degree hubs and the degree
distribution still decays in a power law fashion. The degree
distributions of scale-free networks generated by PA with different
hard cutoff values are shown in Fig.~\ref{fig_pk-pa}(b). As can be
seen in the figure they are slightly different than PA without a
cutoff in terms of exponent except that they have an accumulation of
nodes with degree equal to hard cutoff. PA model, in its original
form, has a constant degree distribution exponent $\gamma$$=$$3$ for
very large networks. However, when the number of connections (or
degree) one node can have is bounded with a finite hard cutoff value
from above it is observed that the absolute value of degree
distribution exponent decreases as in Fig.~\ref{fig_pk-pa}(c). A
more detailed description of the PA model is provided in
Appendix~\ref{app_pa}.

\begin{figure*}
\begin{center}
\begin{tabular}{ccc}
\includegraphics[keepaspectratio=true,angle=0,width=60mm]
{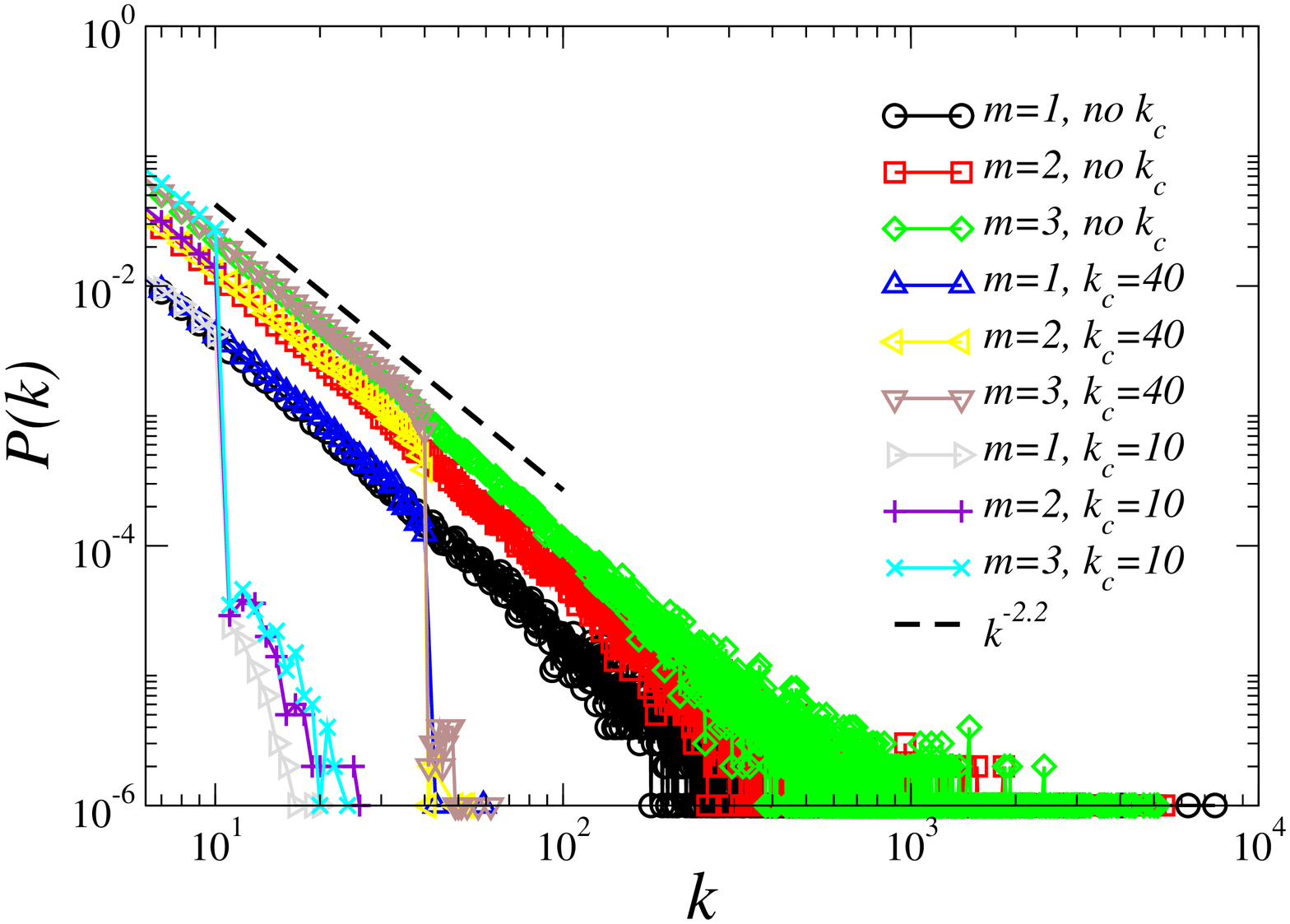} &
\hspace{-2mm}
\includegraphics[keepaspectratio=true,angle=0,width=60mm]
{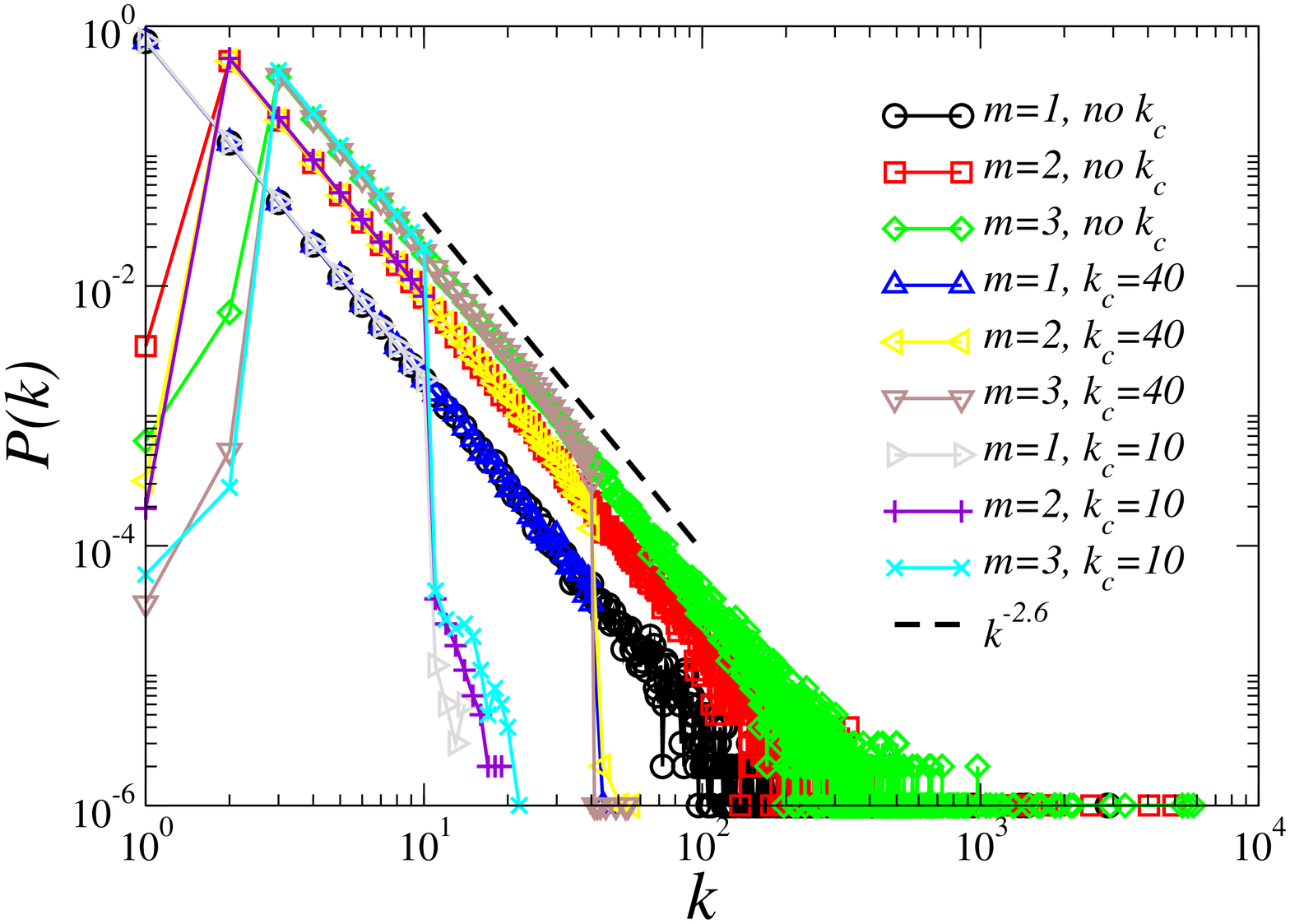} &
\hspace{-2mm}
\includegraphics[keepaspectratio=true,angle=0,width=60mm]
{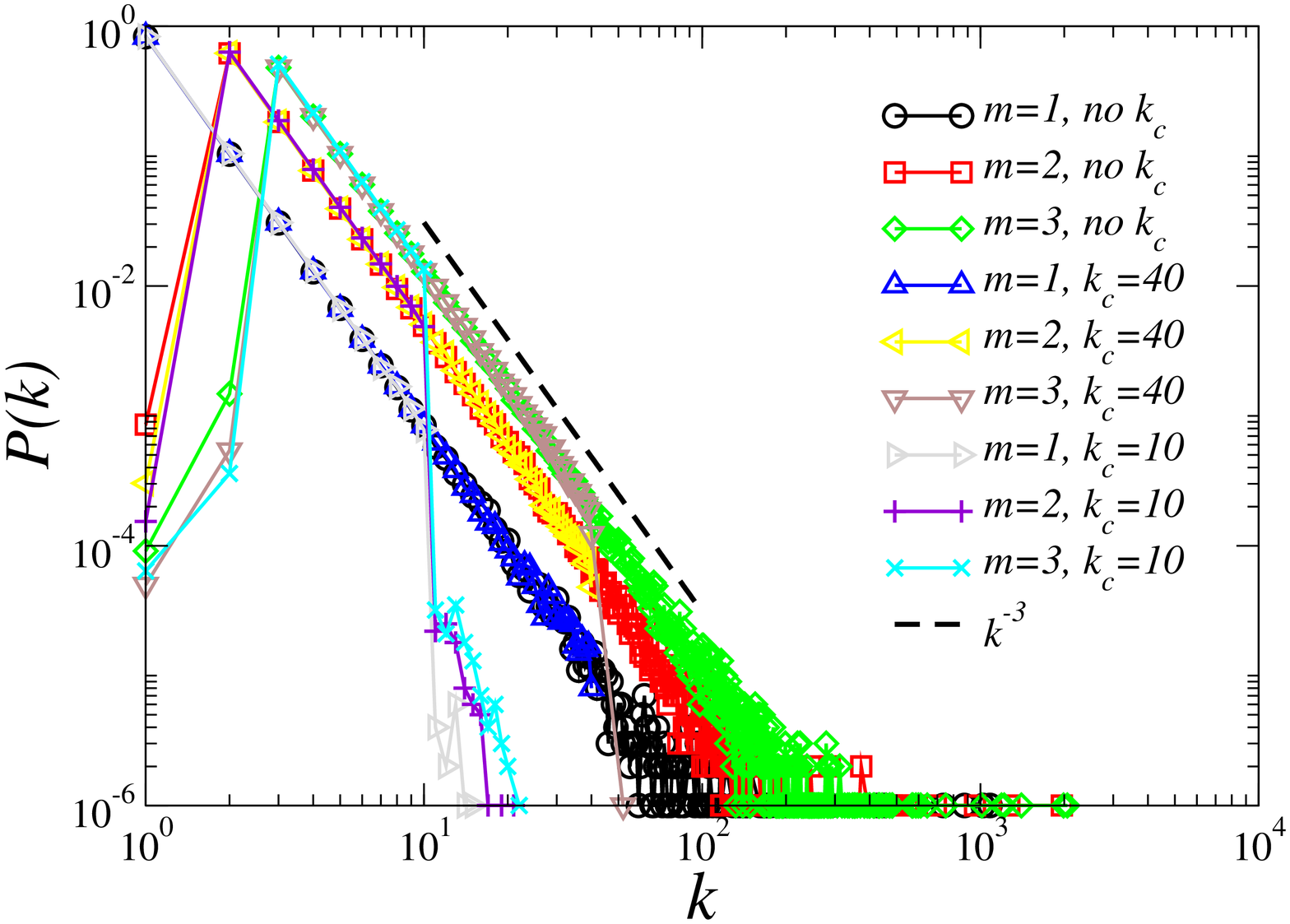} \vspace{-5mm}
\\
\small{(a) $\gamma=2.2$} & \small{(b) $\gamma=2.6$} & \small{(c)
$\gamma=3$}
\end{tabular}
\end{center}
\caption{Degree distributions of CM.} \label{fig_pk-mr}
\end{figure*}

\subsection{Configuration Model (CM)}

Given that the PA model yields lower degree distribution exponents
as the hard cutoff reduces, we were motivated to work on generation
of power-law networks with different exponents. In other words,
instead of dealing with hard cutoffs, it is possible to achieve a
natural cutoff value less than the targeted hard cutoff. In this
manner, the peaks at the hard cutoff value of
Fig.~\ref{fig_pk-pa}(b) can be prevented and a smooth power-law
distribution of degrees can be obtained. For this reason, modified
PA models such as nonlinear preferential attachment
\cite{KRAPIVSKY00,KRAPIVSKY01}, dynamic edge-rewiring,
\cite{albert00}, and fitness models \cite{BIANCONI01_1,BIANCONI01_2}
have been proposed. Here, we use the configuration model (CM) with a
predefined degree distribution to generate a static scale-free
network \cite{molloy95}.

CM \cite{BENDER78,molloy95,molloy98} was introduced as an algorithm
to generate uncorrelated random networks with a given degree
distribution. In CM, the vertices of the graph are assigned a fixed
sequence of degrees $\{k_i\}_{i=0}^{N}, m \leq k_i \leq k_c$, where
typically $k_c$$=$$N$, chosen at random from the desired degree
distribution $P(k)$, and with the additional constraint that the
$\sum_i k_i$ must be even. Then, pairs of nodes are chosen randomly
and connected by undirected edges. This model generates a network
with the expected degree distribution and no degree correlations;
however, it allows self-loops and multiple-connections when it is
used as described above. It was proved in \cite{boguna04} that the
number of multiple connections when the maximum degree is fixed to
the system size, i.e., $k_c$$=$$N$, scales with the system size $N$
as $N^{3-\gamma}\ln N$. Since we work with hard cutoff values
typically less than the natural cutoff the number of multiple links
is much less than the original CM for which $k_c$$=$$N$
\cite{CATANZARO05,SERRANO05}. After this procedure we simply delete
the multiple connections and self-loops from the network which gives
a very marginal error in the degree distribution exponent. Deleting
this discrepancies also causes some very negligible number of nodes
in the network to have degrees less than the fixed minimum degree
($m$) value, even zero, as seen in Fig.~\ref{fig_pk-mr}. One another
characteristic of the CM is that the network is not a connected
network when $m$$=$$1$, i.e., it has disconnected clusters (or
components). For $m$$>$$1$, the network is almost surely connected
having one giant component including all the nodes. The algorithm we
use for CM can be seen in Appendix~\ref{app_cm}.

The main disadvantage of PA and CM methods is that they require
global knowledge about the network, i.e., the degrees of all peers
or the maximum/total degree, which might be usually difficult to
store and share by the nodes for unstructured P2P networks. This
motivated us to modify the PA procedure so that it makes use of
local information as much as possible. In the next section we
explore the local heuristics in creating scale-free networks and
introduce two new attachment models.

\begin{figure*}
\begin{center}
\begin{tabular}{ccc}
\includegraphics[keepaspectratio=true,angle=0,width=60mm]
{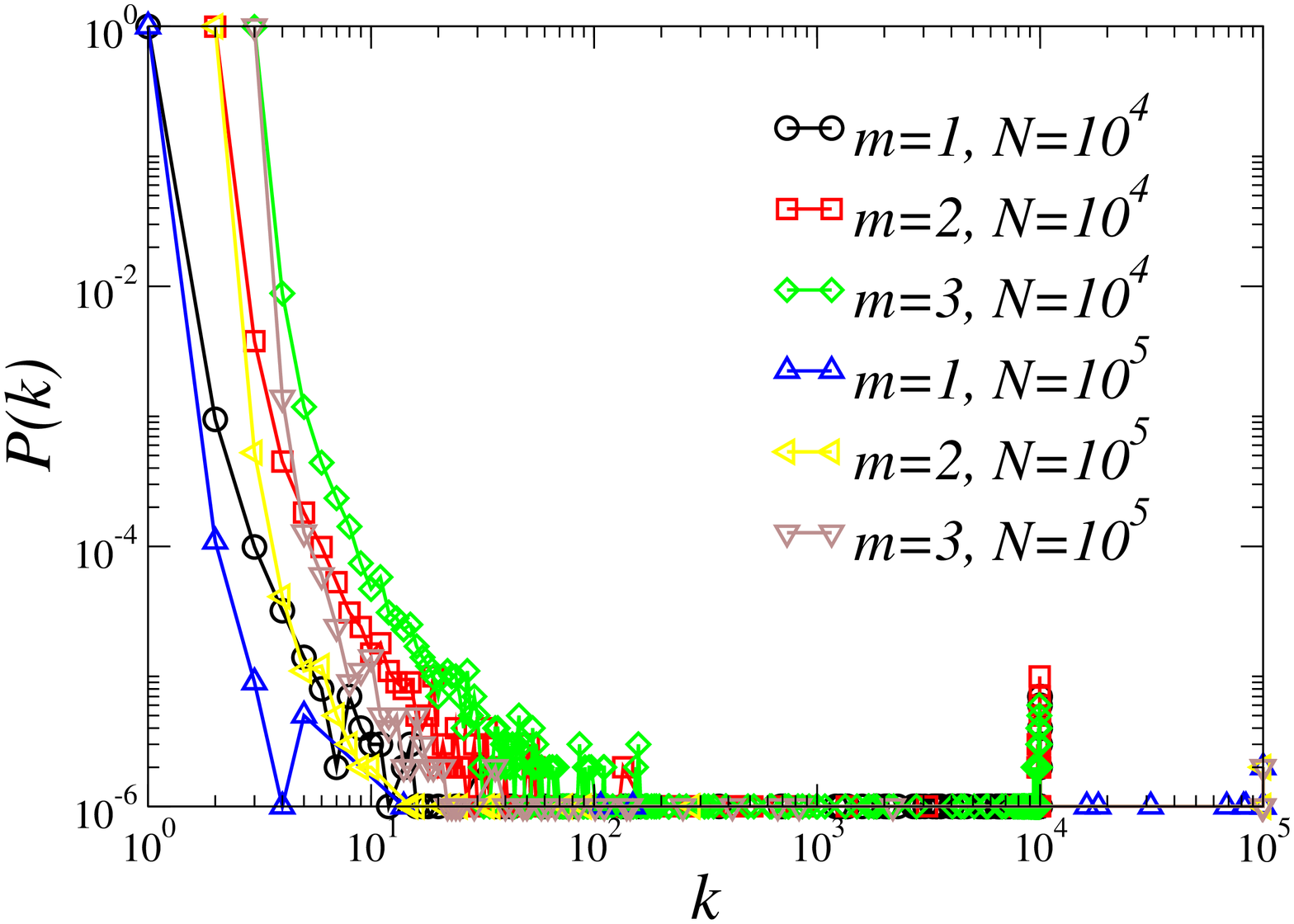} & \hspace{-2mm}
\includegraphics[keepaspectratio=true,angle=0,width=60mm]
{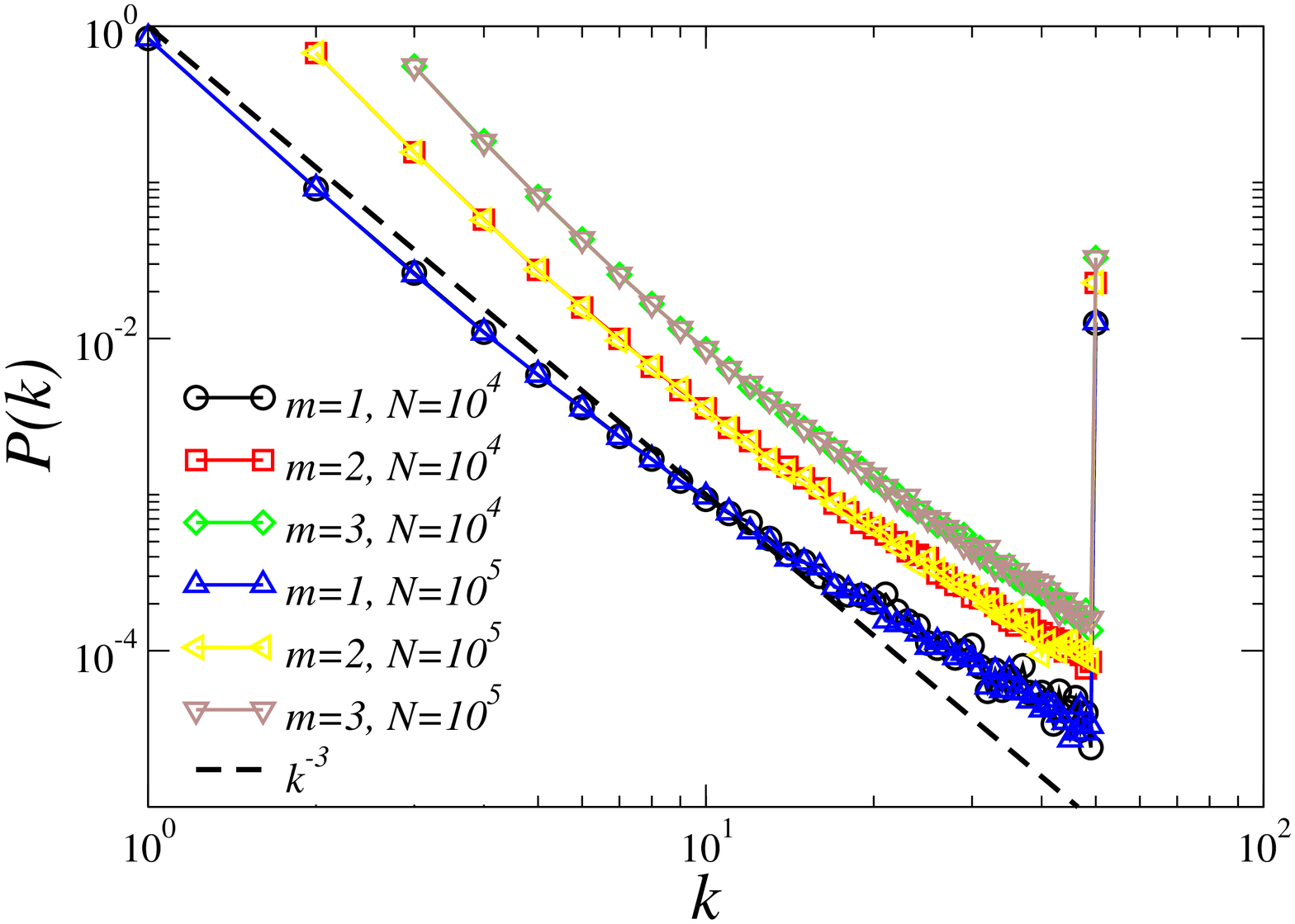} & \hspace{-2mm}
\includegraphics[keepaspectratio=true,angle=0,width=60mm]
{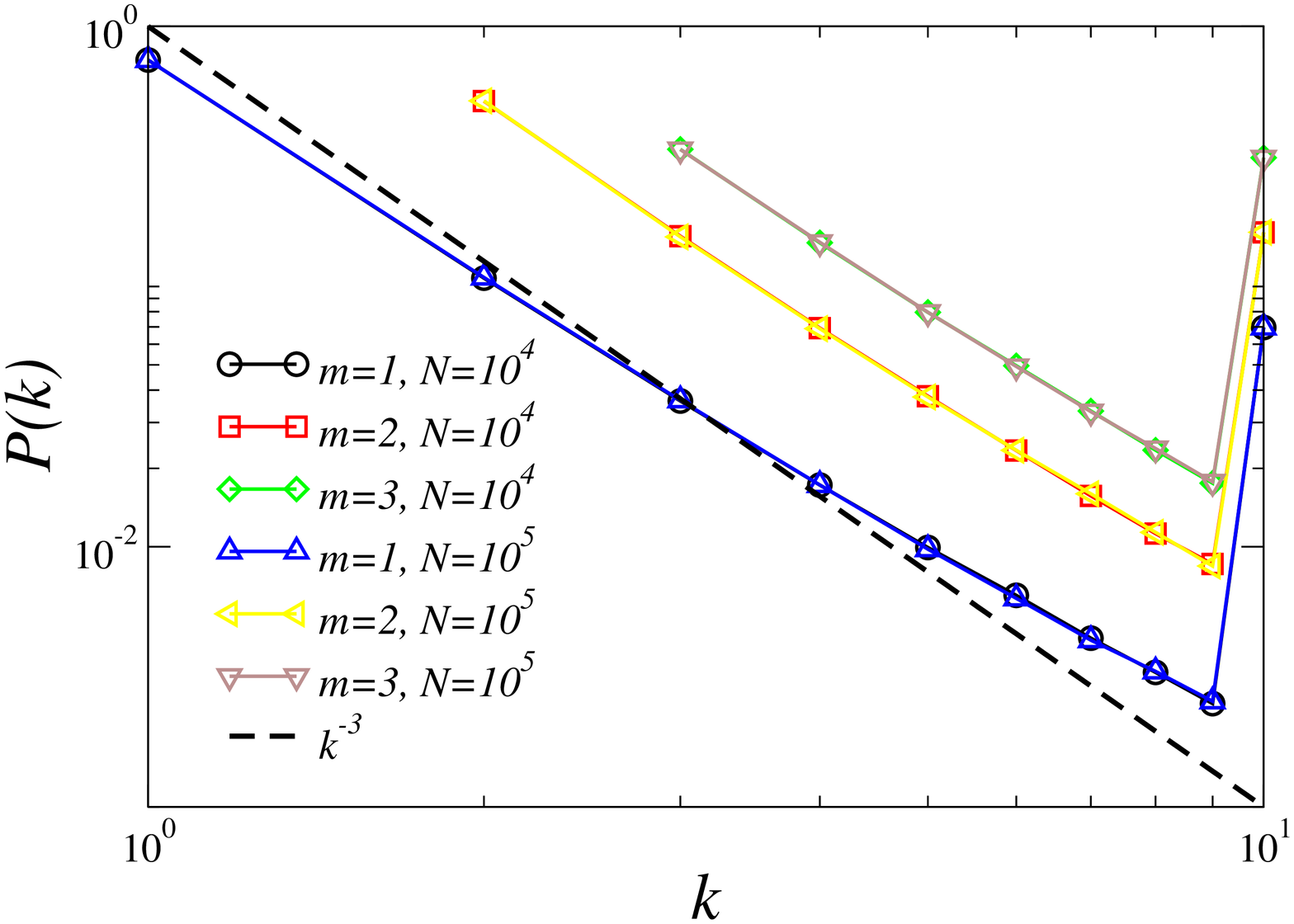} \vspace{-5mm}
\\
\small{(a) no cutoff} & \small{(b) $k_c=50$} & \small{(c) $k_c=10$}
\end{tabular}
\end{center}
\caption{Degree distributions of HAPA model.} \label{fig_pk-hapa}
\end{figure*}

\section{Local Heuristics for Scale-Free Overlay Topology Construction}
\label{sec:local-heuristics}

In the PA model as outlined in the previous section, the new node
has to make random attempts to connect to the existing nodes with a
certain probability depending on the degree of the existing node. To
implement this in a P2P (or any distributed) networks, the new node
has to have information about the global topology (e.g. the current
number of degrees each node has for the PA model), which might be
very hard to maintain in reality. Such global topology information
is needed in the CM method as well.

Thus, in order for a topology construction mechanism to be practical
in P2P networks, it must allow joining of new nodes by just using
locally available information. Of course, the cost of using only
local information is expected to be loss of scale-freeness (or any
other desired characteristics) of the whole overlay topology, which
will result in loss of search efficiency in return. In this section,
we present two practical methods using local heuristics not
necessarily using global information about the latest topology: HAPA
and DAPA.

\subsection{Hop-and-Attempt Preferential Attachment (HAPA)}

In this method, the new node randomly selects an existing node and
attempts to connect. Then it randomly selects a node which is a
neighbor of the previously selected node and attempts to connect.
Thus, the new node hops between the neighboring nodes and attempts
to connect by using the existing links in the network until it fills
all its stubs, i.e., the number of links it has reaches $m$. The
algorithm for HAPA model can be seen in Appendix~\ref{app_hapa}.

This hopping process gives a better chance to the new node to find
the high-degree hubs in the network than the PA does since the hubs
in scale-free networks are only a couple of hops away from the
low-degree nodes and it is less likely to find hubs by random node
selection. So, some nodes in the network (probably they are the
initial nodes and their number is $m+1$ due to network generation
algorithm) become dominant and attract almost all the nodes to
themselves, thus deserve the name \textit{super hubs}. The super
hubs have degrees on the order of network size. It is easily seen
that this procedure makes the topology of the system a star-like
topology if the network is not limited by a cutoff. Naturally,
without a hard cutoff the degree distribution is not a power-law and
the average shortest path/diameter is very small with respect to
scale-free networks generated by PA [see Fig.~\ref{fig_pk-hapa}(a)].
As shown in Figs.~\ref{fig_pk-hapa}(b) and \ref{fig_pk-hapa}(c),
when a hard cutoff is introduced the degree distribution gets closer
to a power law having an exponent $\gamma=3$ but with possibly
exponential factors making a degree exponent calculation very hard.

\begin{figure*}
\begin{center}
\begin{tabular}{ccc}
\includegraphics[keepaspectratio=true,angle=0,width=60mm]
{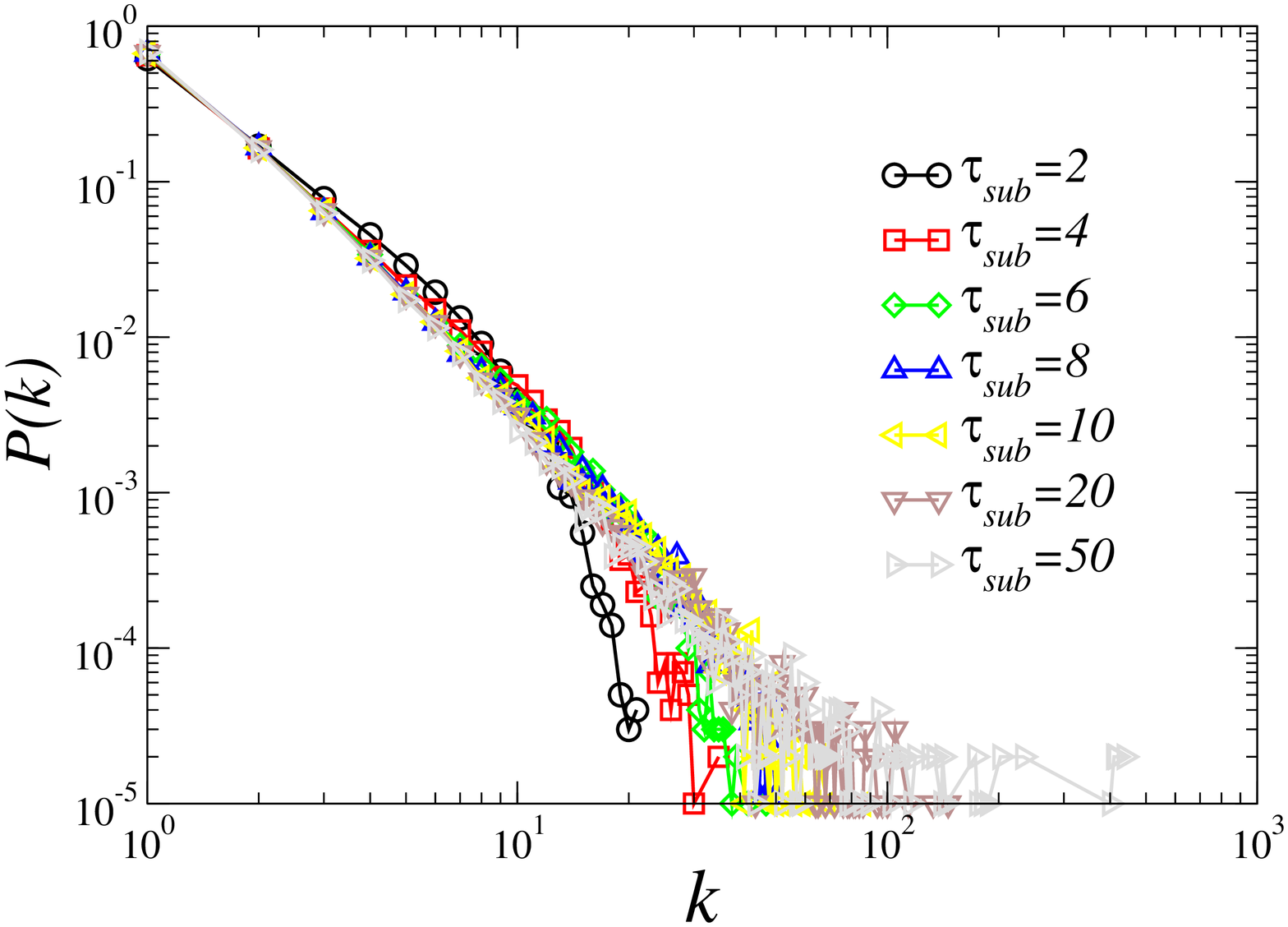} &
\hspace{-2mm}
\includegraphics[keepaspectratio=true,angle=0,width=60mm]
{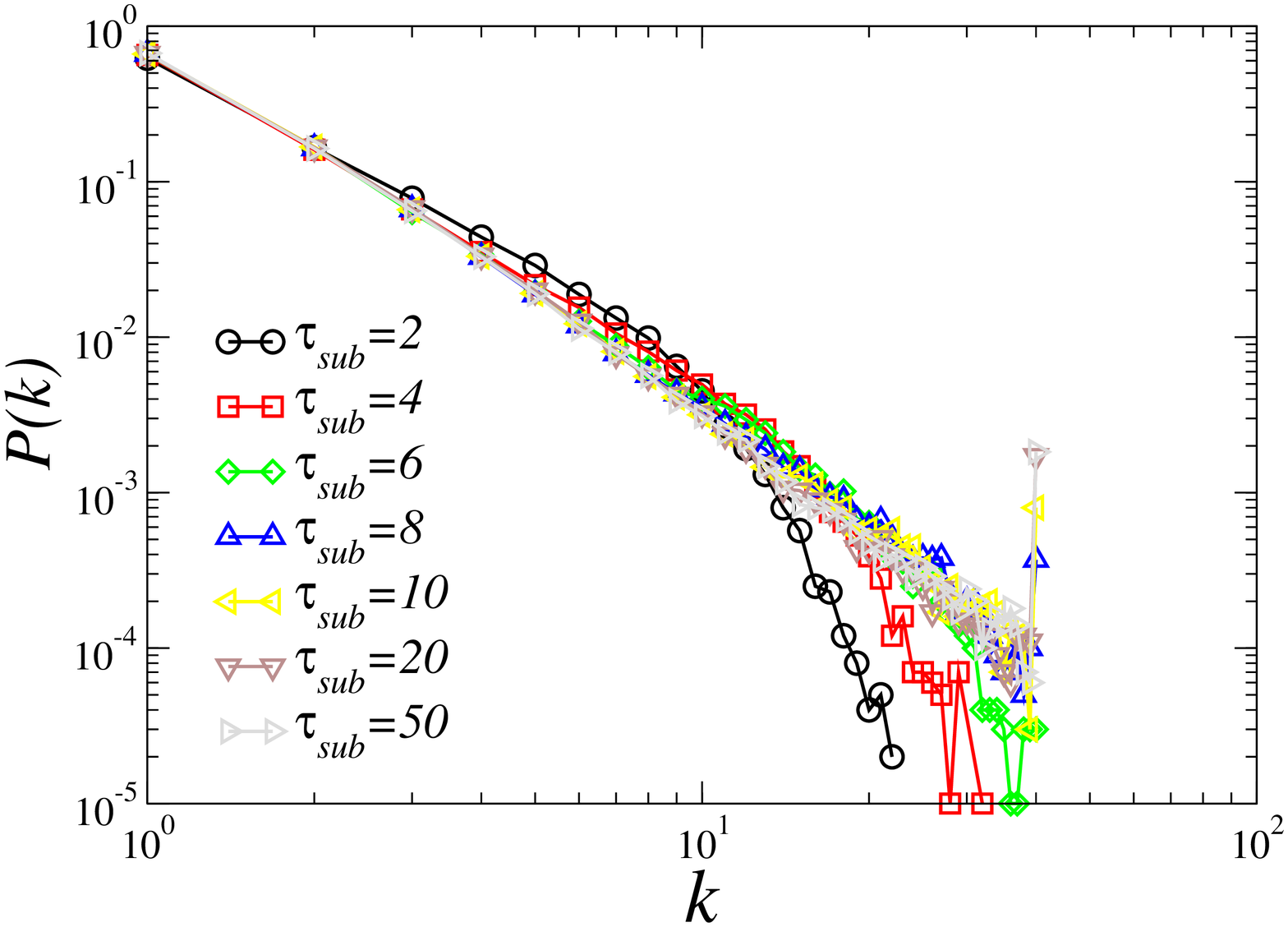} &
\hspace{-2mm}
\includegraphics[keepaspectratio=true,angle=0,width=60mm]
{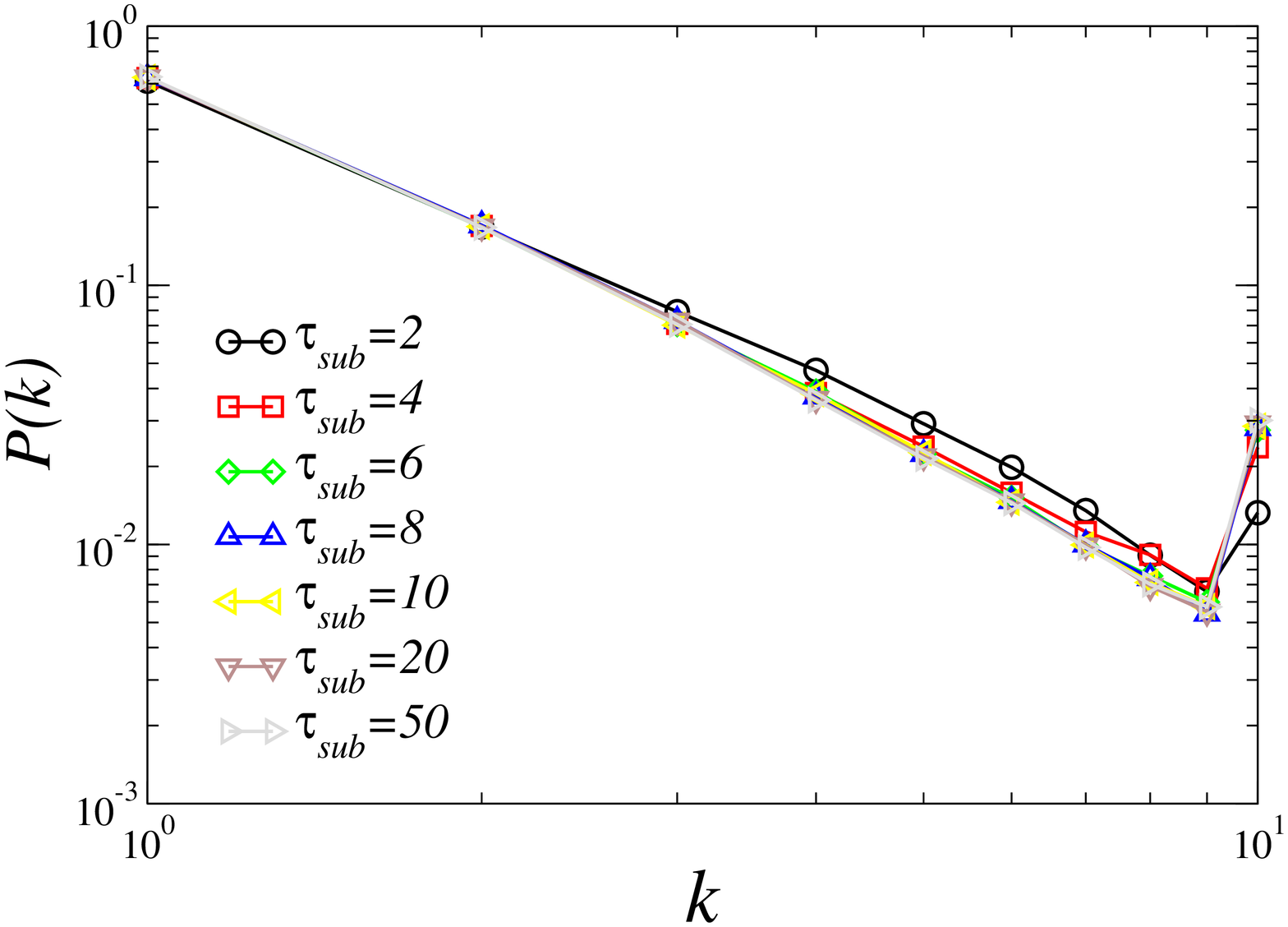}
\vspace{-5mm}
\\
\small{(a) $m=1$, no cutoff} & \small{(b) $m=1$, $k_c=40$} &
\small{(c) $m=1$, $k_c=10$}
\\
\includegraphics[keepaspectratio=true,angle=0,width=60mm]
{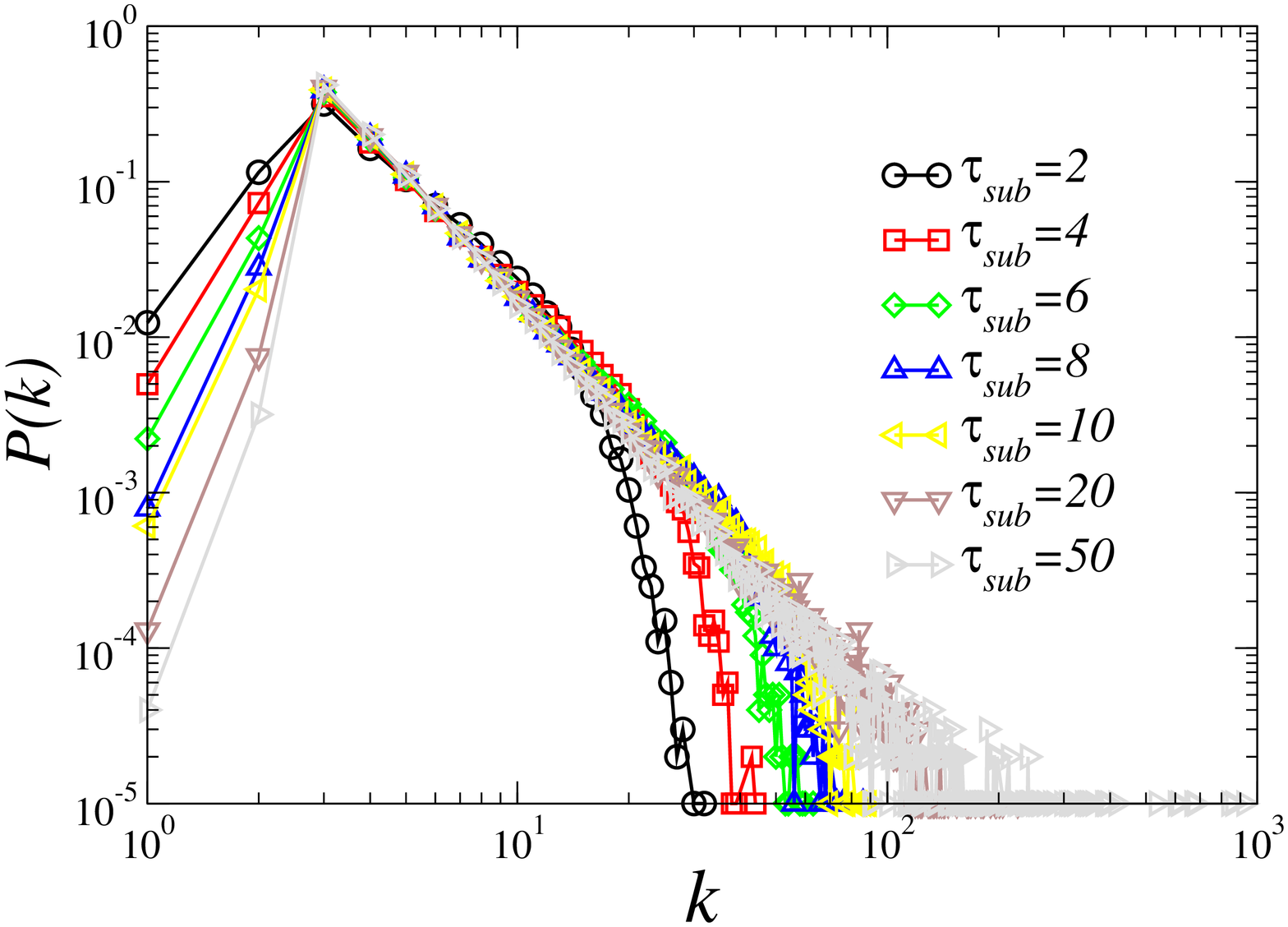} &
\hspace{-2mm}
\includegraphics[keepaspectratio=true,angle=0,width=60mm]
{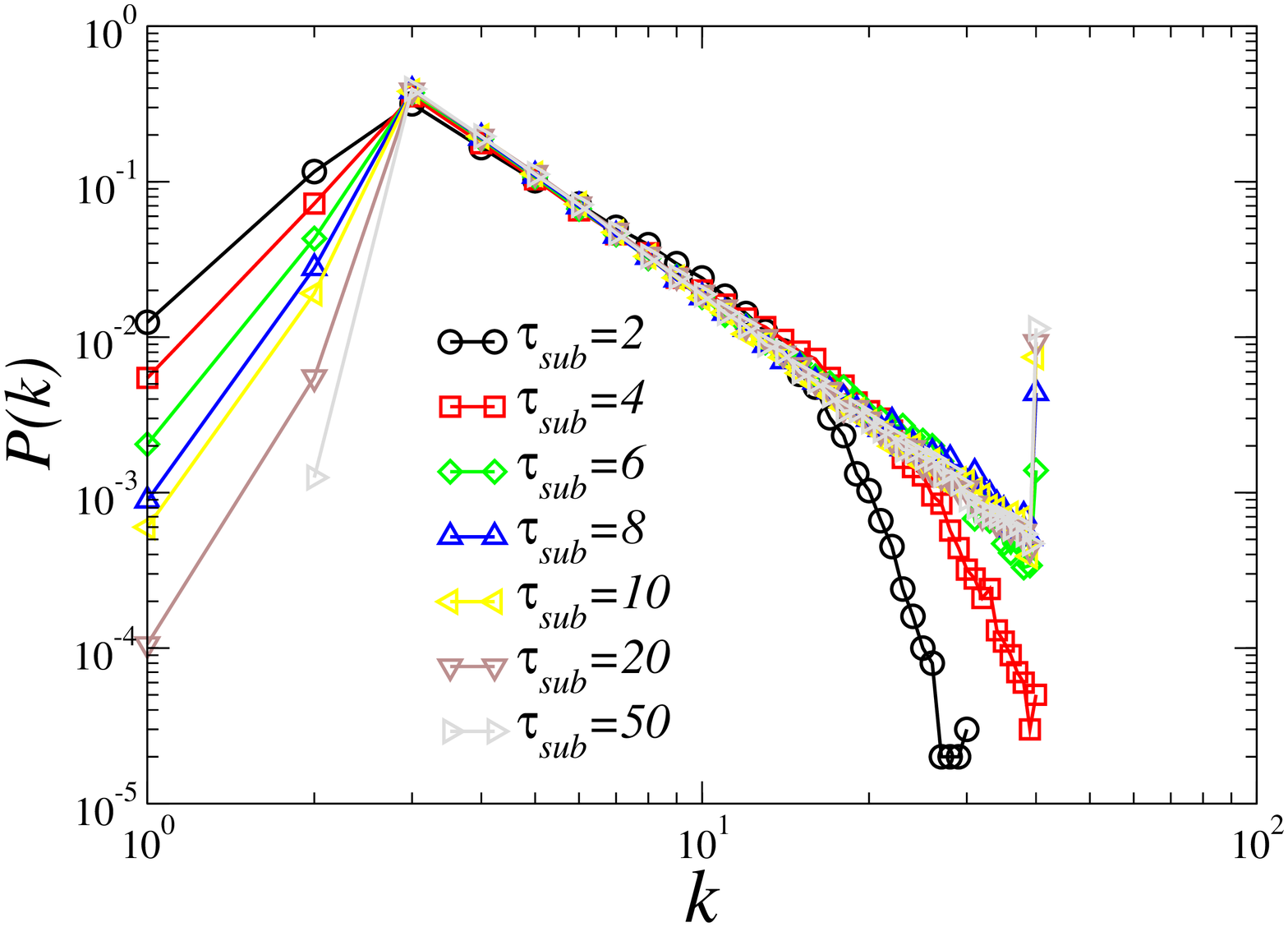} &
\hspace{-2mm}
\includegraphics[keepaspectratio=true,angle=0,width=60mm]
{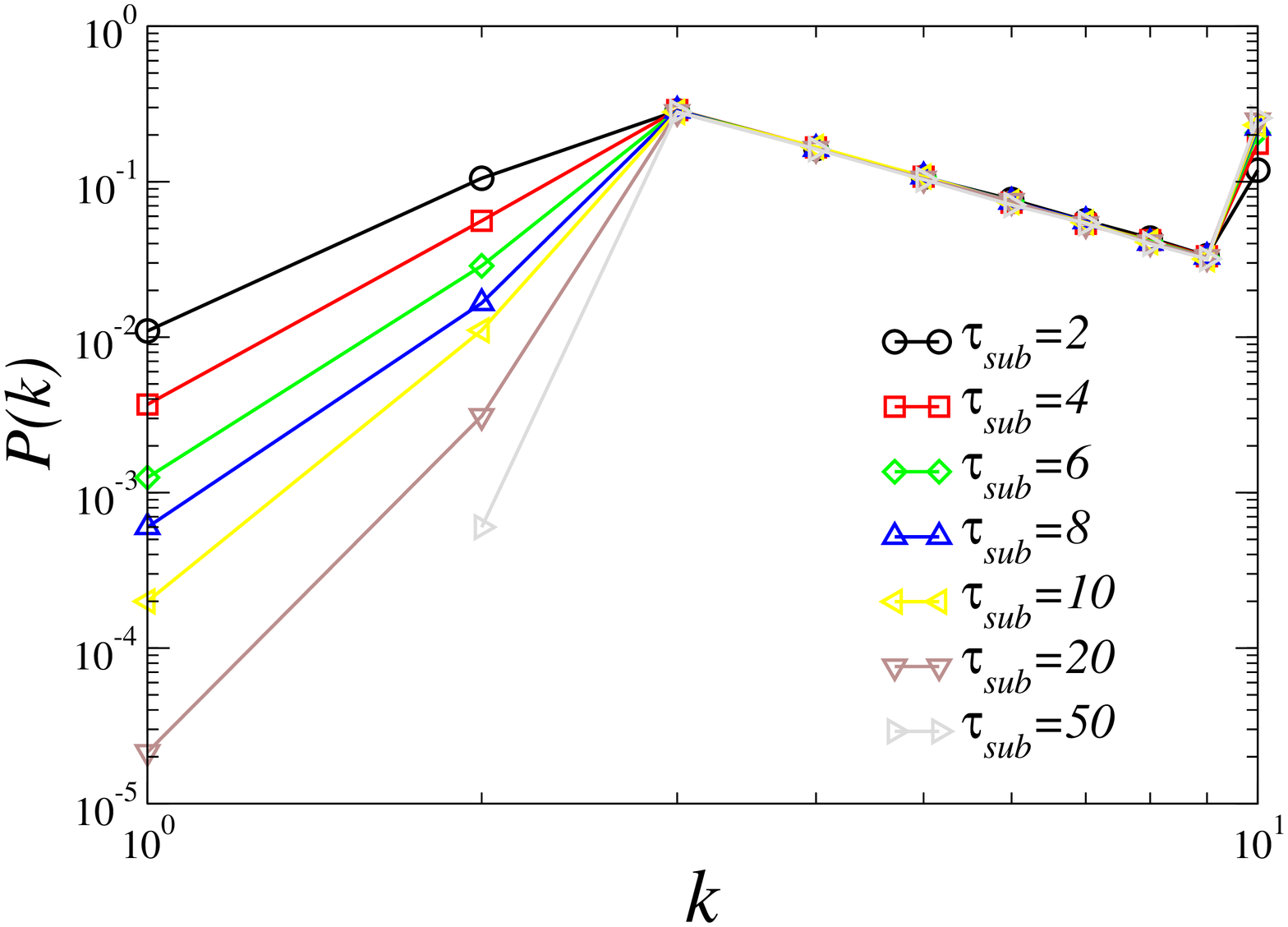}
\vspace{-5mm}
\\
\small{(d) $m=3$, no cutoff} & \small{(e) $m=3$, $k_c=40$} &
\small{(f) $m=3$, $k_c=10$}
\\
& \hspace{-2mm}
\includegraphics[keepaspectratio=true,angle=0,width=60mm]
{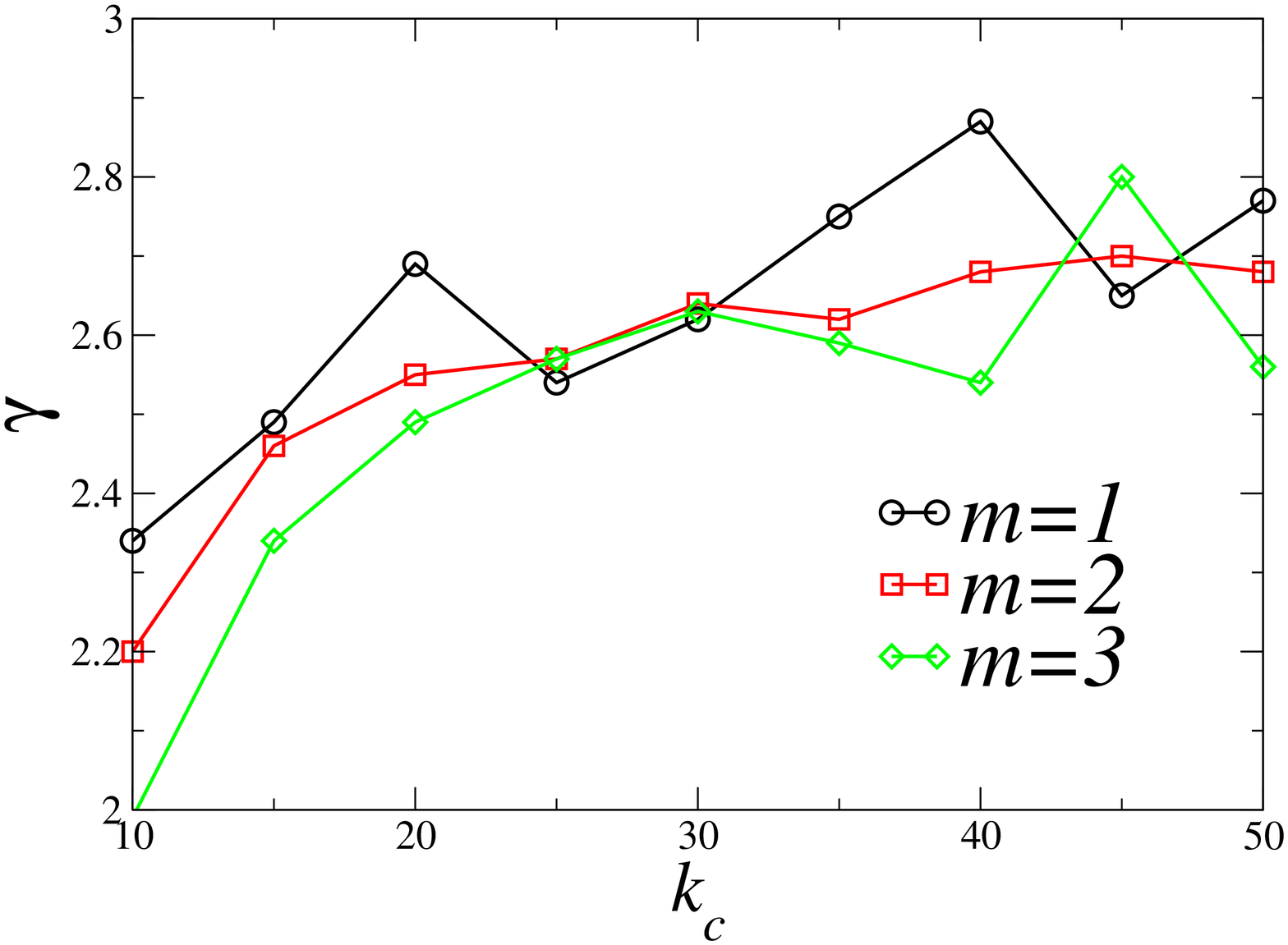} & \hspace{-2mm} \vspace{-5mm}
\\
& \small{(g) Degree exponent vs cutoff.} &
\end{tabular}
\end{center}
\caption{Degree distributions of DAPA model.} \label{fig_pk-dapa}
\end{figure*}

\subsection{Discover-and-Attempt Preferential Attachment (DAPA)}
\label{sec_dapa}

DAPA model imitates the method for finding peers in Gnutella-like
unstructured P2P networks. First, we assume that we have a network
called \textit{substrate} network with a predefined and
preconstructed topology at hand. Then, we construct an overlay
network on this substrate network by using the PA method among the
set of nodes visible/reachable to a specific node (the horizon of
the node) in a number of steps, which we call \emph{local
time-to-live} (or local TTL) and represent with $\tau_{sub}$.

DAPA model starts with a substrate network $G_S$ with $N_S$ nodes.
The topology for the substrate network may be a two-dimensional
regular network (mesh with nodes connected to four neighbors in four
different directions) or a geometric (Euclidean) random network
(GRN) \cite{DALL02} with a giant component. GRN is a random network
with a metric. It is constructed by assigning each node random
coordinates in a $d$-dimensional box of unit volume, i.e., each
coordinate is drawn from a uniform distribution on the unit
interval. Then, any two randomly placed nodes are linked if they are
closer to each other than some distance $R$. Two and three
dimensional GRNs have been widely used in continuum percolation and
real networks modeling (see references in \cite{DALL02}). GRNs have
Poissonian degree distributions in the form
$P(k)=e^{-\overline{k}}\overline{k}^k/k!$, where $\overline{k}$ is
the average degree. In 2D, if $R$$>$$R_{critical}$$=$$0.012$ for
$N$$=$$10^4$ corresponding to critical average degree
$\overline{k}$$=$$4.52$, the GRN has a giant component
\cite{DALL02}. Throughout the paper we use GRN as a substrate
network with an average degree $\overline{k}$$=$$10$ and size $N=2
\times 10^4$ because GRN is topologically closer to real life nodes
in the Internet than a regular or highly random network.

In this model, initially, a few nodes are randomly selected from the
substrate network and added to the previously empty overlay network,
$G_O$, then these nodes are connected to each other in $G_O$. At
each step one random node is chosen in the substrate network and let
it send a query to its neighborhood reachable in $\tau_{sub}$ hops
to get a list of peers in its horizon. Then by using the rules of
preferential attachment the new node connects to $m$ peers with
probability proportional to their degrees divided by the total
degrees of the peers in its horizon. If the number of peers in the
horizon is less than $m$, then the new node connects to peers it can
find. The nodes which can find at least one peer in their horizon is
added to the $G_O$ and becomes a peer. A peer belonging to $G_O$ can
not be selected again to look for new peers. This process is
continued until the number of peers in $G_O$ reaches the desired
number $N_O$. The algorithm for DAPA model is presented in
Appendix~\ref{app_dapa}.

The degree distribution of the network generated by DAPA model
exhibits some interesting characteristics. For small values of
$\tau_{sub}$, the nodes are shortsighted, i.e., they cannot see
enough peers in their short horizon causing the degree distribution
to be an exponential. For high enough $\tau_{sub}$ values, the
degree distribution changes into a power-law. Thus, one can go from
an exponential to a scale-free network by playing with the measure
of locality ($\tau_{sub}$), as can be seen in
Fig.~\ref{fig_pk-dapa}. As the hard cutoff gets smaller the
difference between the degree distributions becomes invisible. For
higher values of $m$ (i.e., $m$$>$$1$), it is possible to find peers
with degree less than $m$ as in Figs.~\ref{fig_pk-dapa}(d-f), since
some nodes cannot find enough peers in their horizon to fill all
their stubs. The degree distribution exponent has a similar behavior
to PA as we change the hard cutoff value, i.e., as the cutoff
decreases the exponent increases [see Fig.~\ref{fig_pk-dapa}(g)].
The data in Fig.~\ref{fig_pk-dapa}(g) is very noisy and the data
points contain quite large error bars because they are obtained from
very scattered degree distribution tails.

A comparison of different network generation models in terms of
locality can be seen in Table \ref{tab:pa}. When a peer is to join
the current overlay topology, the PA and CM do need global
information about the current topology whilst HAPA and DAPA methods
use local information partially or fully, respectively. Therefore,
HAPA and DAPA methods are more practical in the context of
unstructured P2P networks.

\renewcommand{\baselinestretch}{1}
\begin{table}
    \caption{Comparison of Different Network Generation Procedures}
    \label{tab:pa}
    \begin{center}
\vspace{-5mm}
    \begin{tabular}{|c|c|}
    \hline
        Procedure & Usage of Global Information \\
    \hline
        PA & Yes \\
        CM & Yes \\
        HAPA & Partial \\
        DAPA & No \\
    \hline
    \end{tabular}
\vspace{-5mm}
    \end{center}
\end{table}
\renewcommand{\baselinestretch}{1.75}

\subsection{Effect of Hard Cutoffs on Topological Characteristics}

Depending on the way one applies hard cutoffs to an initially
scale-free topology results in different topological
characteristics, such as the degree distribution, diameter (or
expected search efficiency). Table~\ref{tab:diameter-behavior}
summarizes the inter-relationship between these three dimensions
that PA models were studied in literature.

When we applied hard cutoffs to the regular PA topologies, their
degree distribution looked like a power-law distribution except that
a major jump happens in the frequency of nodes having degree equal
to the hard cutoff [see Fig.~\ref{fig_pk-pa}(b)]. Unlike the
original PA topologies without any cutoff [see
Fig.~\ref{fig_pk-pa}(a)], these topologies exhibit different
power-law exponents when the jump on the hard cutoffs is taken into
account. We measured the estimated power-law exponent for these PA
topologies with hard cutoffs and plotted Fig.~\ref{fig_pk-pa}(c),
which shows the power-law exponent of the degree distribution with
respect to the hard cutoff that was applied to the original PA
topology. As expected, Fig.~\ref{fig_pk-pa}(c) shows that the degree
distribution exponent $\gamma$ degrades to lower values when harder
cutoffs are applied, suggesting that search efficiency (in
connection with the diameter size) will also degrade for smaller
cutoffs.

The CM does not allow changes in the degree distribution exponent
because the degree sequence is drawn from a predefined distribution
generated by using a specific degree exponent [see
Fig.~\ref{fig_pk-mr}]. The only change in degree distribution
exponent is due to the deletion of self loops and multiple
connections and can be considered negligible. It is also observed
that applying harder (smaller) cutoffs to the degrees decreases the
probability to have self loops and multiple connections.

In HAPA model, it is not even possible to say that we still have
power-law degree distributions. Without a hard cutoff the degree
distribution decreases very fast as the degree increases and there
are a few nodes with degree on the order of system size, i.e.,
star-like topology [see Fig.~\ref{fig_pk-hapa}(a)]. Applying a
cutoff destroys the star-like topology and changes the degree
distribution into a one similar to power-law with exponential cutoff
dependent correction as can be seen in Figs.~\ref{fig_pk-hapa}(b,c).

DAPA model is qualitatively very similar to PA for high enough
$\tau_{sub}$ values [see Fig.~\ref{fig_pk-dapa}]. The small
$\tau_{sub}$ makes the network an exponential one. By tuning this
parameter, one can change the degree distribution from exponential
to power-law. As in the PA model, applying harder cutoff increases
the degree distribution exponent as can be seen from
Fig.~\ref{fig_pk-dapa}(g).

\begin{figure}
\centering
\includegraphics[keepaspectratio=true,angle=0,width=88mm]{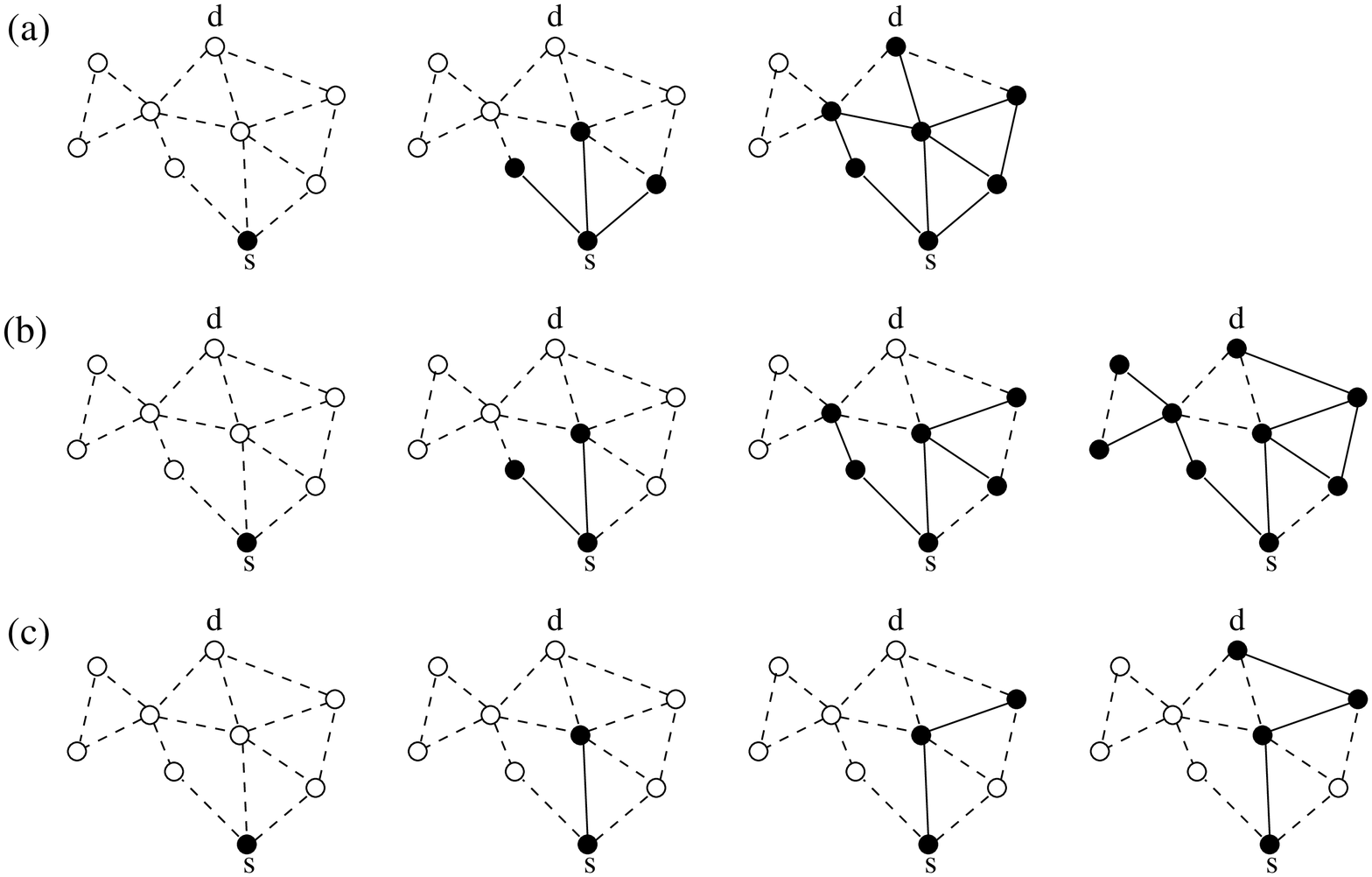}
\vspace{-4mm} \caption{Search strategies to find the destination
node d, starting from the source node s. (a) Flooding (FL) or
broadcast search with $\tau$$=$$2$. (b) Normalized flooding (NF)
search with $\tau$$=$$3$. (c) Random walk (RW) search with
$\tau$$=$$3$. } \label{fig_search}
\end{figure}

\section{Simulations}
\label{sec:simulations}

In P2P networks that do not have a central server including Gnutella
and Freenet files are found by forwarding queries to one's neighbors
until the target is found. In the previous sections, in addition to
studying well-known techniques like Preferential Attachment (PA) and
Configuration Model (CM) for scale-free topology construction, we
introduced new algorithms (i.e. HAPA and DAPA) with the same purpose
within the context of unstructured P2P networks. Here, we study a
number of message-passing algorithms that can be efficiently used to
search items in P2P networks utilizing the power-law (the presence
of hubs) degree distribution in sample networks generated by our
topology construction algorithms. These algorithms are completely
decentralized and do not use any kind of global knowledge on the
network. We consider three different search algorithms:
\emph{flooding} (FL), \emph{normalized flooding} (NF), and
\emph{random walk} (RW). Goals of our simulation experiments
include:
\begin{itemize}
\item{\emph{Effect of hard cutoffs on search efficiency:}} Applying hard
cutoffs on power-law topologies reduces the degree distribution
exponent, which should affect the search efficiency (i.e.
\emph{number of hits per unit time}) on such topologies. We are
interested in observing this effect for the three search algorithms
on the topologies constructed by our topology construction
algorithms.

\item{\emph{Topology construction with global vs. local
information:}} Though we showed in the previous section that using
local information when a peer is joining yields a less scale-free
topology, the effect of this on search efficiency still needs to be
shed light on. Our simulations aim to investigate this too.

\item{\emph{Messaging complexity:}} One side-effect of changing
topology characteristics is that it will affect the messaging
complexity (i.e. \emph{number of messages per search request}) of
the search algorithms. We would like to observe this effect as well.
\end{itemize}

\subsection{Search Algorithms}

\begin{figure}
\centering \subfigure[FL results for PA model.]{\label{fig_fl-pa}
\includegraphics[keepaspectratio=true,angle=0,width=70mm]
{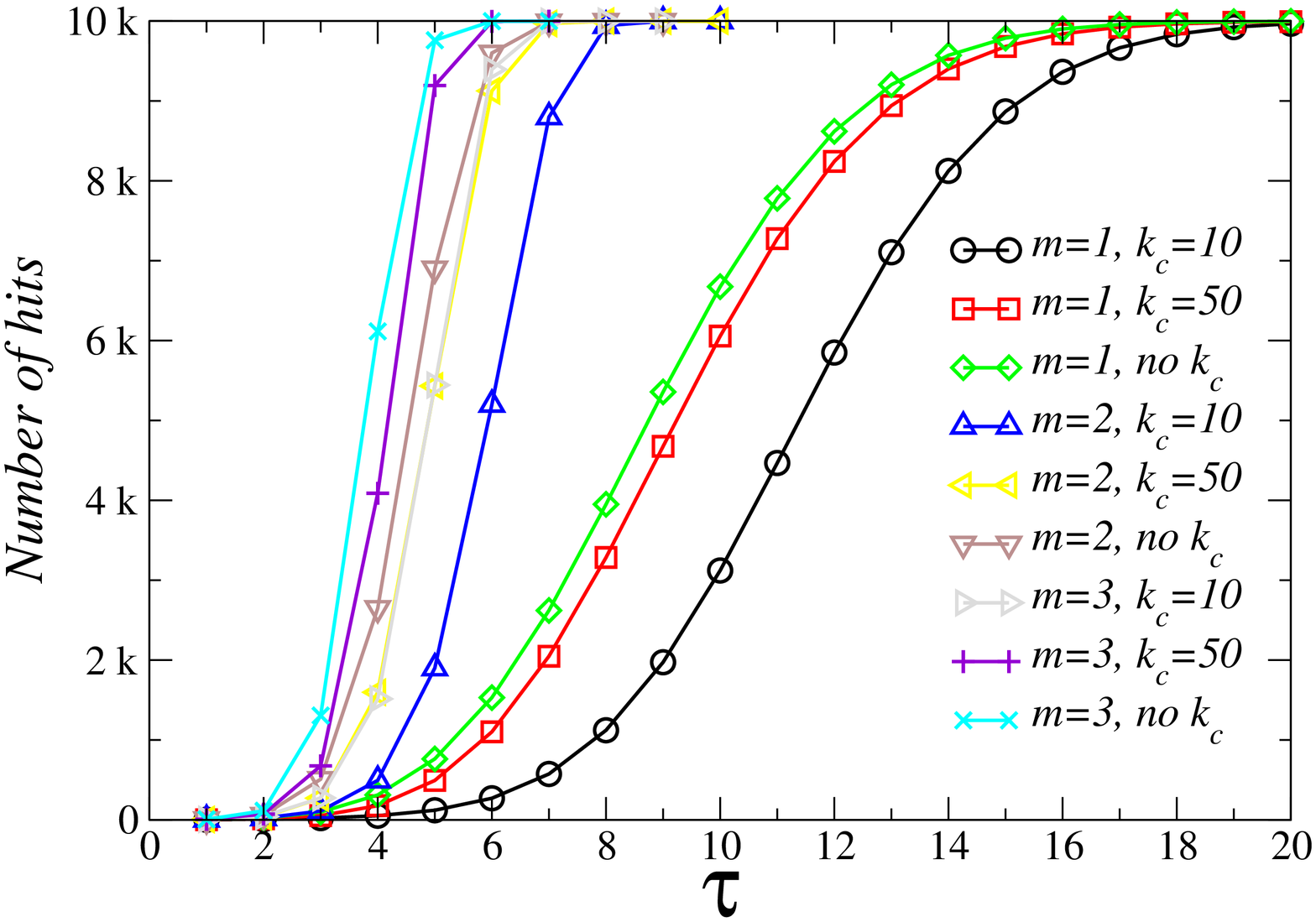}}
\hspace{5mm} \subfigure[FL results for HAPA model.]{
\label{fig_fl-hapa}
\includegraphics[keepaspectratio=true,angle=0,width=70mm]
{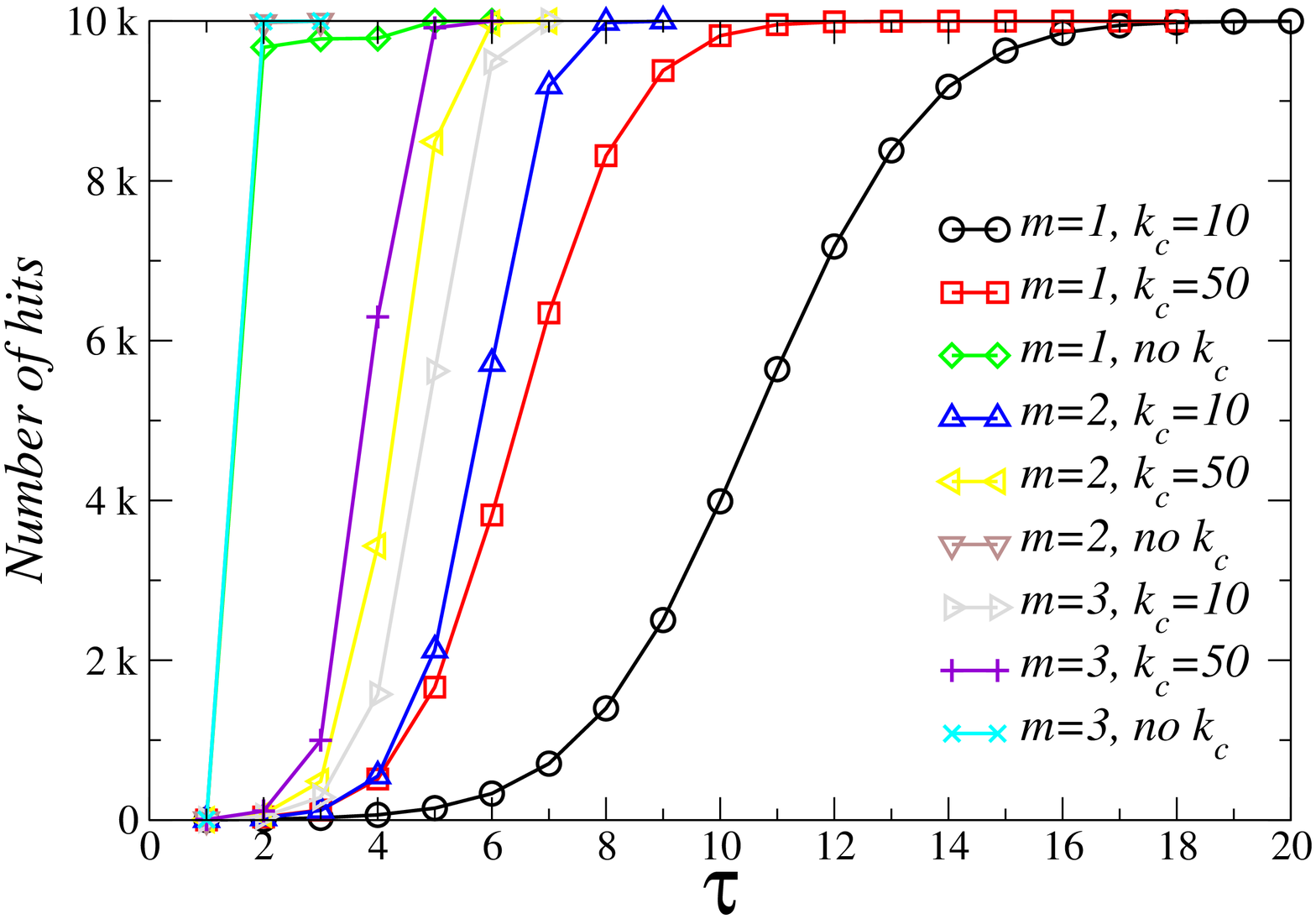}}
\caption{FL results for PA and HAPA models.}
\end{figure}

\subsubsection{Flooding (FL)}

FL is the most common search algorithm in unstructured P2P networks.
In search by FL, the source node $s$, sends a message to all its
nearest neighbors. If the neighbors do not have the requested item,
they send on to their nearest neighbors excluding the source node
[see Fig.~\ref{fig_search}(a)]. This process is repeated a certain
number of times, which is usually called as \textit{time-to-live
(TTL)} and we represent it with $\tau$ in this paper. After a
message is forwarded an amount of time equal to $\tau$, it is
discarded. Independent floods by the nodes make the FL algorithm
parallel. On the other hand, in this algorithm a large number of
messages is created since the destination node cannot stop the
search. This corresponds to a complete sweep of all the nodes within
a $\tau$ hop distance from the source. The delivery time in search
by FL is measured of intermediate links traversed, and is equal to
the shortest path length. Since the average shortest path for
small-world networks, including scale-free ones generated by the PA
model, is proportional to the logarithm of system size $N$ or even
slower, the average delivery time ($T_N$) is logarithmic as well,
i.e.,
\begin{equation}
T_N = \log(N).
\end{equation}

The main disadvantage with FL is that it requires a very large
amount of messaging traffic because most of the nodes are visited
and forced to exchange messages, which makes search by FL
unscalable. Another disadvantage is that FL has poor granularity,
i.e., each additional step in the search significantly increases the
number of nodes visited \cite{LCCLS-2002}. Yet, search efficiency of
FL (i.e., number of hits per search) provides a way of determining
how other realistic and scalable search algorithms can perform in
comparison to the best possible, i.e., the search efficiency of FL.

\begin{figure*}
\begin{center}
\begin{tabular}{ccc}
\includegraphics[keepaspectratio=true,angle=0,width=60mm]
{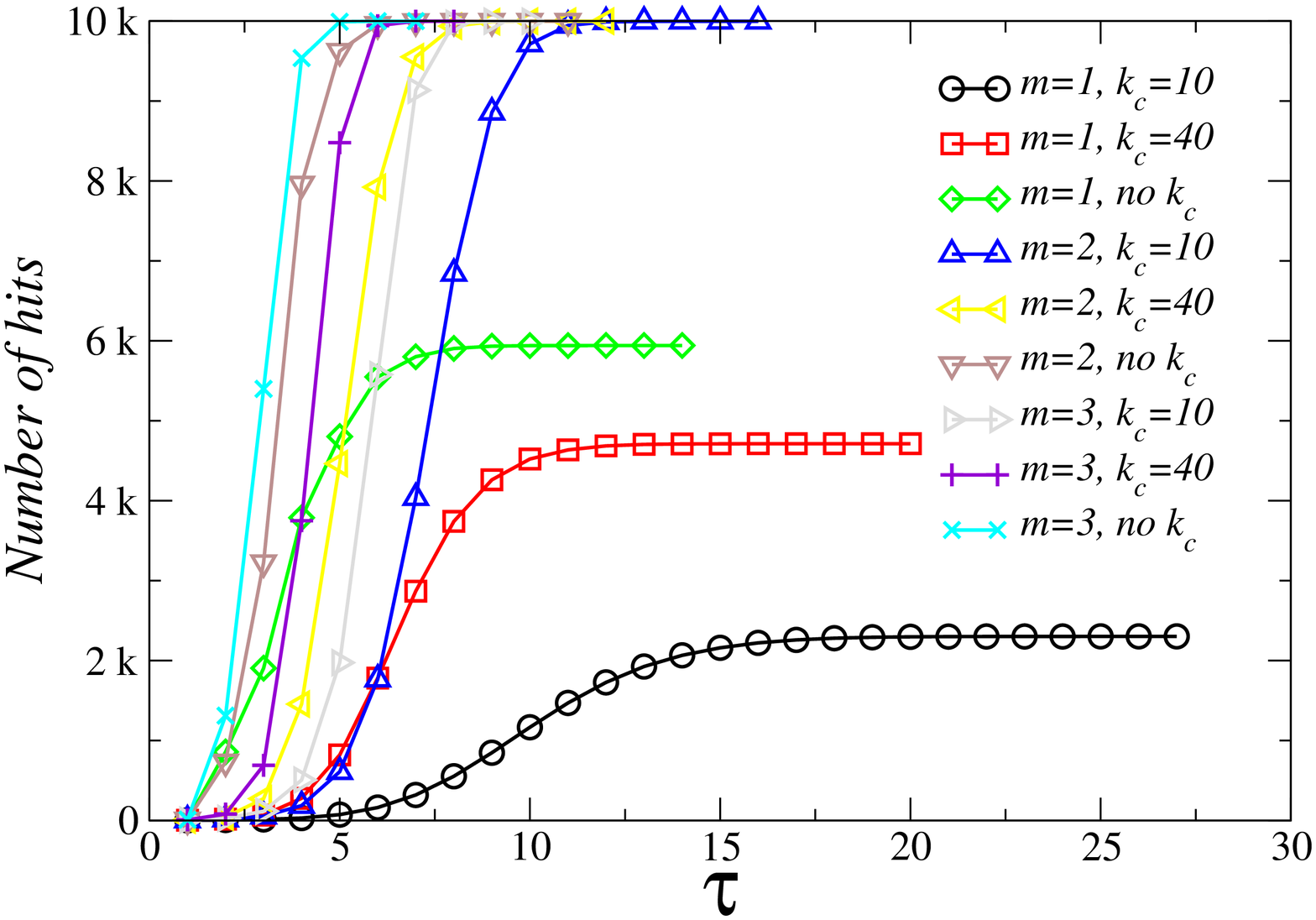} &
\hspace{-2mm}
\includegraphics[keepaspectratio=true,angle=0,width=60mm]
{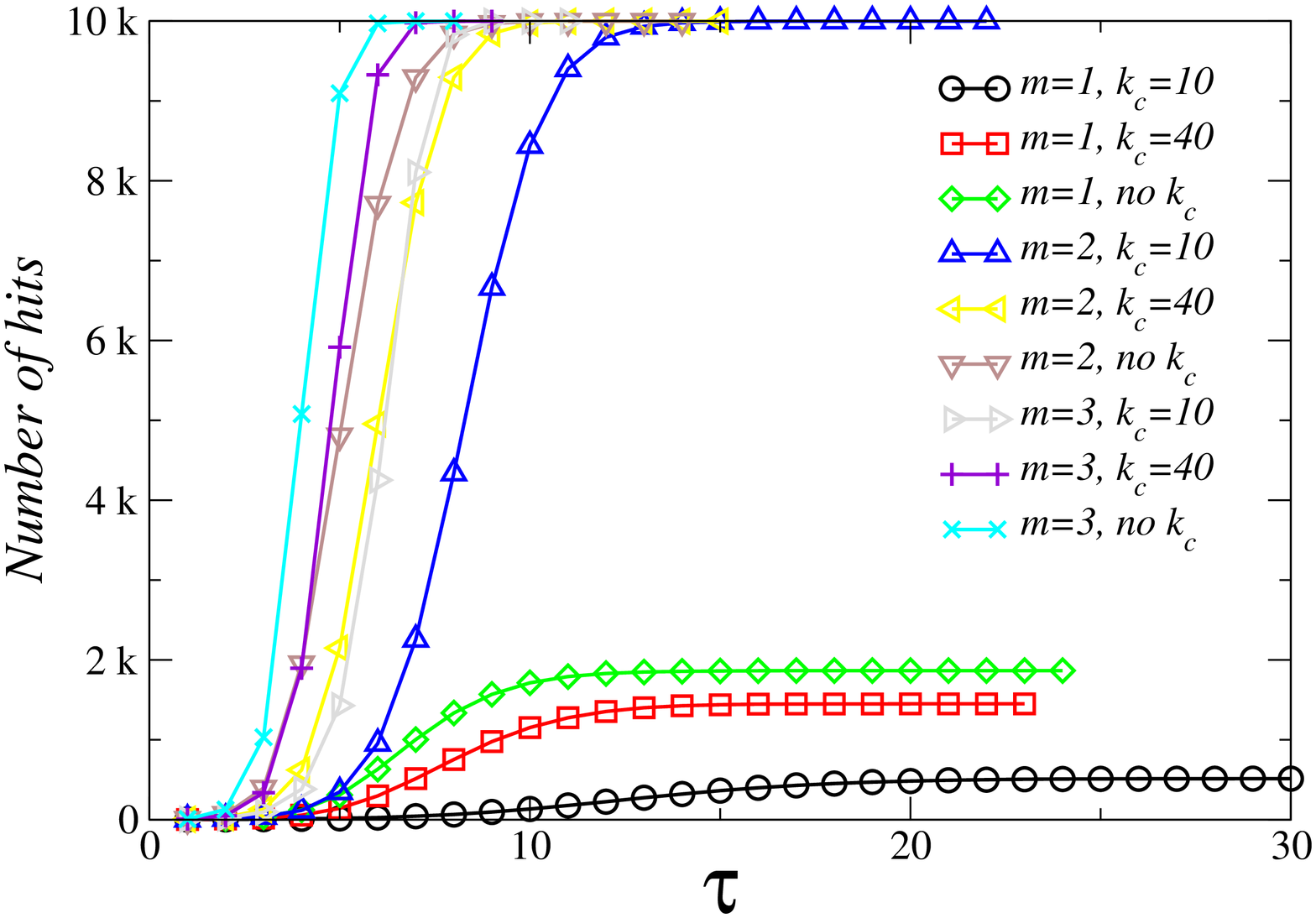} &
\hspace{-2mm}
\includegraphics[keepaspectratio=true,angle=0,width=60mm]
{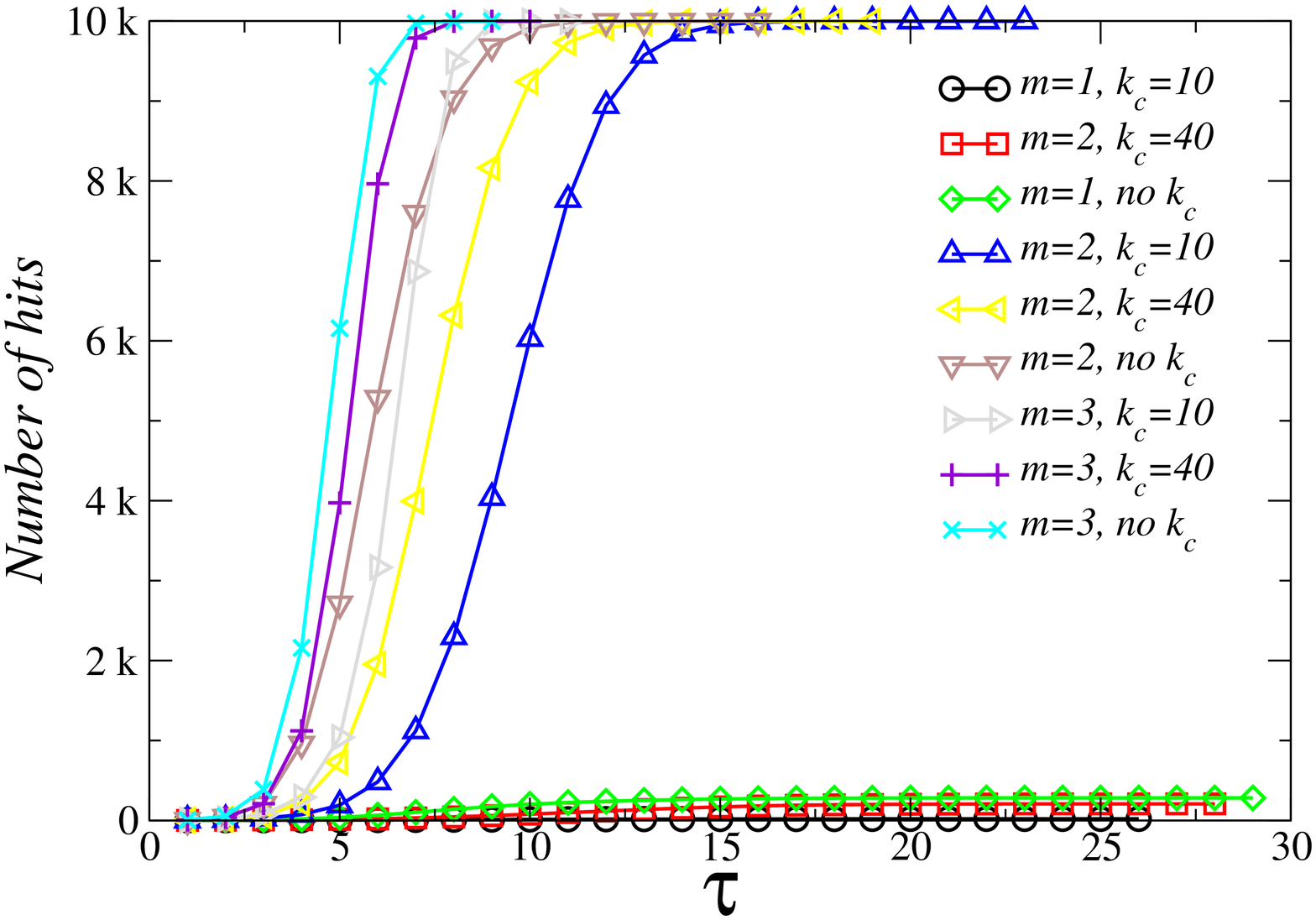}
\vspace{-5mm}
\\
\small{(a) $\gamma=2.2$} & \small{(b) $\gamma=2.6$} & \small{(c) $\gamma=3$}
\end{tabular}
\end{center}
\caption{FL results for the CM.} \label{fig_fl-mr}
\end{figure*}

\begin{figure*}
\begin{center}
\begin{tabular}{ccc}
\includegraphics[keepaspectratio=true,angle=0,width=60mm]
{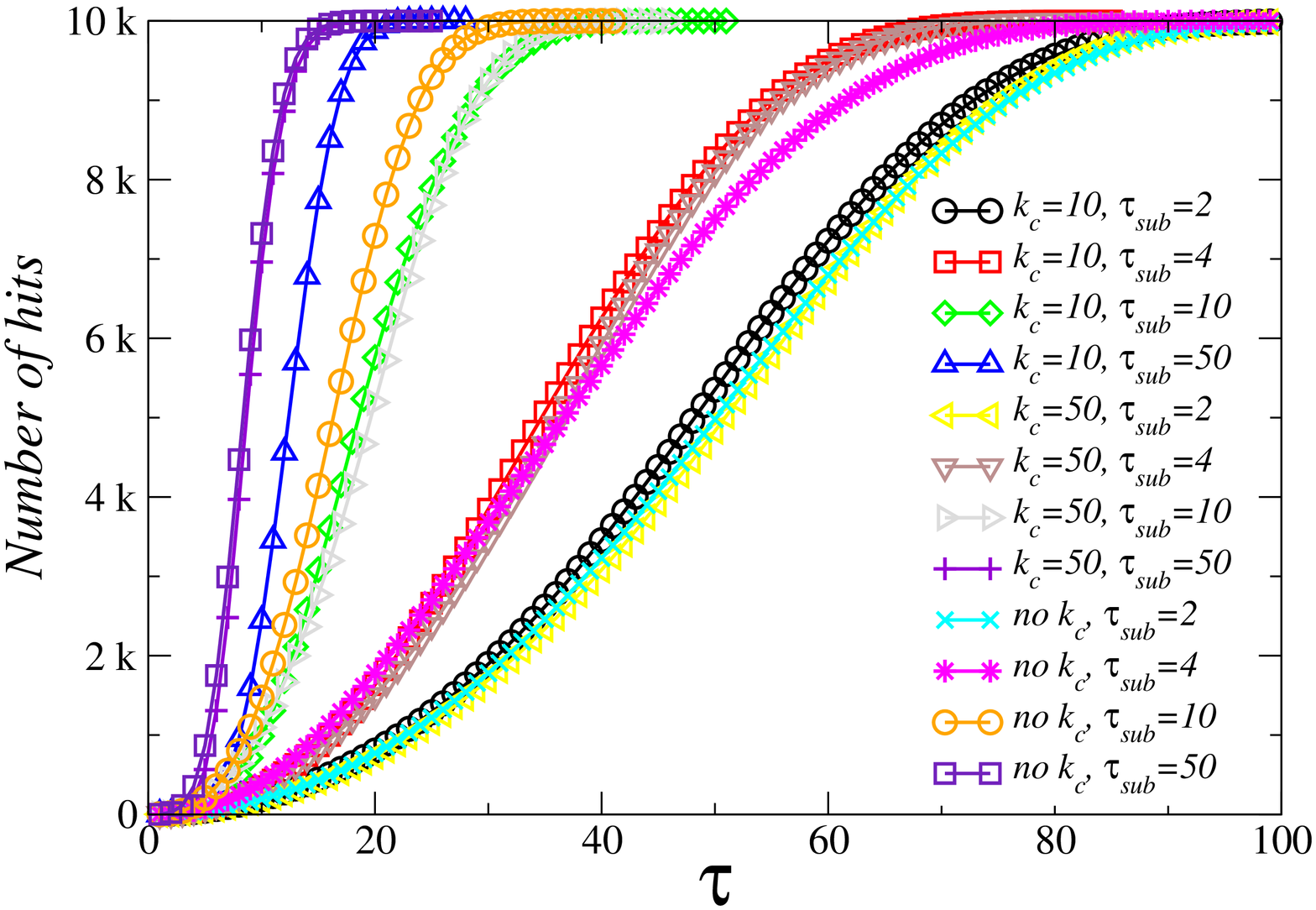} &
\hspace{-2mm}
\includegraphics[keepaspectratio=true,angle=0,width=60mm]
{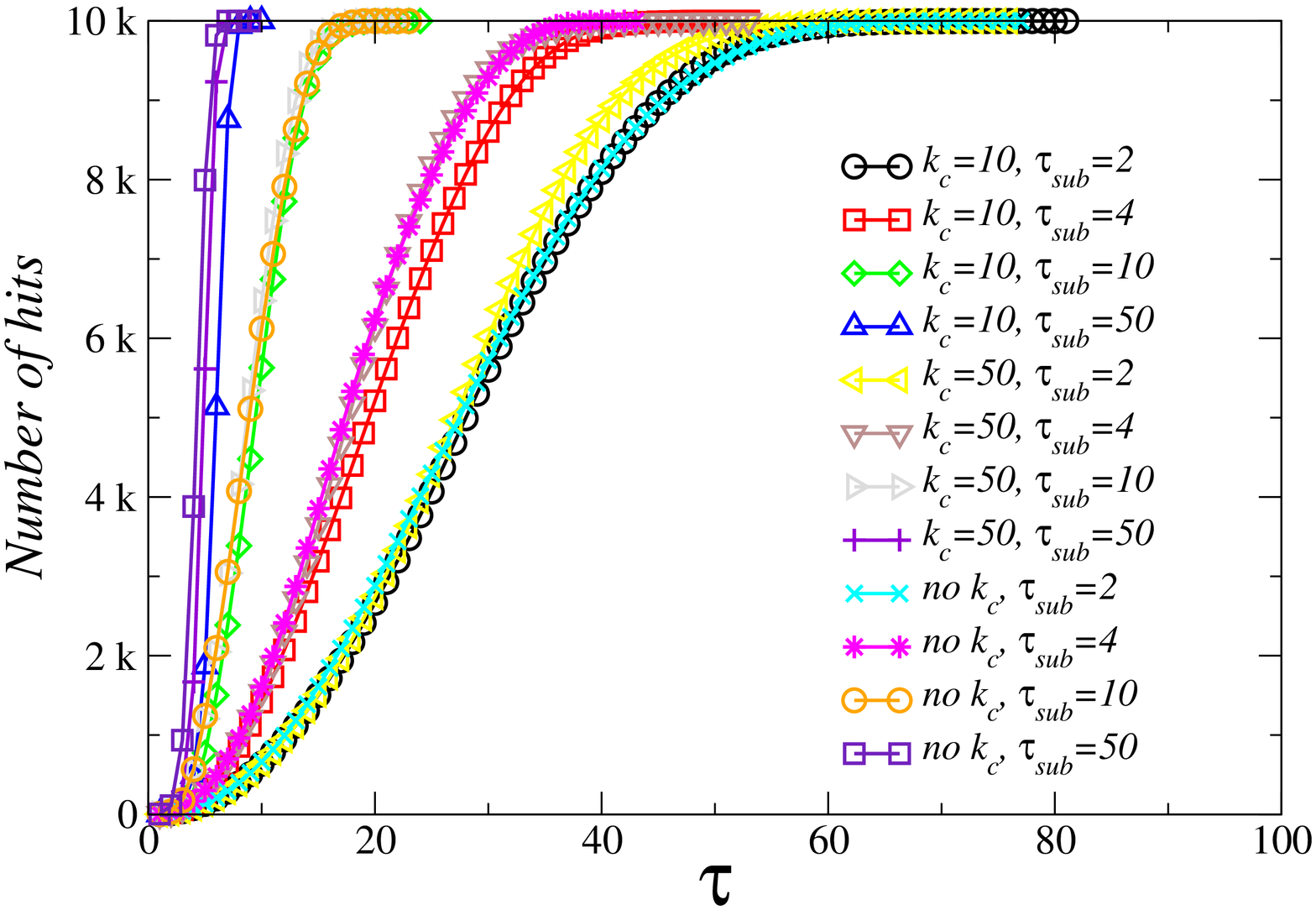} &
\hspace{-2mm}
\includegraphics[keepaspectratio=true,angle=0,width=60mm]
{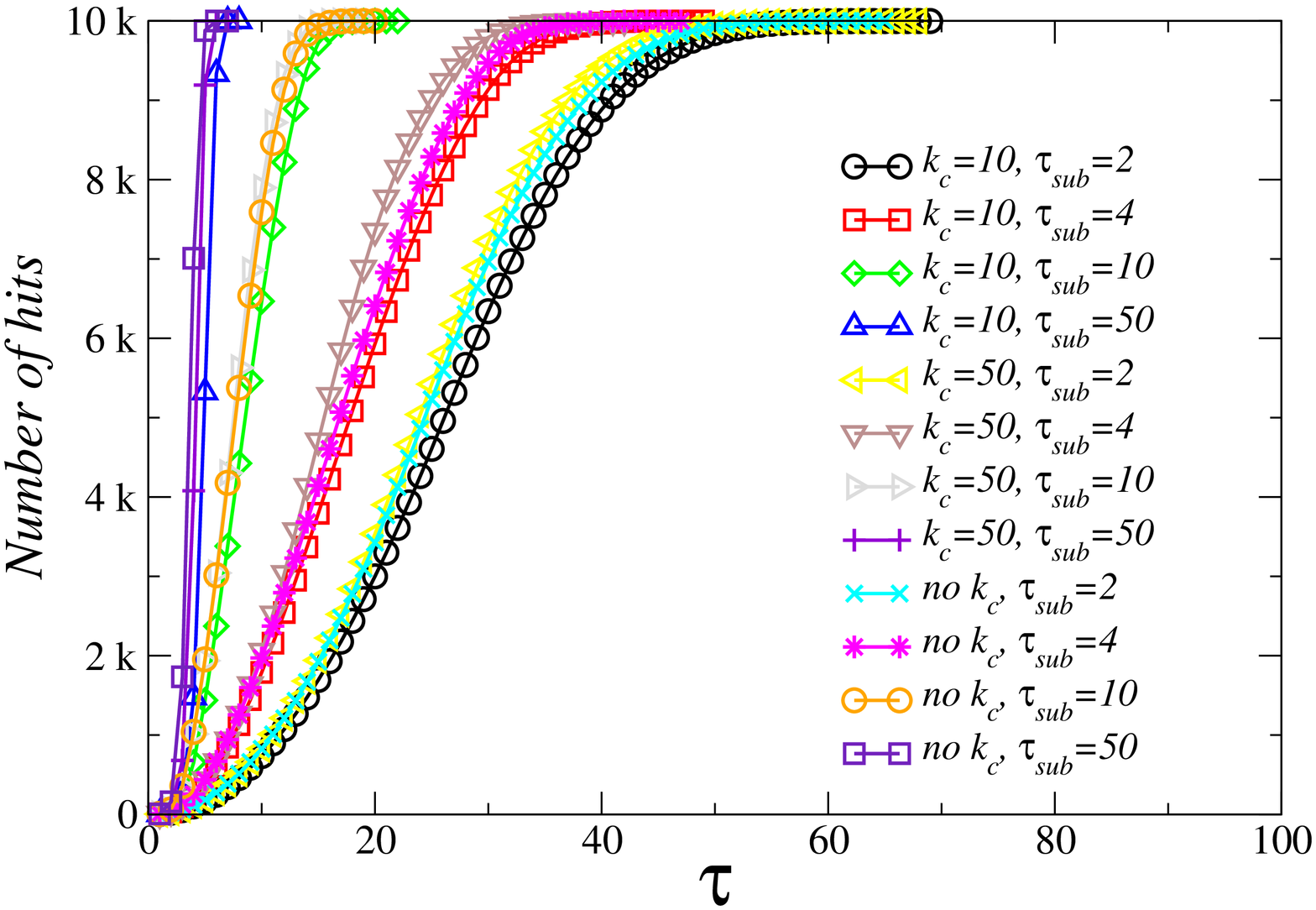}
\vspace{-5mm}
\\
\small{(a) $m=1$} & \small{(b) $m=2$} & \small{(c) $m=3$}
\end{tabular}
\end{center}
\caption{FL results for the DAPA model.} \label{fig_fl-dapa}
\end{figure*}

\subsubsection{Normalized Flooding (NF)}

In search by FL, when large degree nodes (hubs) are reached, the
number of neighbors for the next step in FL increases dramatically
leading to a poor granularity. This also causes a lot of shared
edges reducing the performance in terms of number of messages per
distinct number of discovered nodes. To overcome this problem,
search by NF algorithm was introduced in \cite{GMS05}. In NF, the
minimum degree in the network $k_{min}$ is an important factor. NF
search algorithm proceeds as follows: When a node of degree
$k_{min}$ receives a message, the node forwards the message to all
of its neighbors excluding the node forwarded the message in the
previous step. When a node with larger degree receives the message,
it forwards the message only to randomly chosen $k_{min}$ of its
neighbors except the one which forwarded the message. The NF
procedure is illustrated in Fig.~\ref{fig_search}(b). In this simple
network with $k_{min}$$=$$2$, the source node sends a message to its
randomly chosen two neighbors and these neighbors forward the
message to their randomly chosen two neighbors. In the third step,
the message reaches its destination.

The NF search algorithm is based on the minimum degree in the
network. The fixed minimum degree is equal to $m$ by definition in
PA and HAPA, whereas in CM an DAPA it is not guaranteed that the
minimum degree will be $m$. In CM, deletion of self-loops and
multiple links reduce the minimum degree to values less than $m$
down to 1. In DAPA, however, the minimum degree might be less than
$m$ because of the short range of horizon for some peers which are
geographically far from other peers. But still since the ratio of
nodes with degree less than $m$ is small we ignored them and ran NF
algorithm based on the predefined minimum degree value $m$.

\begin{figure*}
\begin{center}
\begin{tabular}{ccc}
\includegraphics[keepaspectratio=true,angle=0,width=60mm]
{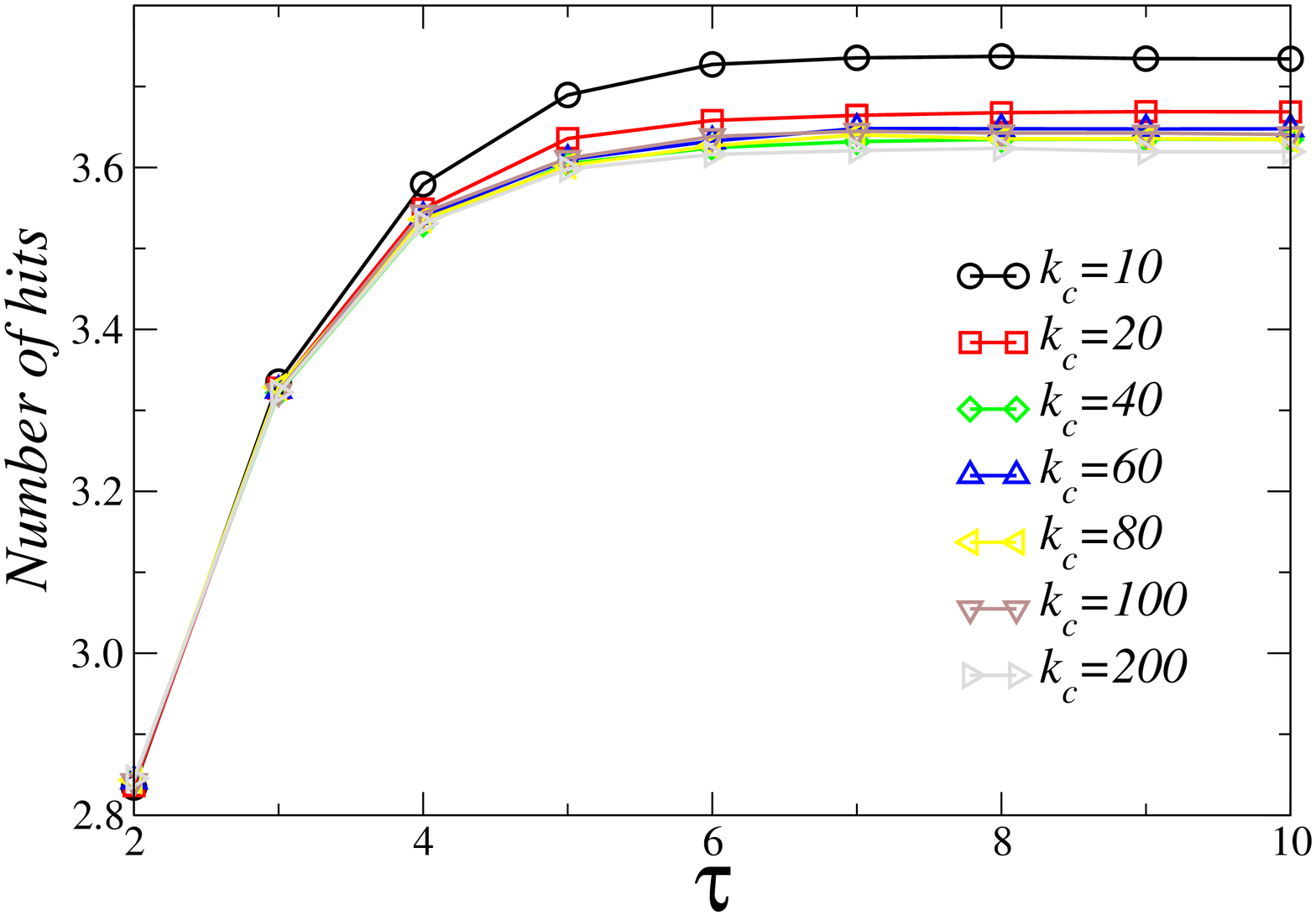} &
\hspace{-2mm}
\includegraphics[keepaspectratio=true,angle=0,width=60mm]
{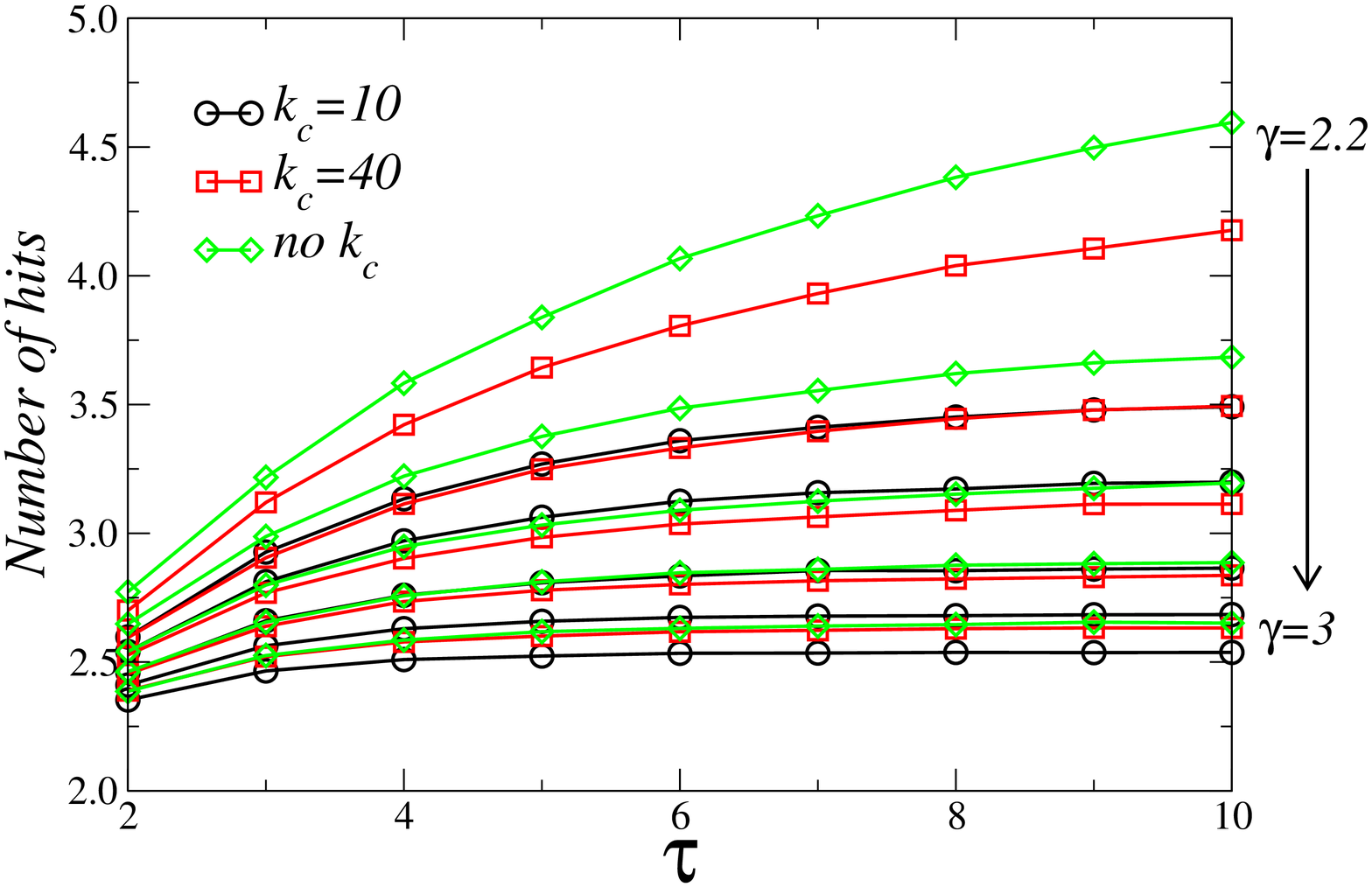} &
\hspace{-2mm}
\includegraphics[keepaspectratio=true,angle=0,width=60mm]
{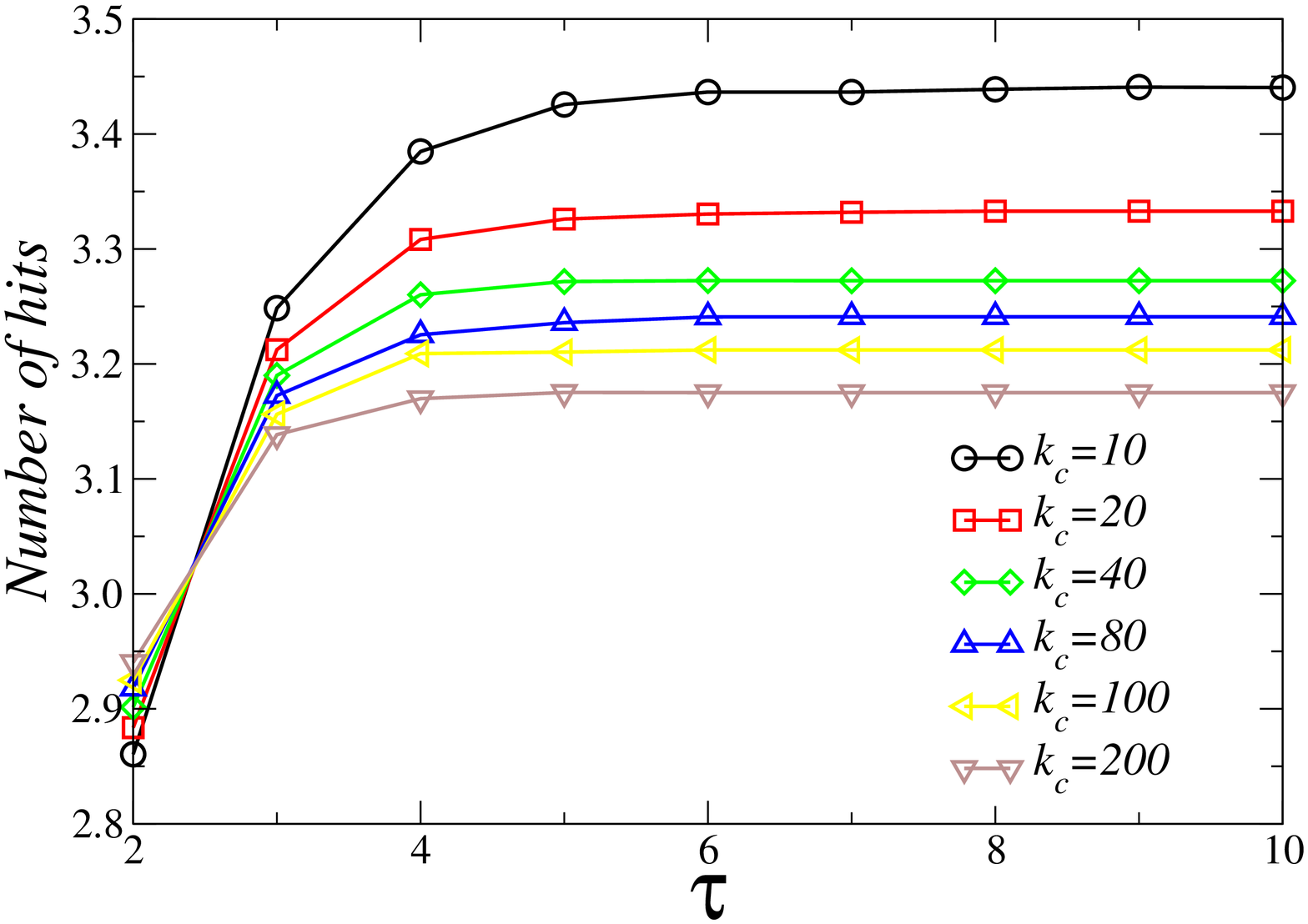}
\vspace{-5mm}
\\
\small{(a) $m=1$} & \small{(b) $m=1$} & \small{(c) $m=1$}
\\
\includegraphics[keepaspectratio=true,angle=0,width=60mm]
{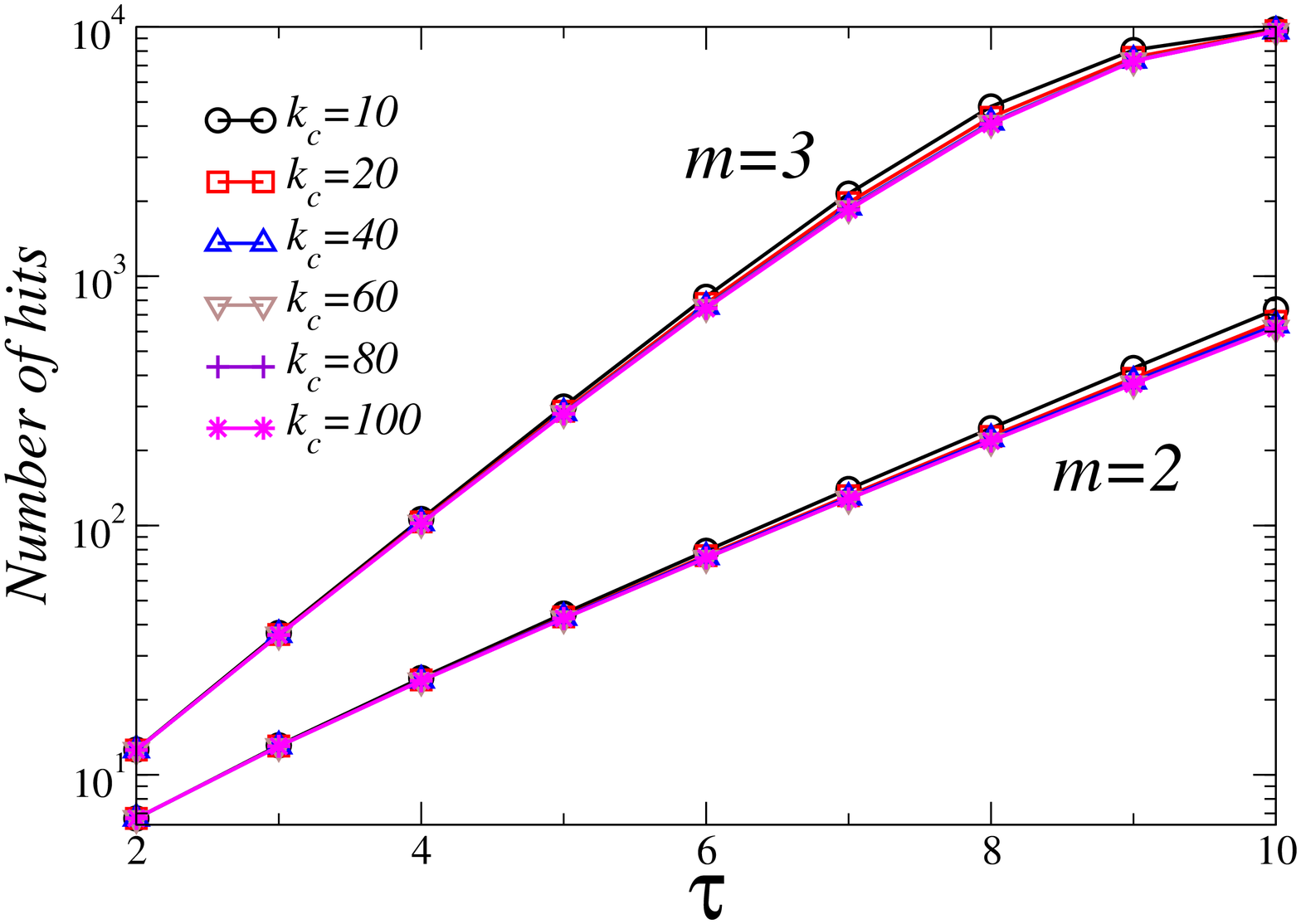} &
\hspace{-2mm}
\includegraphics[keepaspectratio=true,angle=0,width=60mm]
{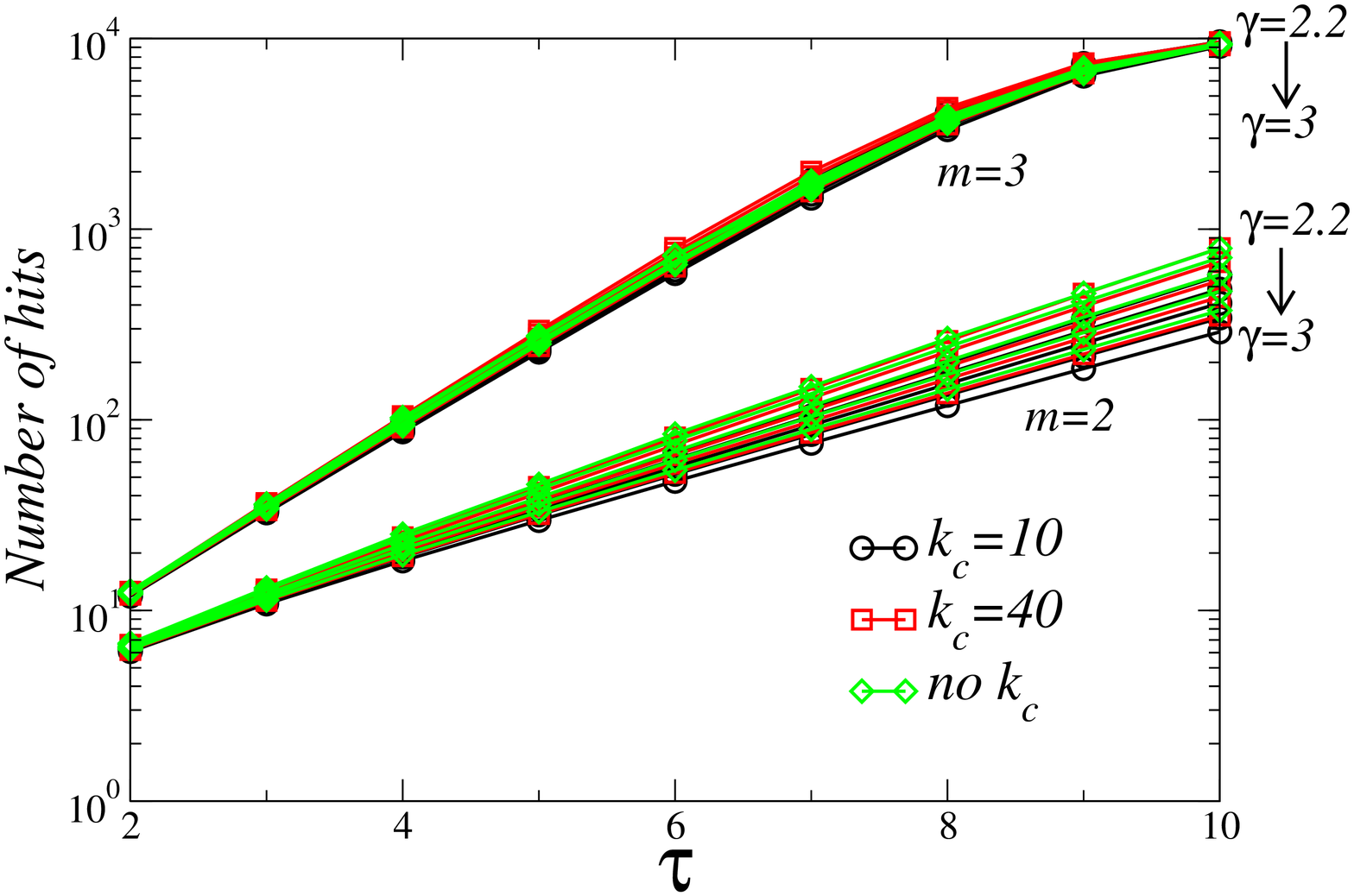} &
\hspace{-2mm}
\includegraphics[keepaspectratio=true,angle=0,width=60mm]
{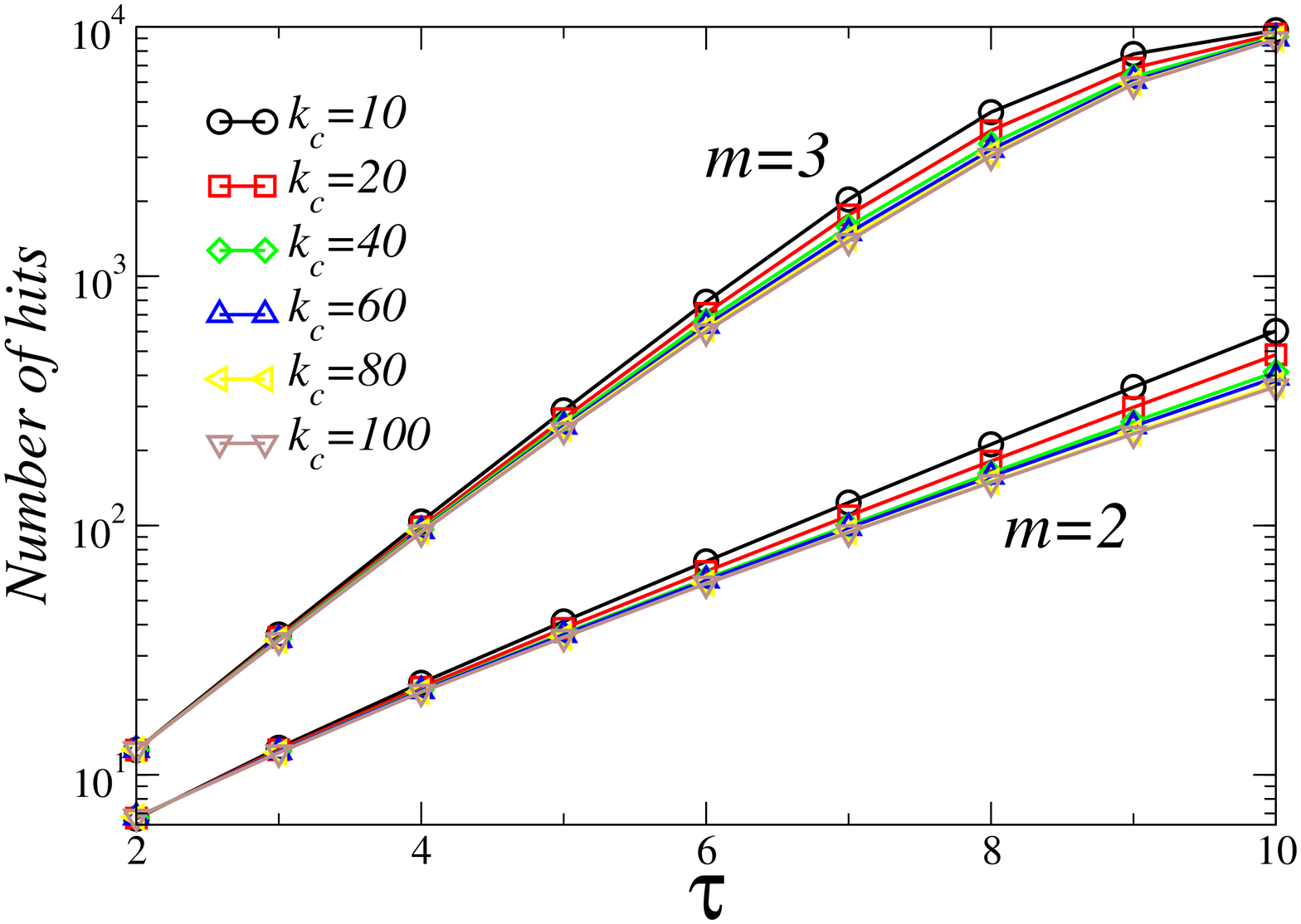}
\vspace{-5mm}
\\
\small{(d) $m=2$ and $m=3$} & \small{(e) $m=2$ and $m=3$} &
\small{(f) $m=2$ and $m=3$}
\\
\small{PA model} & \small{CM} & \small{HAPA model}
\end{tabular}
\end{center}
\caption{NF results for PA, CM, and HAPA models.} \label{fig_nf}
\end{figure*}

\begin{figure*}
\begin{center}
\begin{tabular}{ccc}
\includegraphics[keepaspectratio=true,angle=0,width=60mm]
{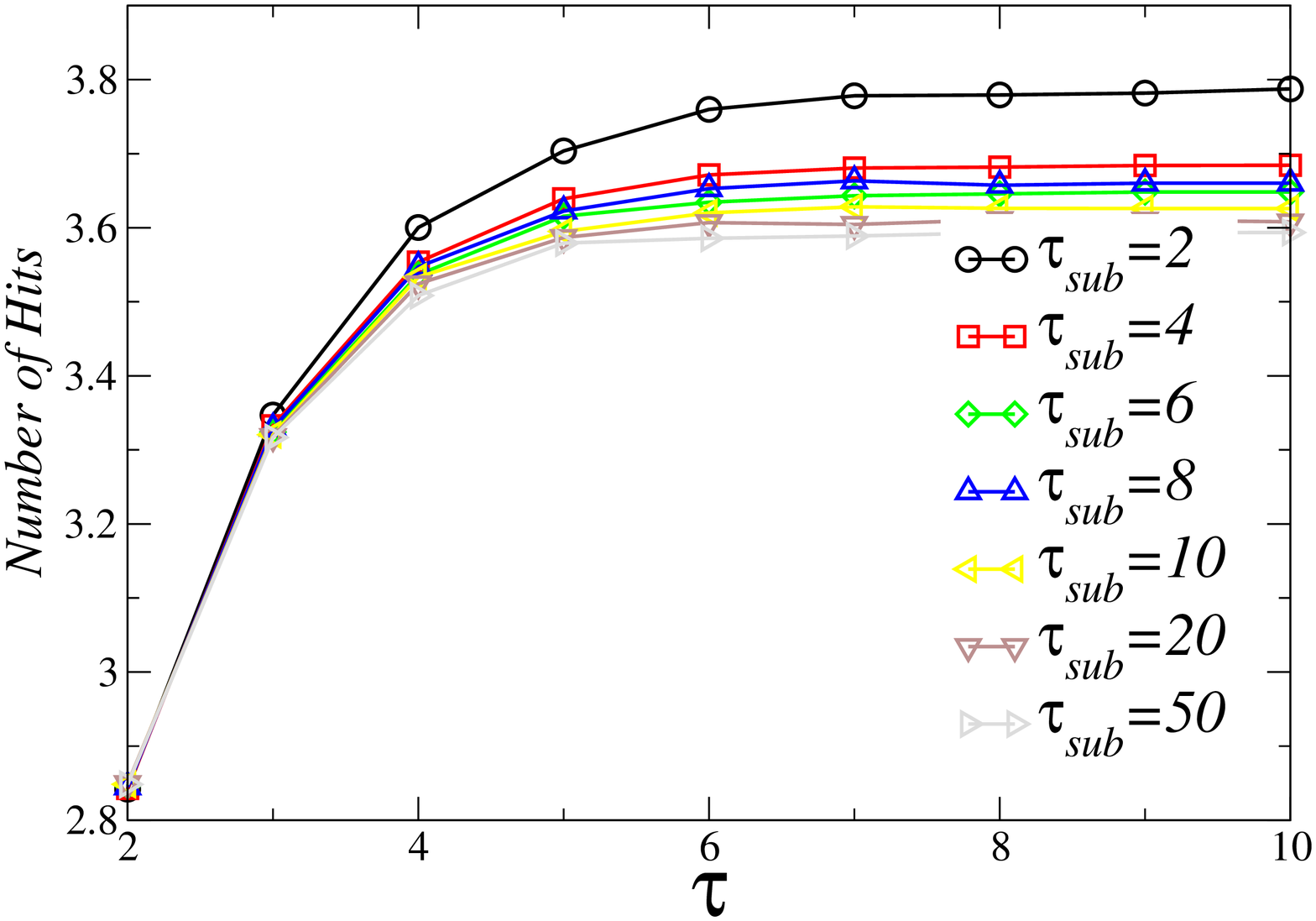}
& \hspace{-2mm}
\includegraphics[keepaspectratio=true,angle=0,width=60mm]
{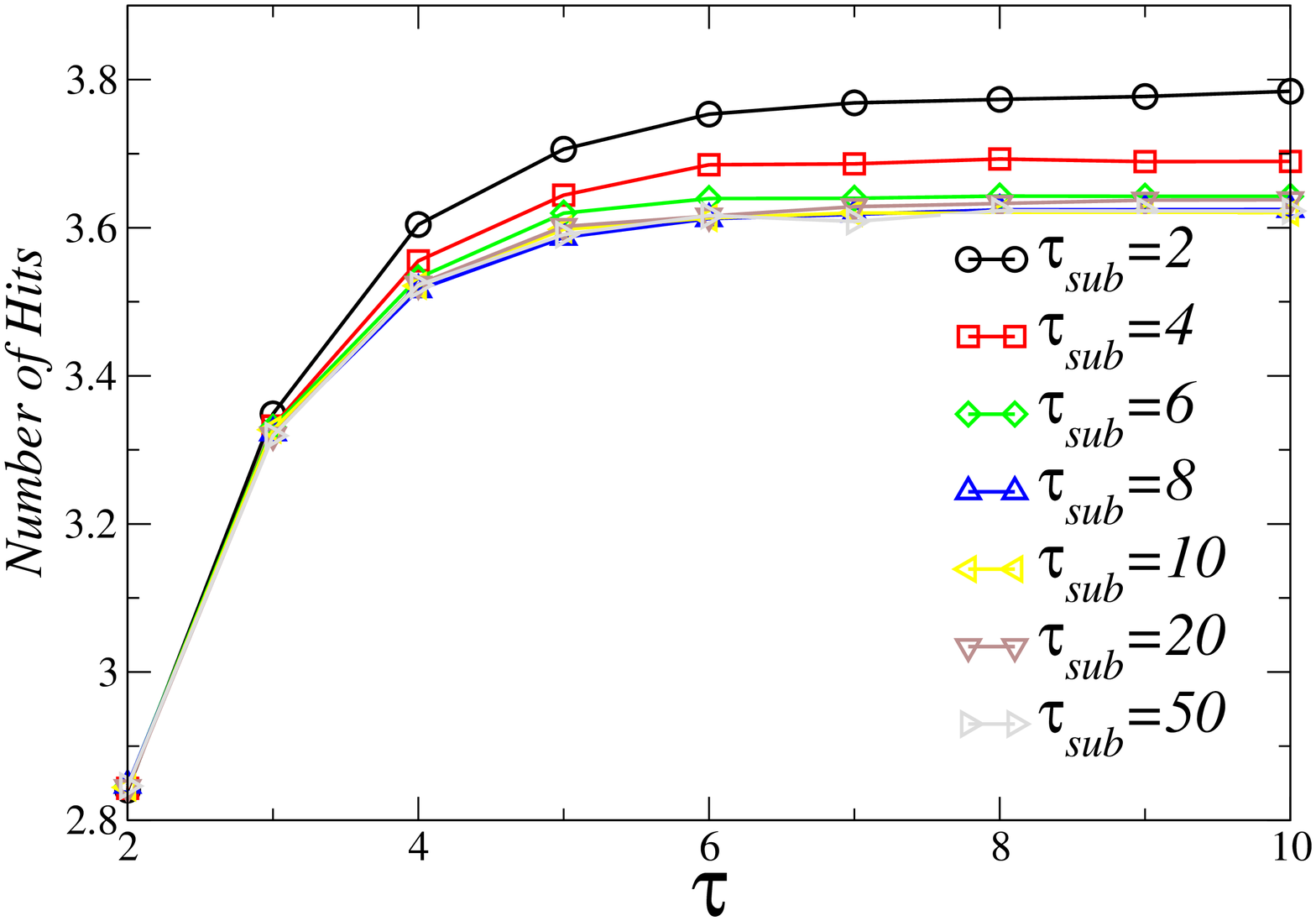}
& \hspace{-2mm}
\includegraphics[keepaspectratio=true,angle=0,width=60mm]
{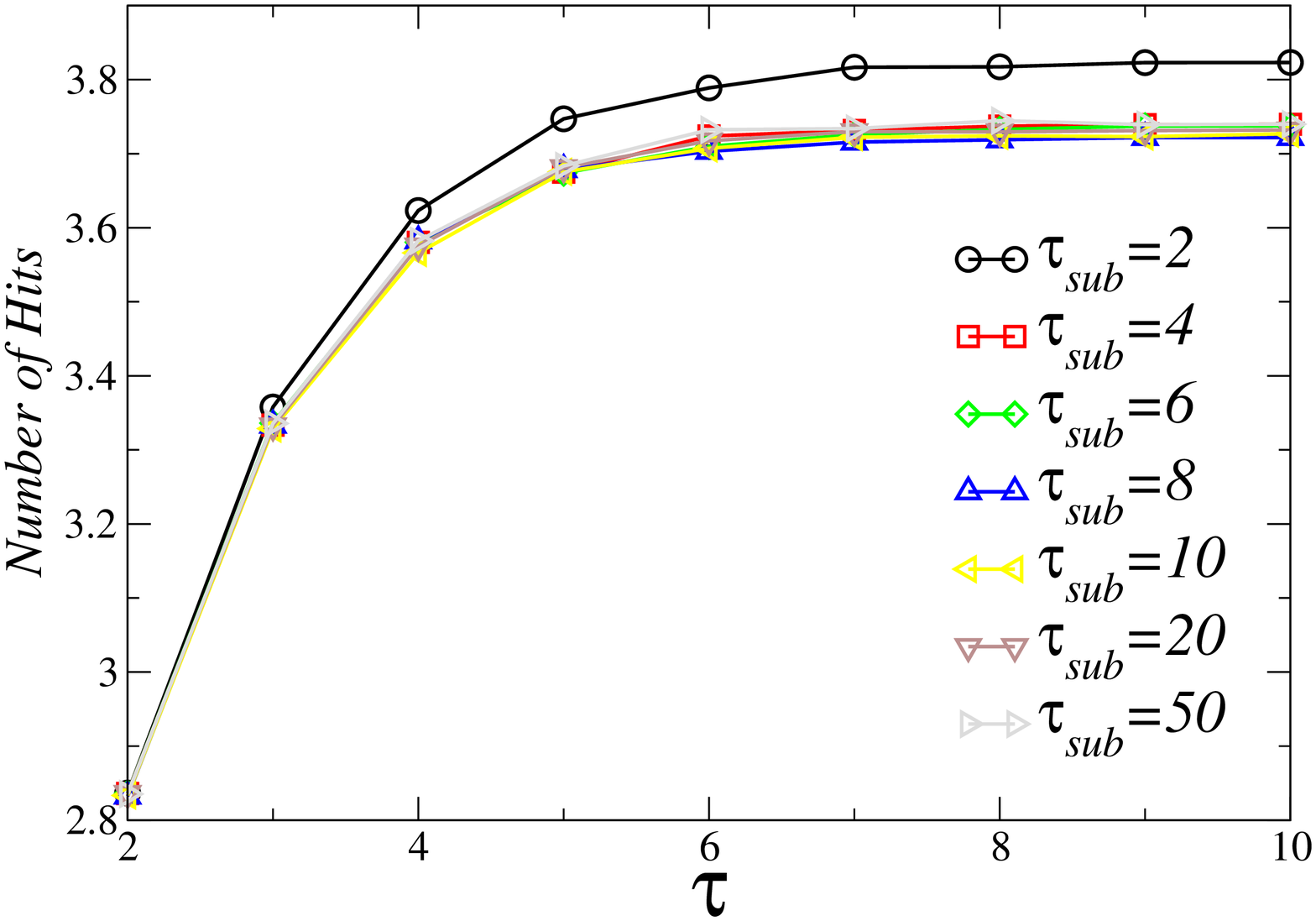}
\vspace{-5mm}
\\
\small{(a) $m=1$, no cutoff} & \small{(b) $m=1$, $k_c=50$} &
\small{(c) $m=1$, $k_c=10$}
\\
\includegraphics[keepaspectratio=true,angle=0,width=60mm]
{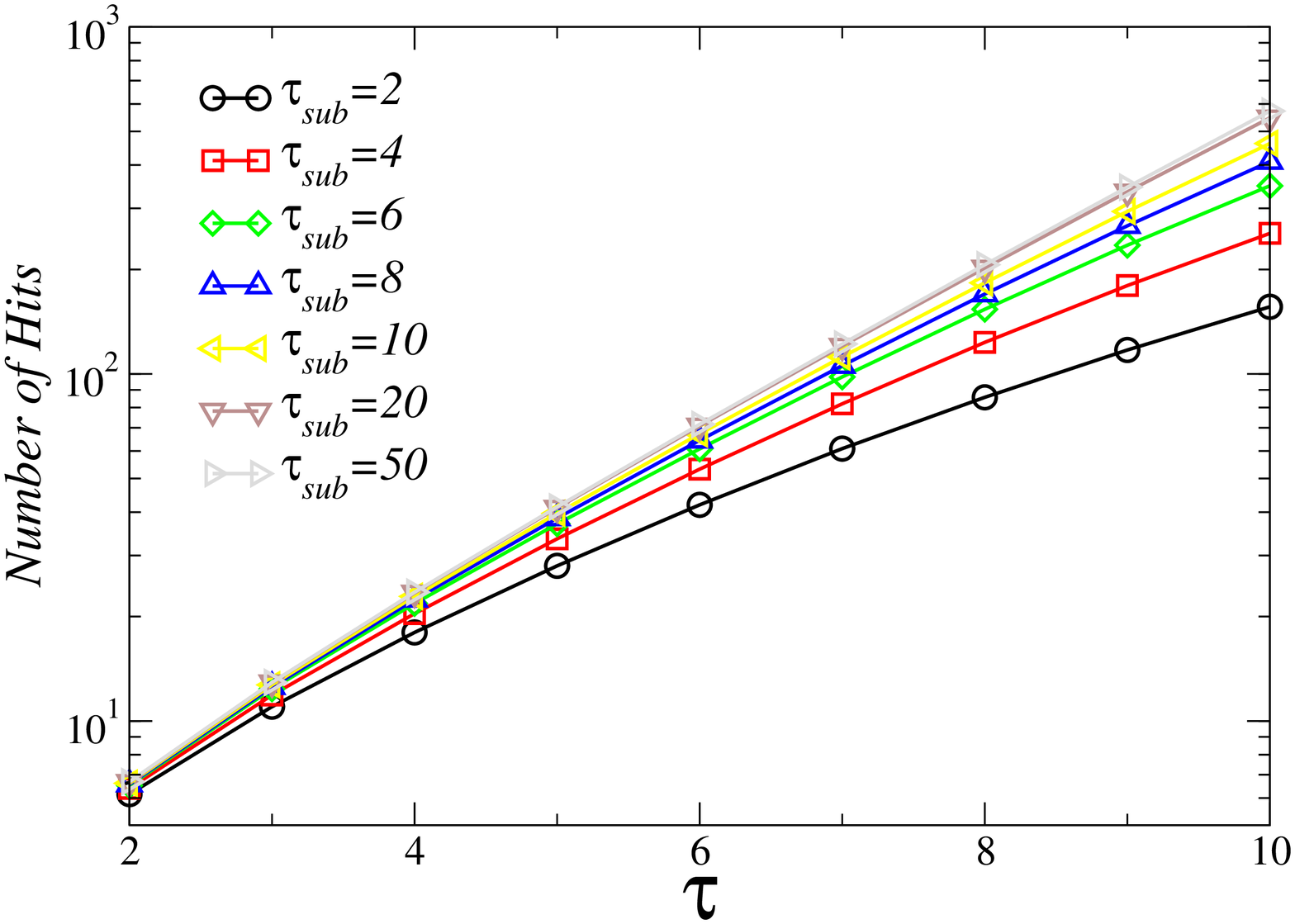}
& \hspace{-2mm}
\includegraphics[keepaspectratio=true,angle=0,width=60mm]
{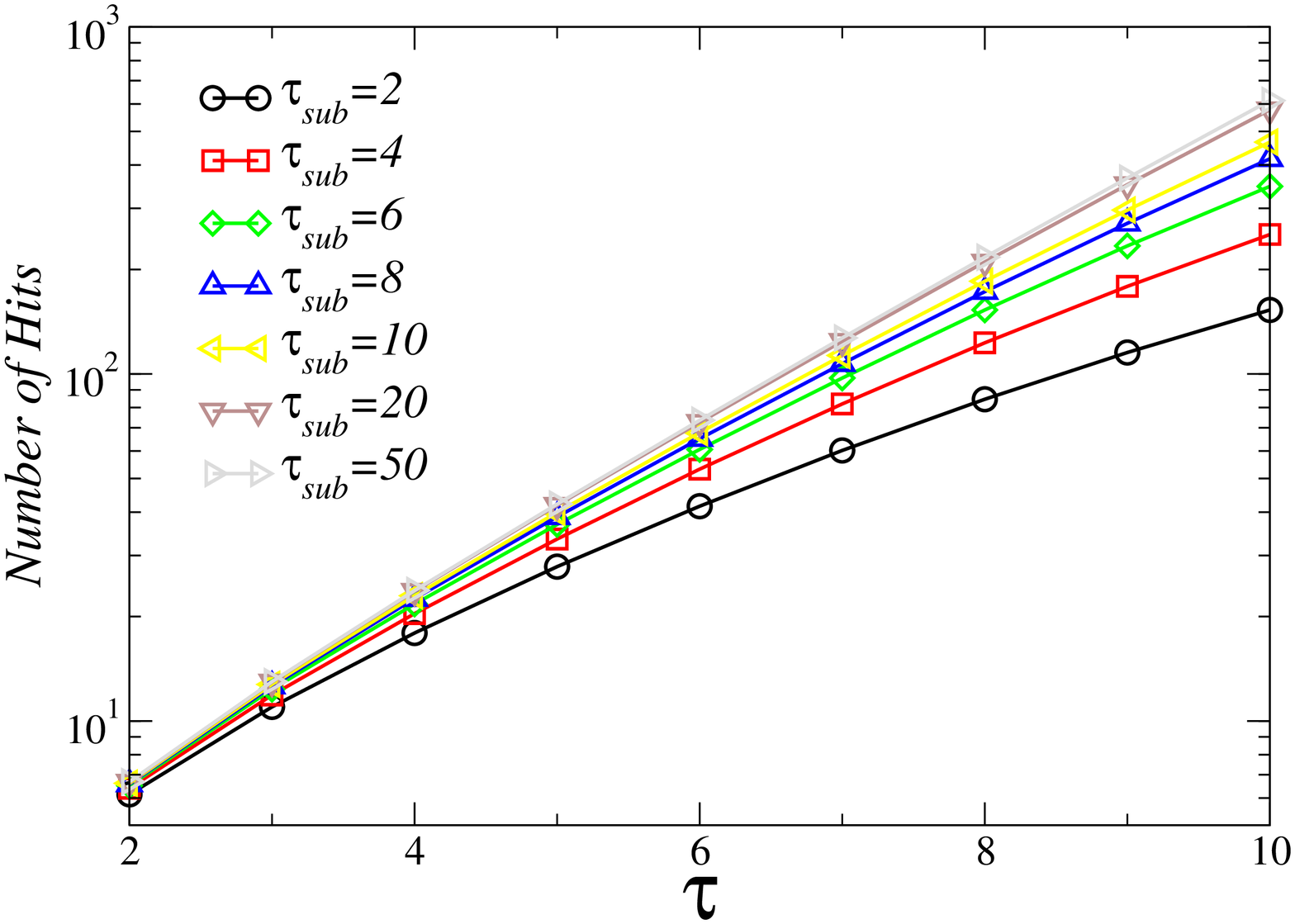}
& \hspace{-2mm}
\includegraphics[keepaspectratio=true,angle=0,width=60mm]
{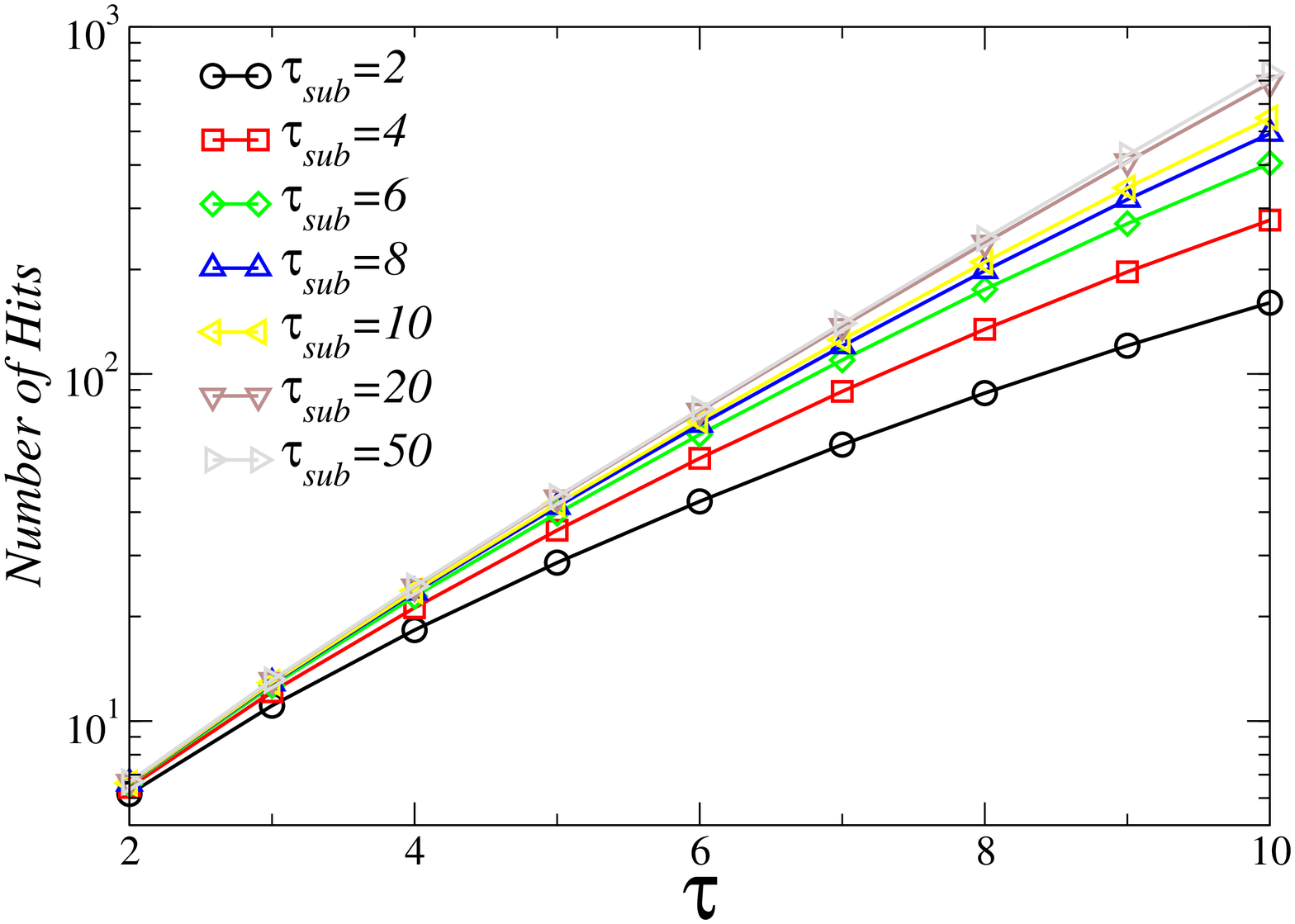}
\vspace{-5mm}
\\
\small{(d) $m=2$, no cutoff} & \small{(e) $m=2$, $k_c=50$} &
\small{(f) $m=2$, $k_c=10$}
\\
\includegraphics[keepaspectratio=true,angle=0,width=60mm]
{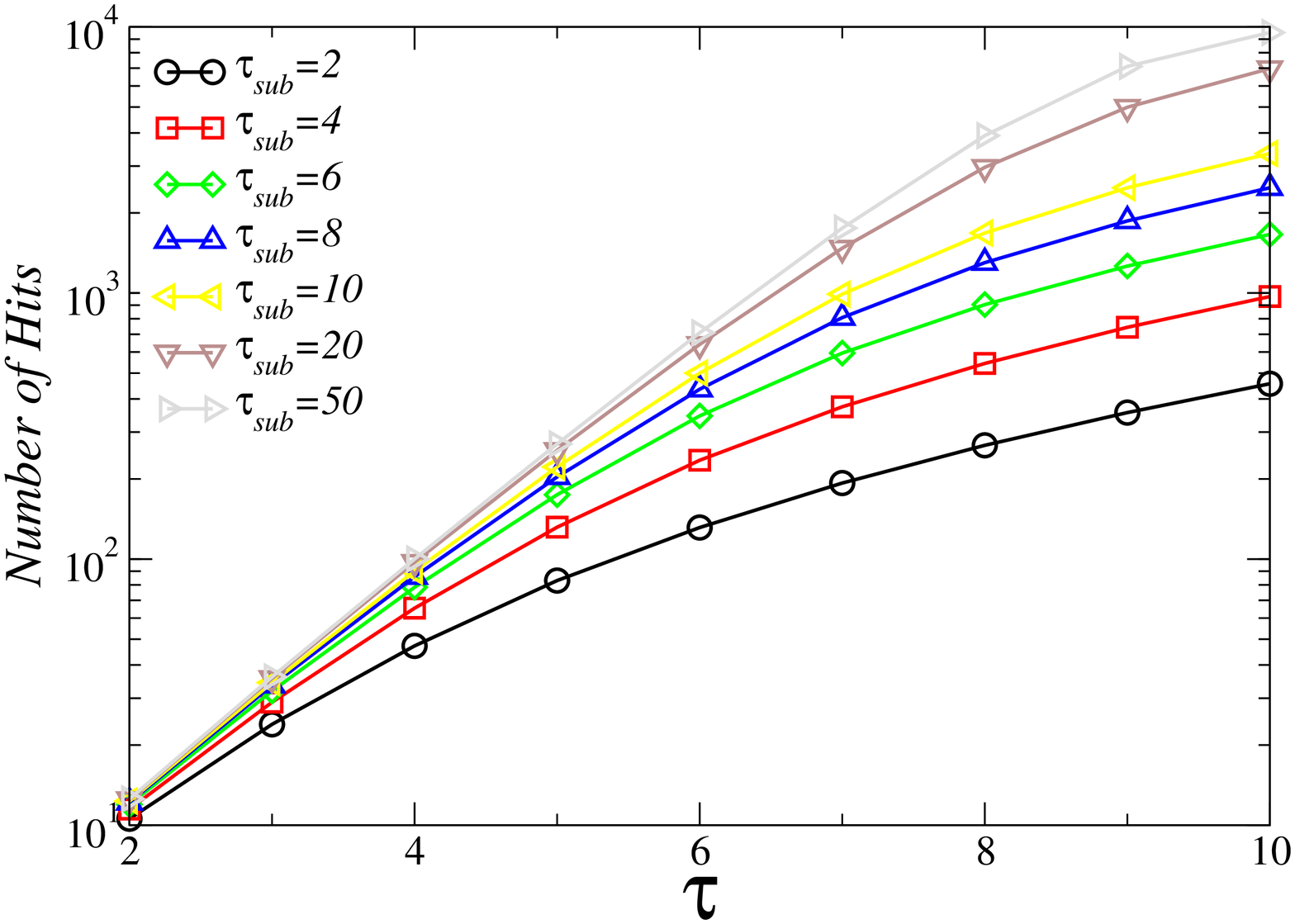}
& \hspace{-2mm}
\includegraphics[keepaspectratio=true,angle=0,width=60mm]
{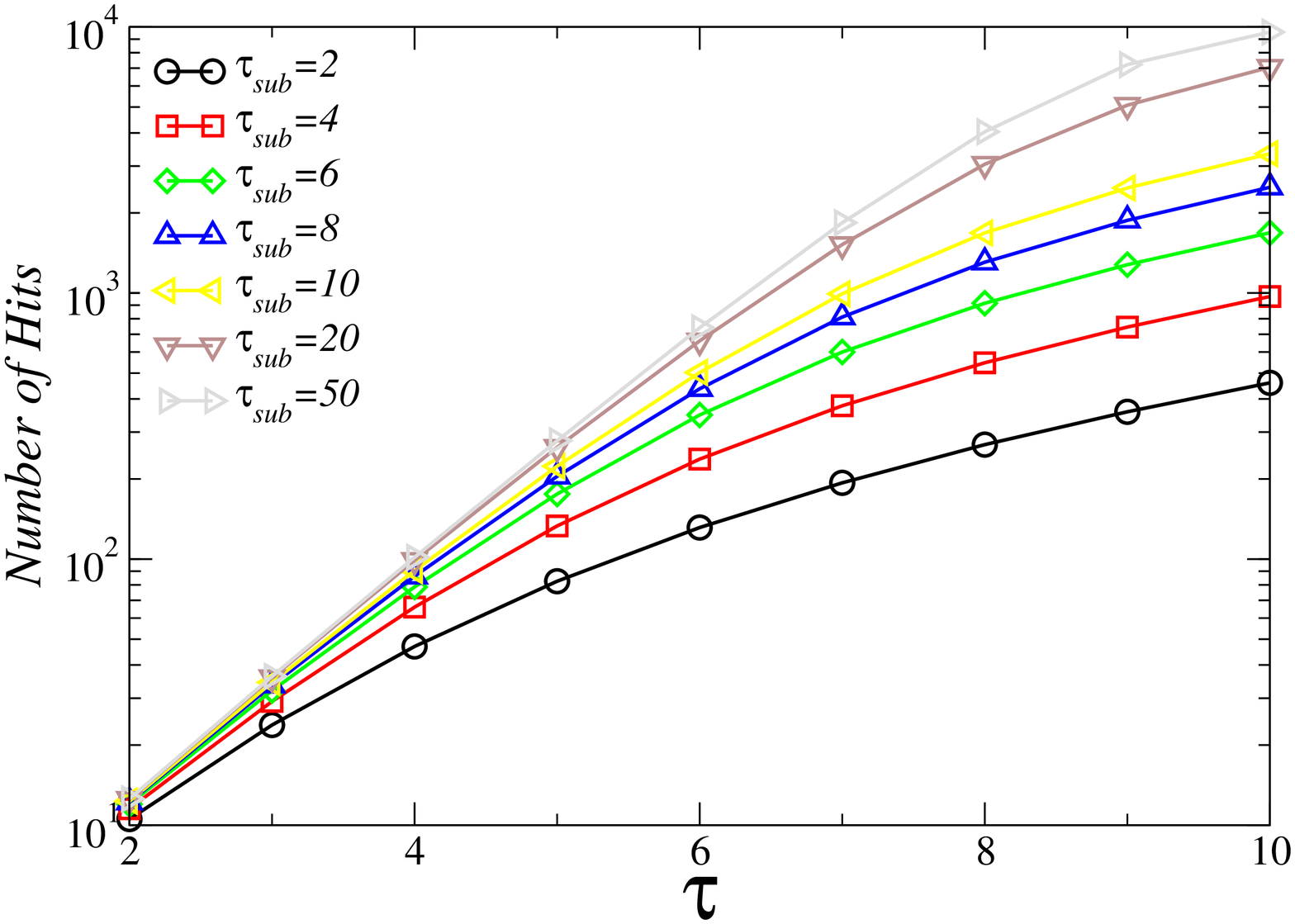}
& \hspace{-2mm}
\includegraphics[keepaspectratio=true,angle=0,width=60mm]
{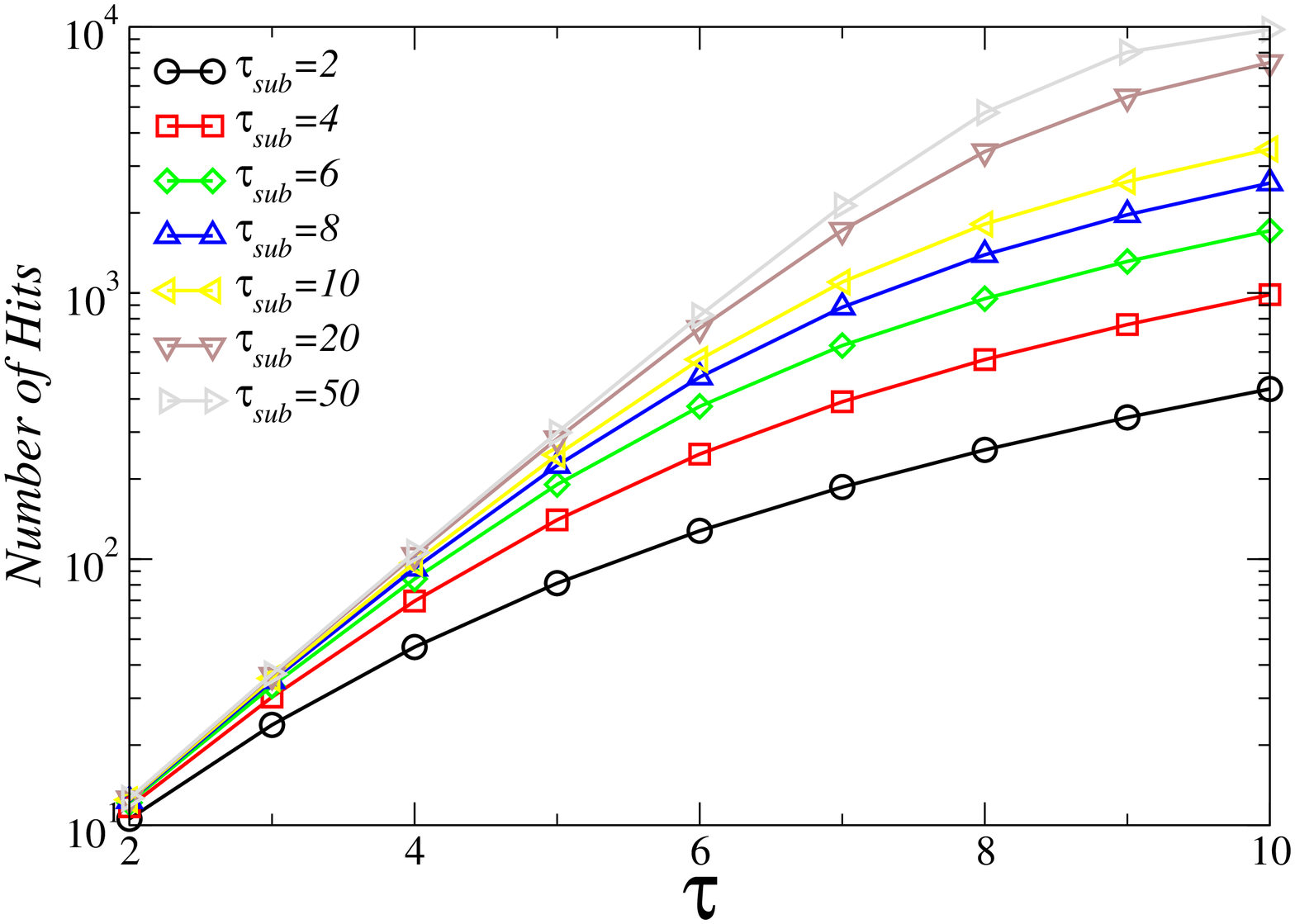}
\vspace{-5mm}
\\
\small{(g) $m=3$, no cutoff} & \small{(h) $m=3$, $k_c=50$} &
\small{(i) $m=3$, $k_c=10$}
\end{tabular}
\end{center}
\caption{NF results for DAPA model.} \label{fig_nf_da}
\end{figure*}

\subsubsection{Random Walk (RW)}

RW or multiple RWs have been used as an alternative search algorithm
to achieve even better granularity than NF. In RW, the message from
the source node is sent to a randomly chosen neighbor. Then, this
random neighbor takes the message and sends it to one of its random
neighbors excluding the node from which it got the message. This
continues until the destination node is reached or the total number
of hops is equal to $\tau$. A schematic of RW can be seen in
Fig.~\ref{fig_search}(c). RW can also be seen as a special case of
FL where only one neighbor is forwarded the search query, providing
the other extreme situation of the tradeoff between delivery time
and messaging complexity. RW search is inherently serial
(sequential), which causes a large increase in the delivery time
\cite{LCCLS-2002,GMS04}. In particular, computer simulations
performed on a generalized scale-free network with degree exponent
$\gamma$$=$$2.1$, which is equal to the value observed in P2P
networks, yield the result \cite{ADAMIC01}:
\begin{equation}
T_N = N^{0.79}.
\end{equation}

Although the RW search is worse than a FL search in scale-free
networks in terms of the time needed to locate a given node, the
average total traffic in the network is equal to $T_N$, and
therefore scales sublinearly with $N$, better than the linear growth
of FL search.

\begin{figure*}
\begin{center}
\begin{tabular}{ccc}
\includegraphics[keepaspectratio=true,angle=0,width=60mm]
{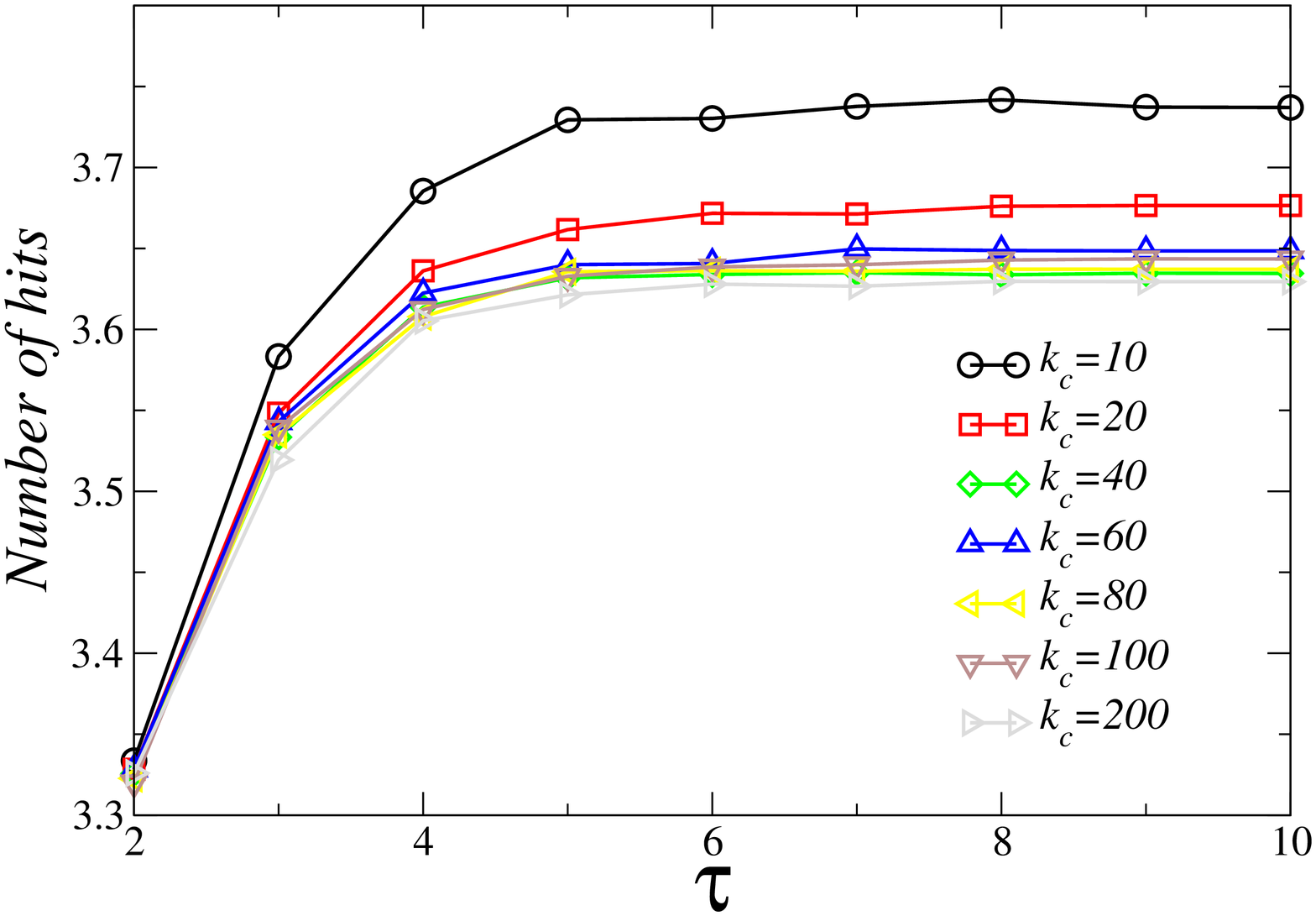}
& \hspace{-2mm}
\includegraphics[keepaspectratio=true,angle=0,width=60mm]
{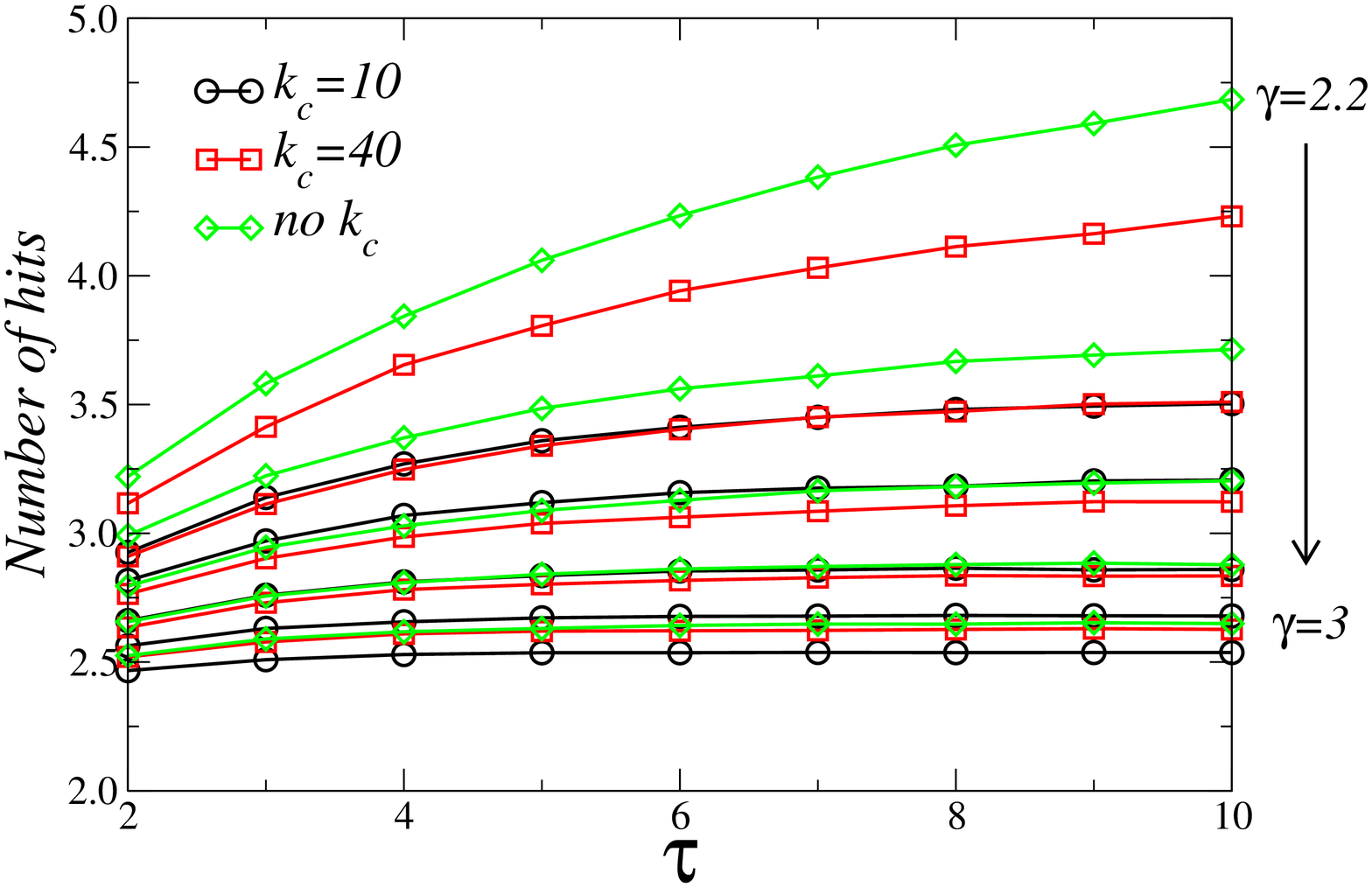}
& \hspace{-2mm}
\includegraphics[keepaspectratio=true,angle=0,width=60mm]
{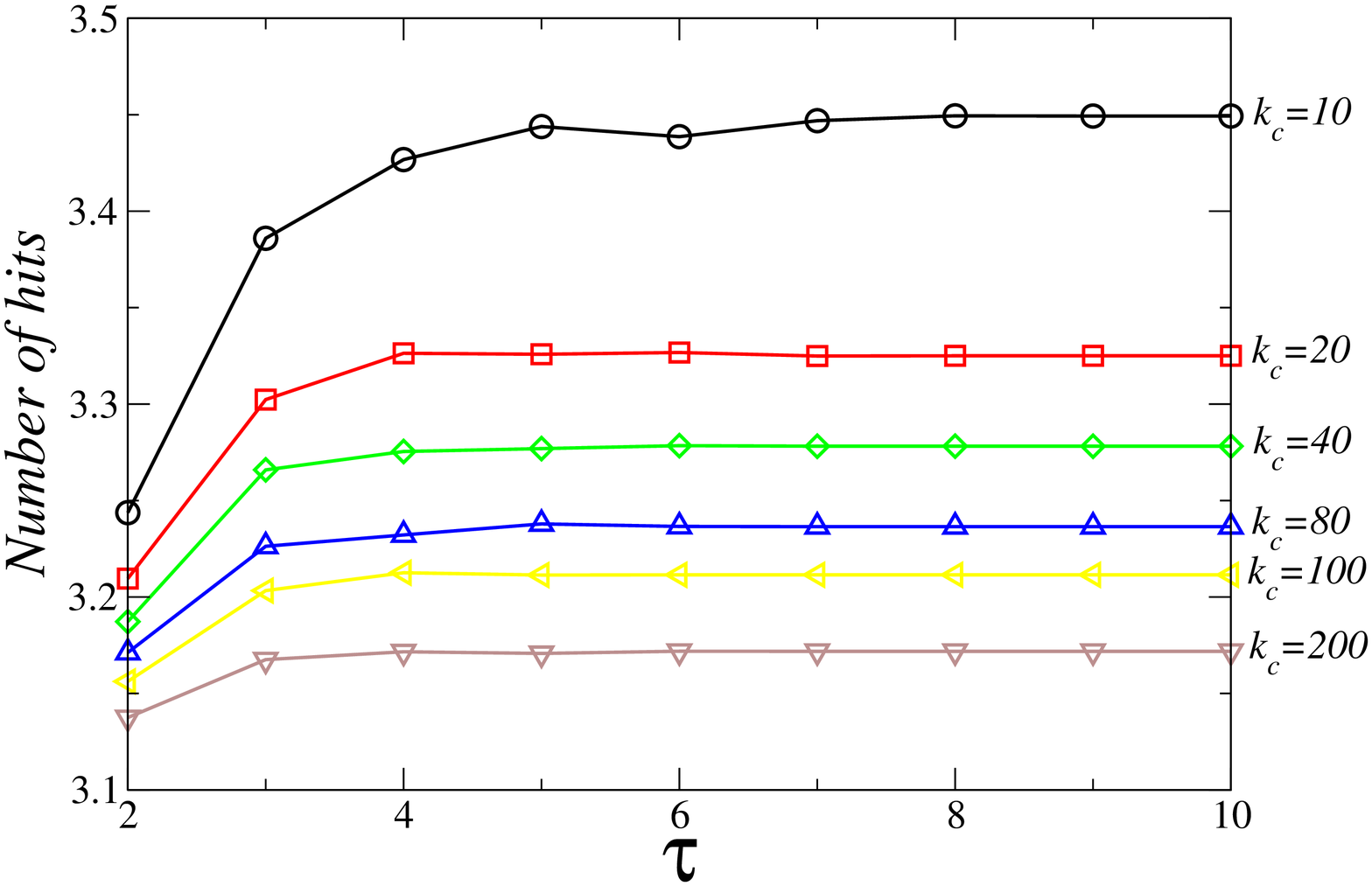} \vspace{-5mm}
\\
\small{(a) $m=1$} & \small{(b) $m=1$} & \small{(c) $m=1$}
\\
\includegraphics[keepaspectratio=true,angle=0,width=60mm]
{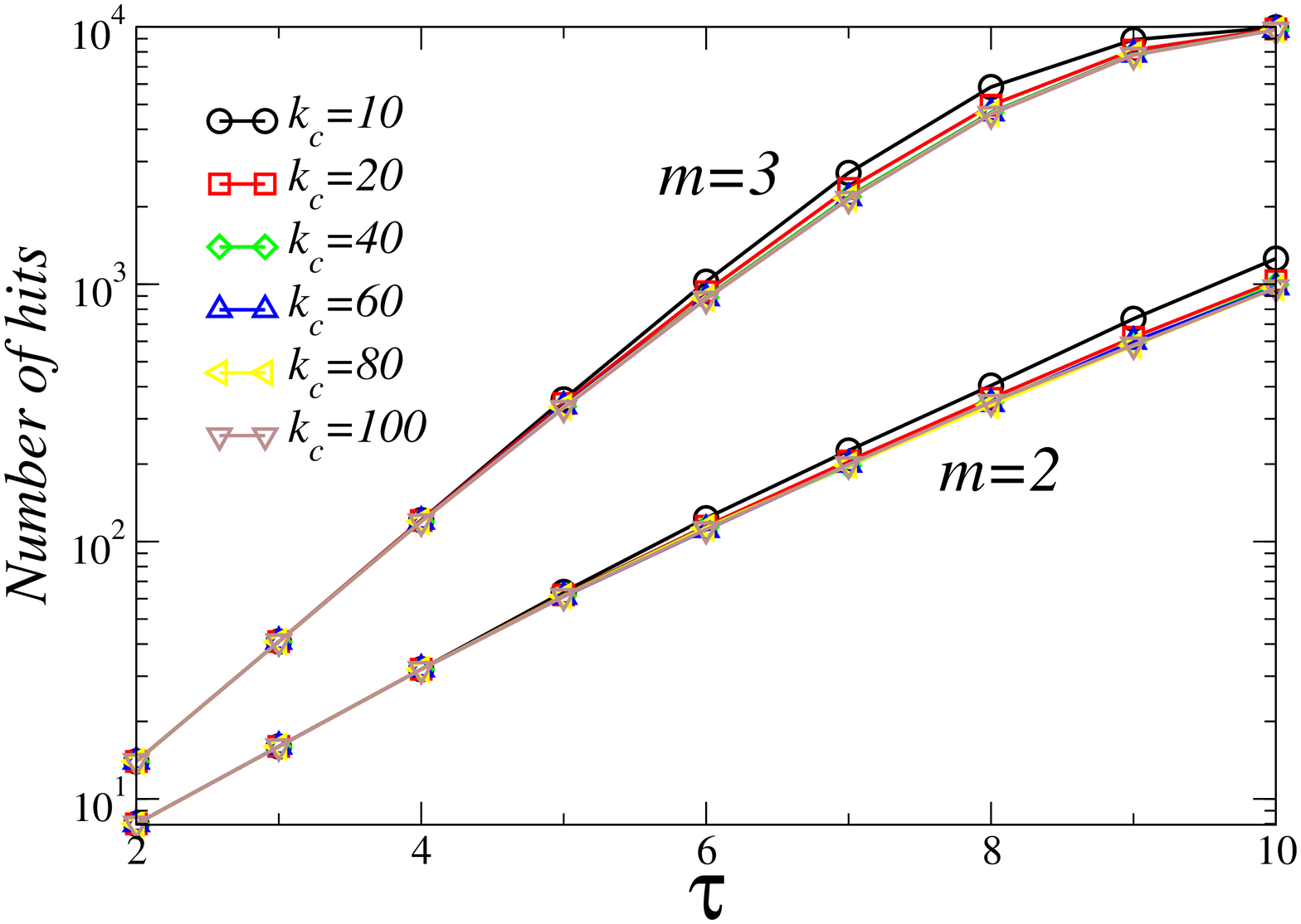}
& \hspace{-2mm}
\includegraphics[keepaspectratio=true,angle=0,width=60mm]
{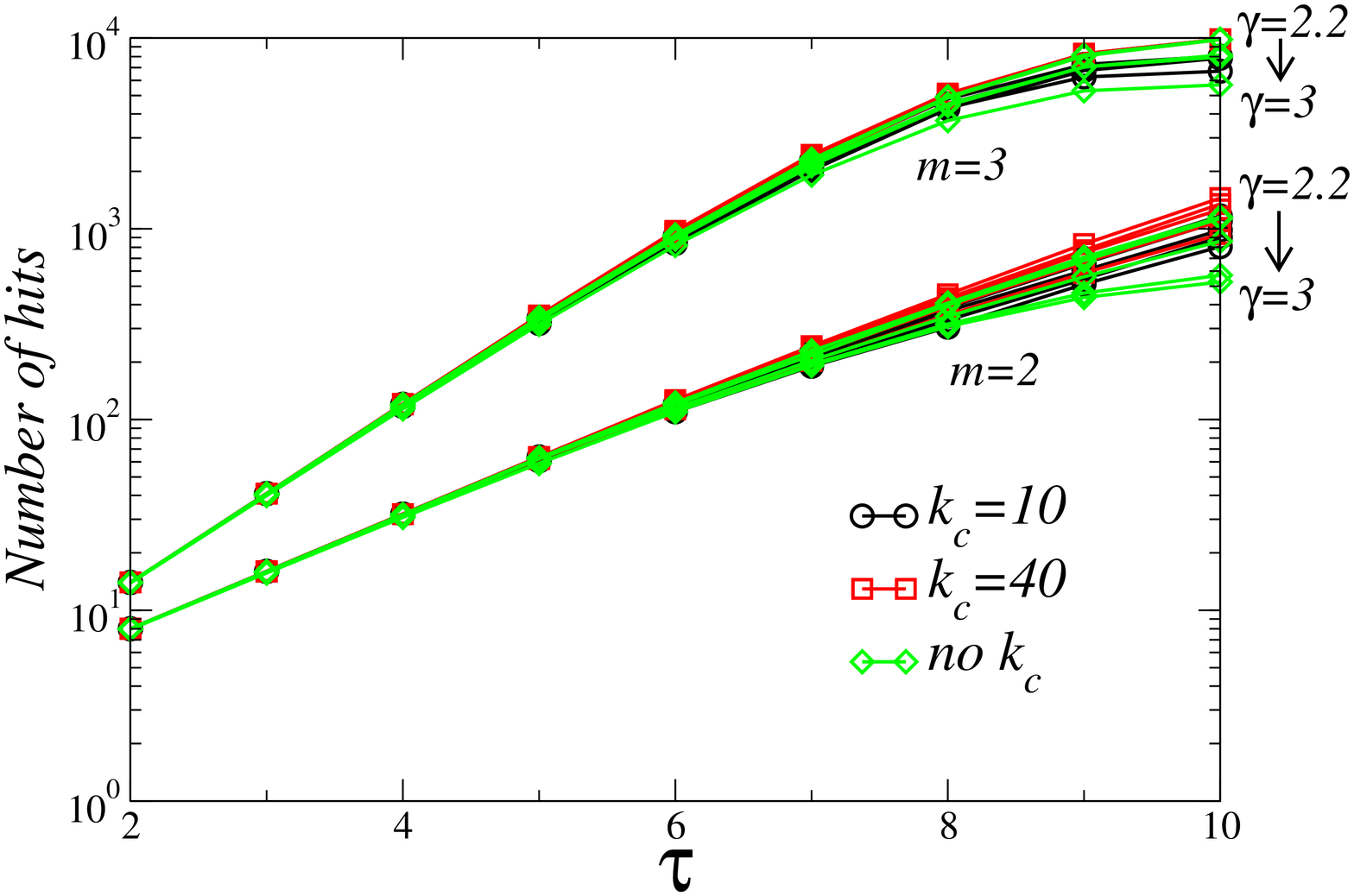}
& \hspace{-2mm}
\includegraphics[keepaspectratio=true,angle=0,width=60mm]
{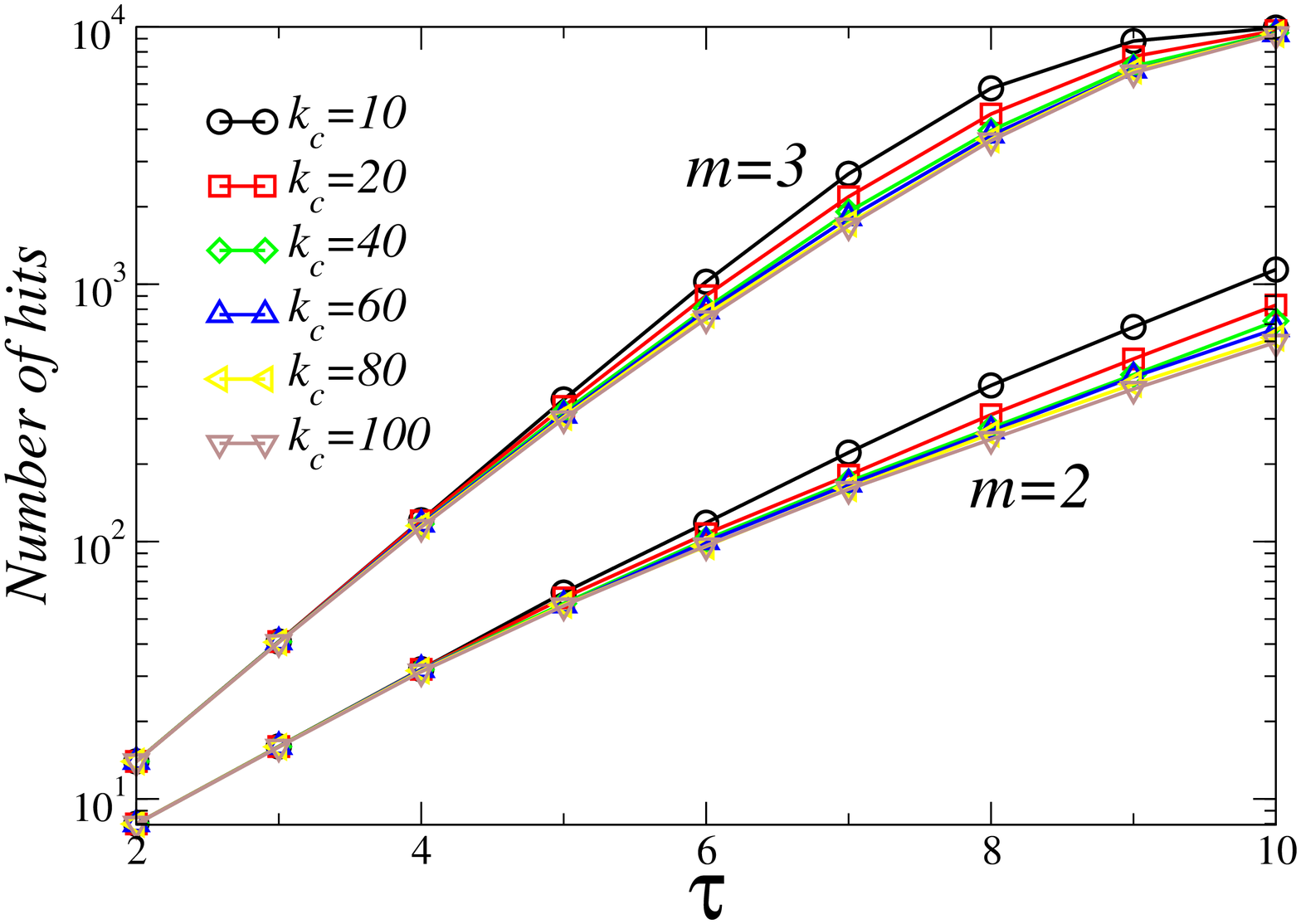} \vspace{-5mm}
\\
\small{(d) $m=2$ and $m=3$} & \small{(e) $m=2$ and $m=3$} &
\small{(f) $m=2$ and $m=3$}
\\
\small{PA model} & \small{CM} & \small{HAPA model}
\end{tabular}
\end{center}
\caption{RW results for PA, CM, and HAPA models.} \label{fig_rw}
\end{figure*}

\begin{figure*}
\begin{center}
\begin{tabular}{ccc}
\includegraphics[keepaspectratio=true,angle=0,width=60mm]
{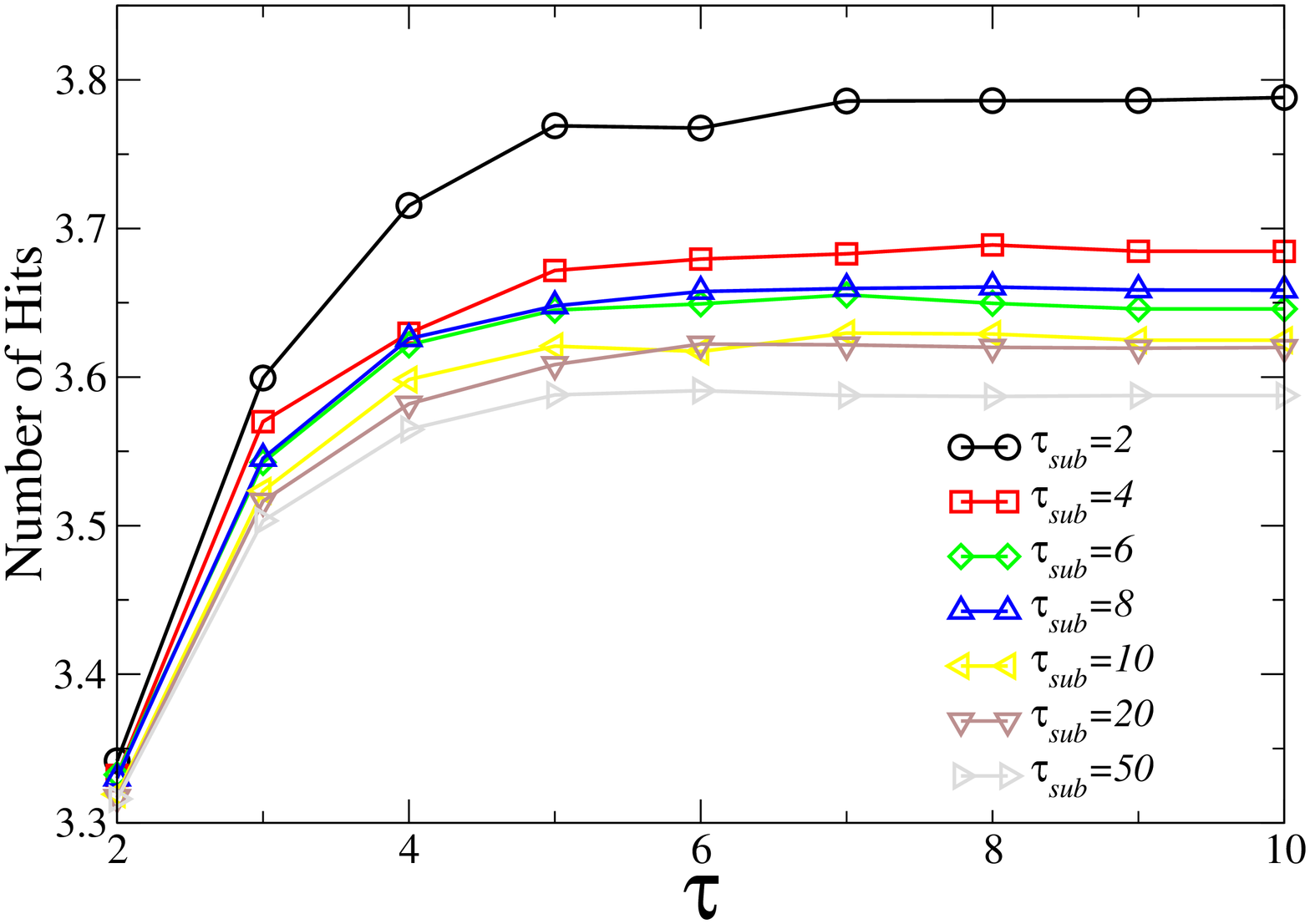}
& \hspace{-2mm}
\includegraphics[keepaspectratio=true,angle=0,width=60mm]
{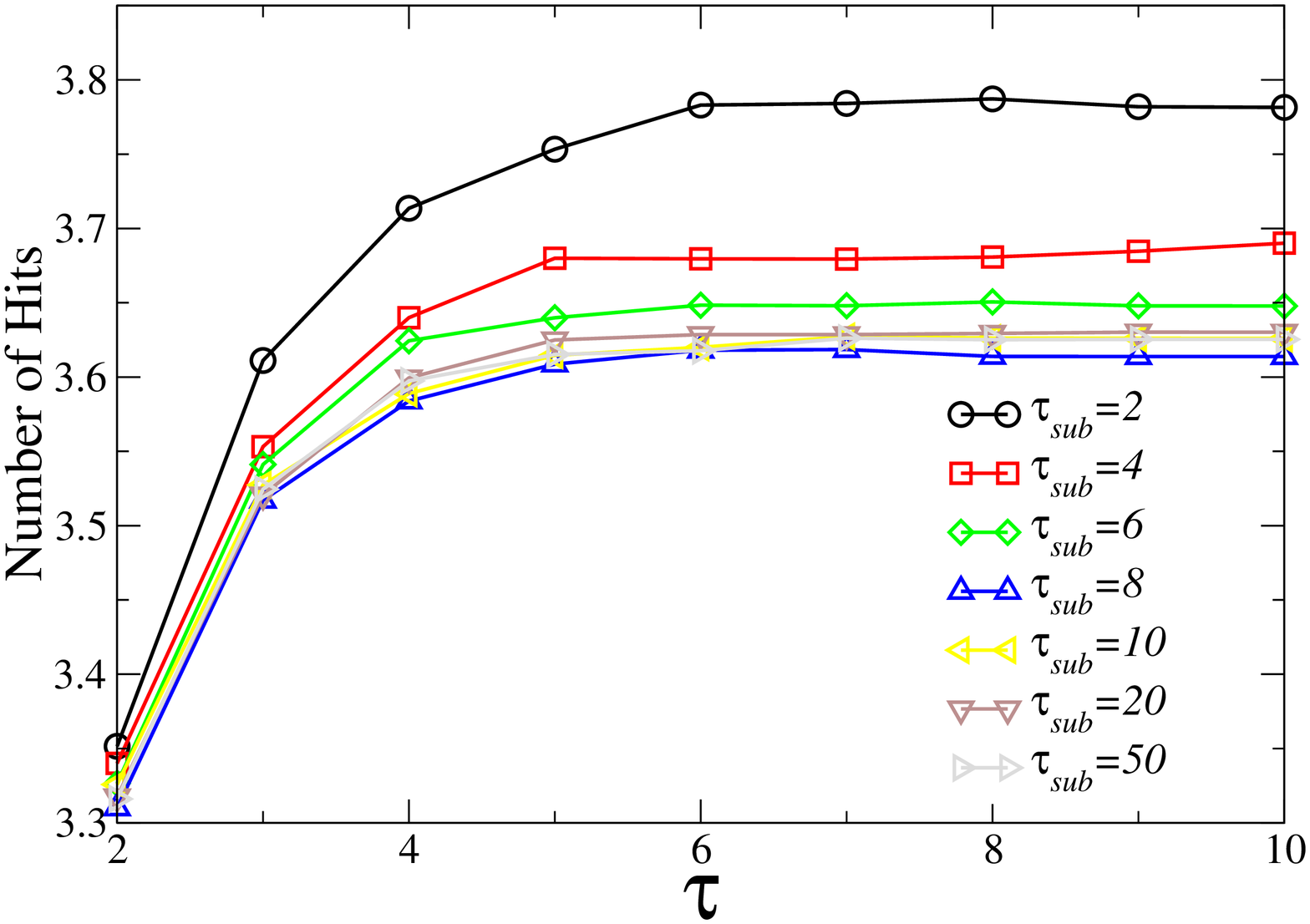}
& \hspace{-2mm}
\includegraphics[keepaspectratio=true,angle=0,width=60mm]
{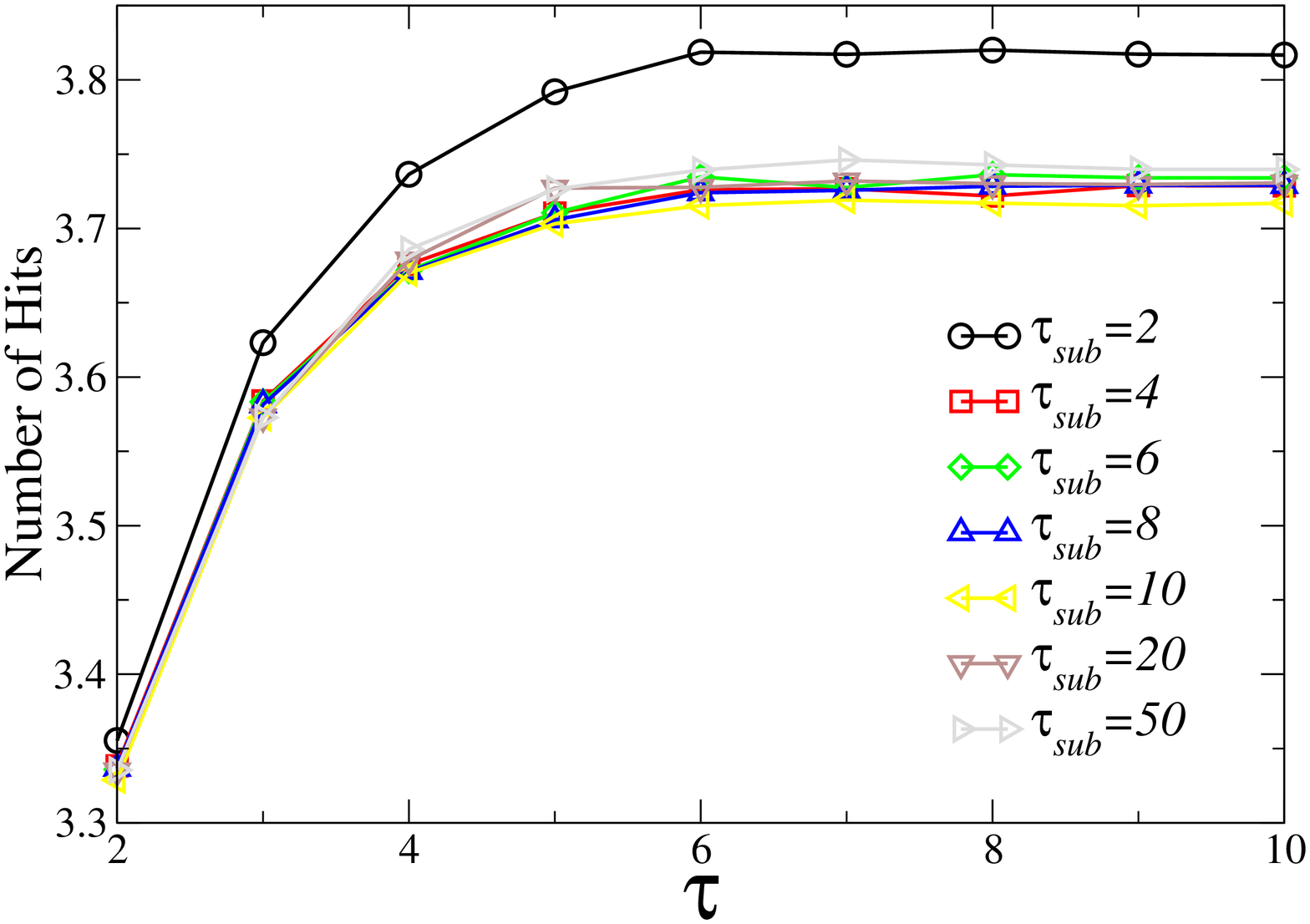}
\vspace{-5mm}
\\
\small{(a) $m=1$, no cutoff} & \small{(b) $m=1$, $k_c=50$} &
\small{(c) $m=1$, $k_c=10$}
\\
\includegraphics[keepaspectratio=true,angle=0,width=60mm]
{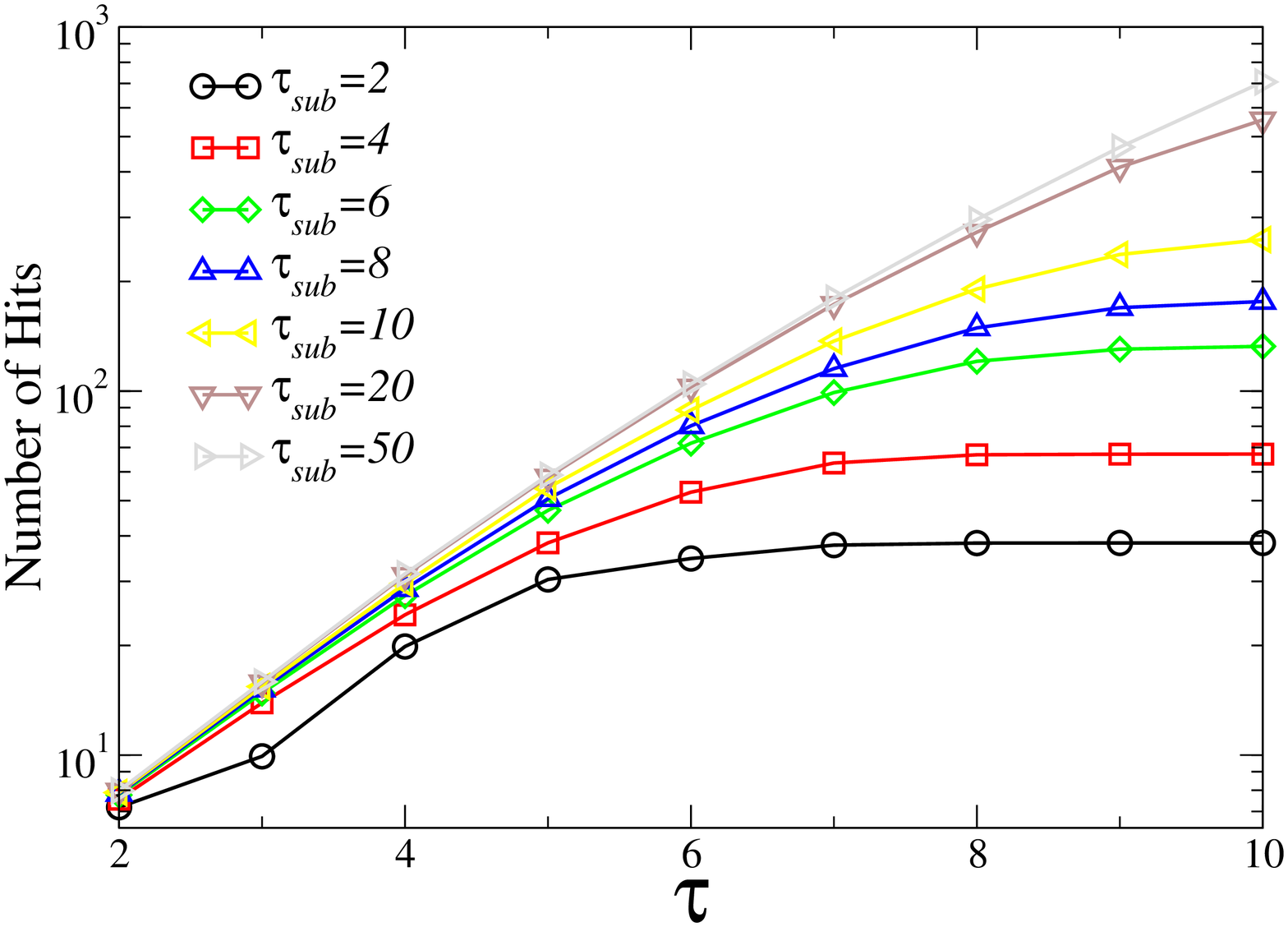}
& \hspace{-2mm}
\includegraphics[keepaspectratio=true,angle=0,width=60mm]
{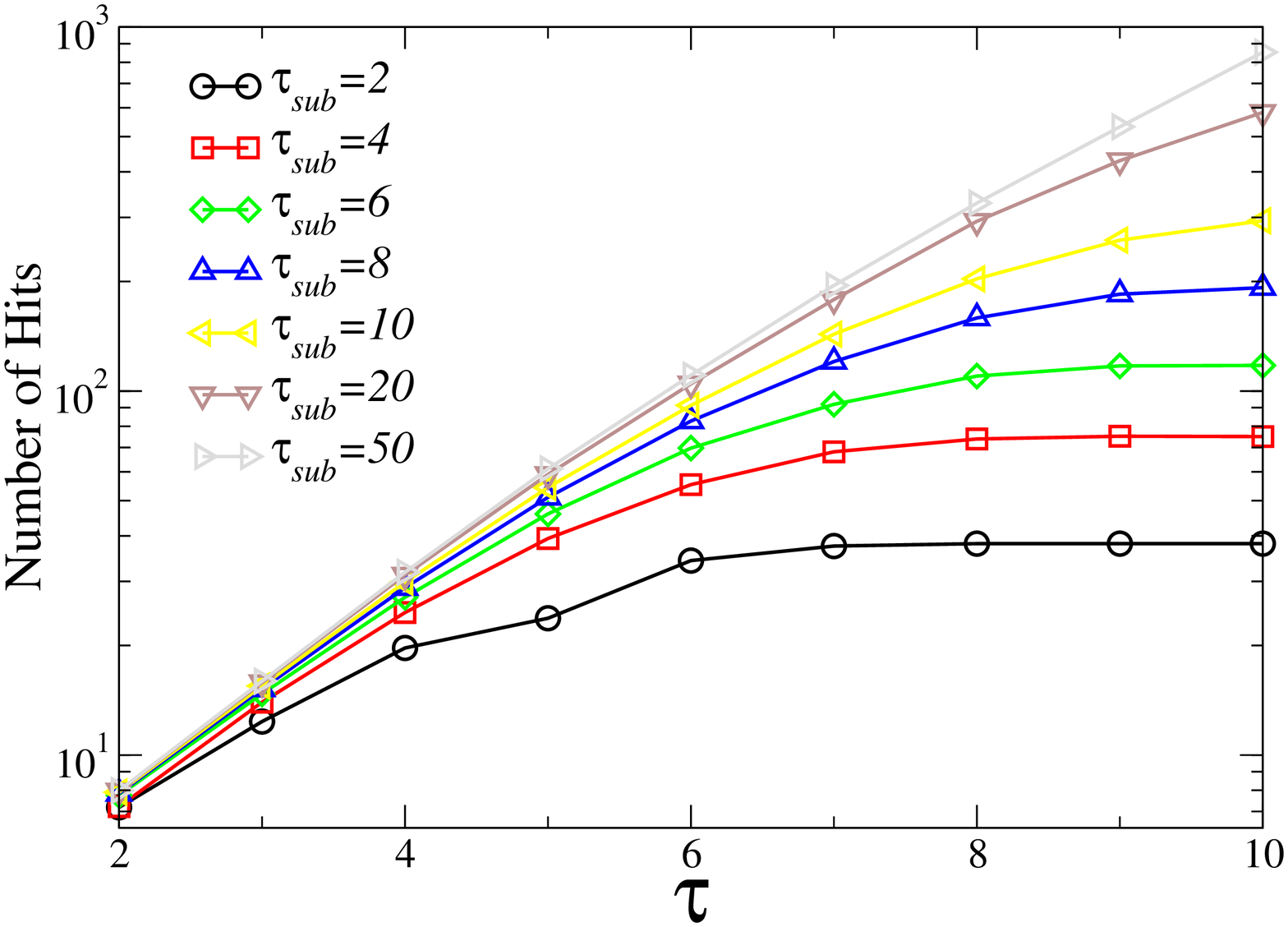}
& \hspace{-2mm}
\includegraphics[keepaspectratio=true,angle=0,width=60mm]
{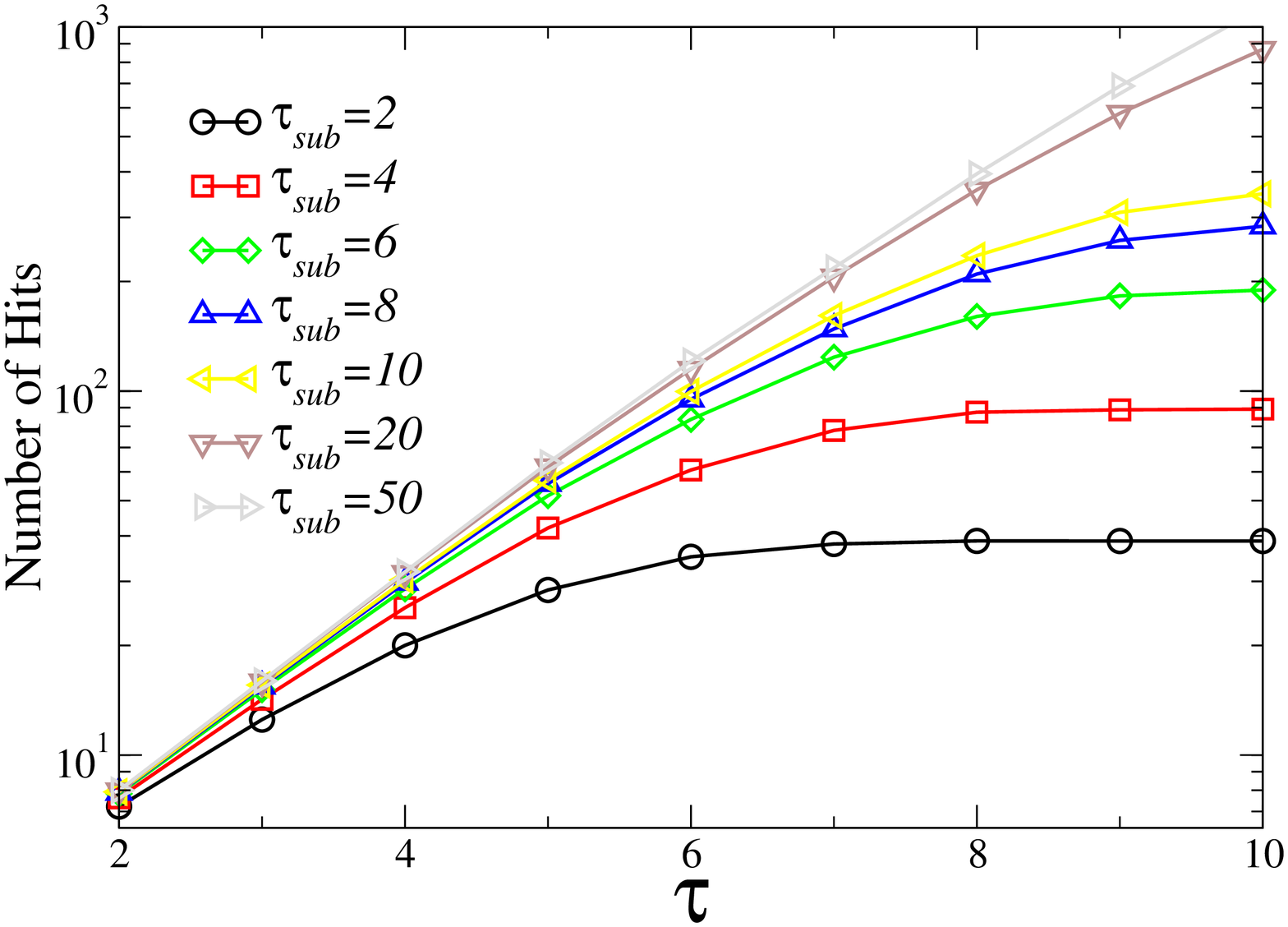}
\vspace{-5mm}
\\
\small{(d) $m=2$, no cutoff} & \small{(e) $m=2$, $k_c=50$} &
\small{(f) $m=2$, $k_c=10$}
\\
\includegraphics[keepaspectratio=true,angle=0,width=60mm]
{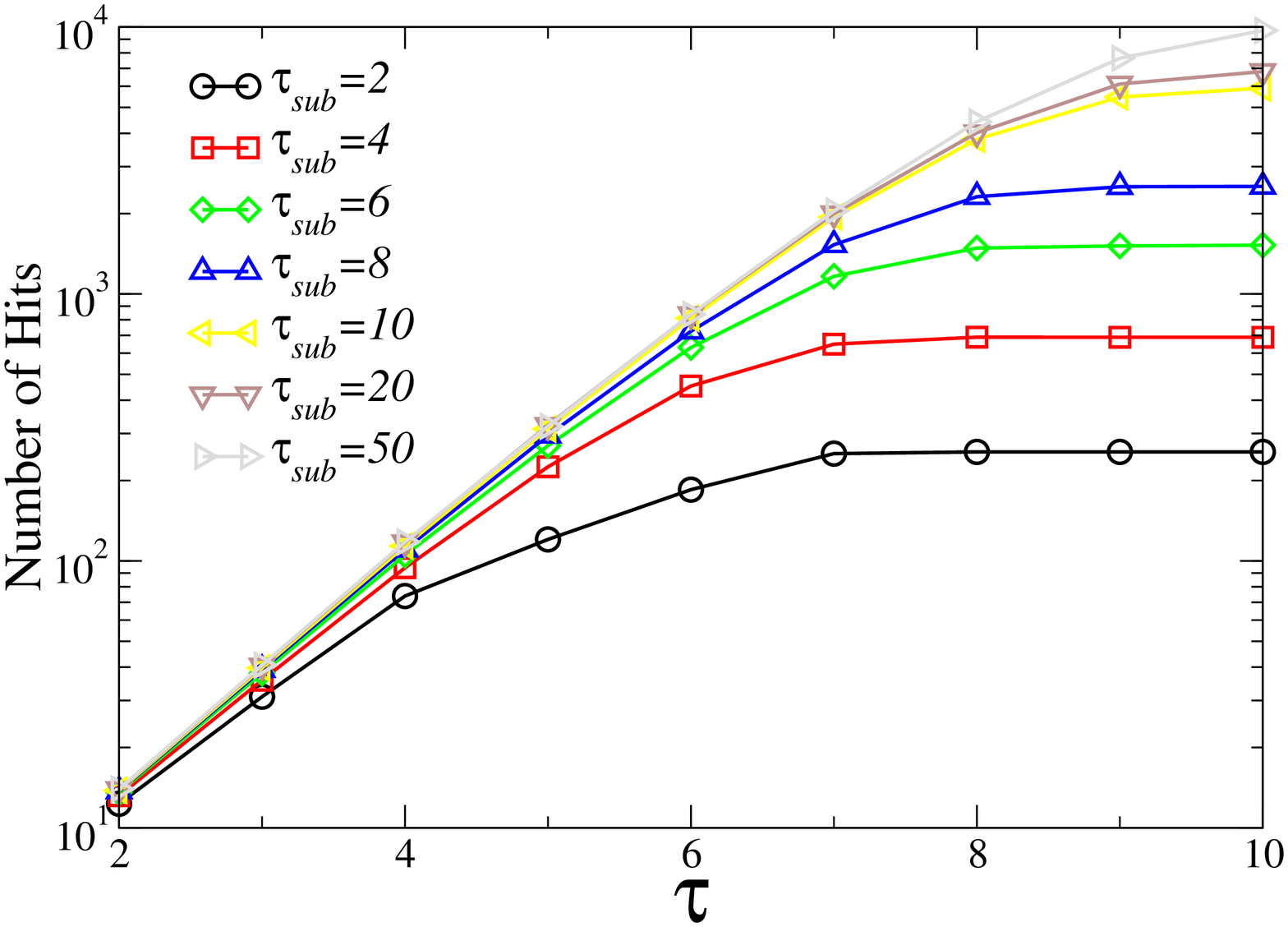}
& \hspace{-2mm}
\includegraphics[keepaspectratio=true,angle=0,width=60mm]
{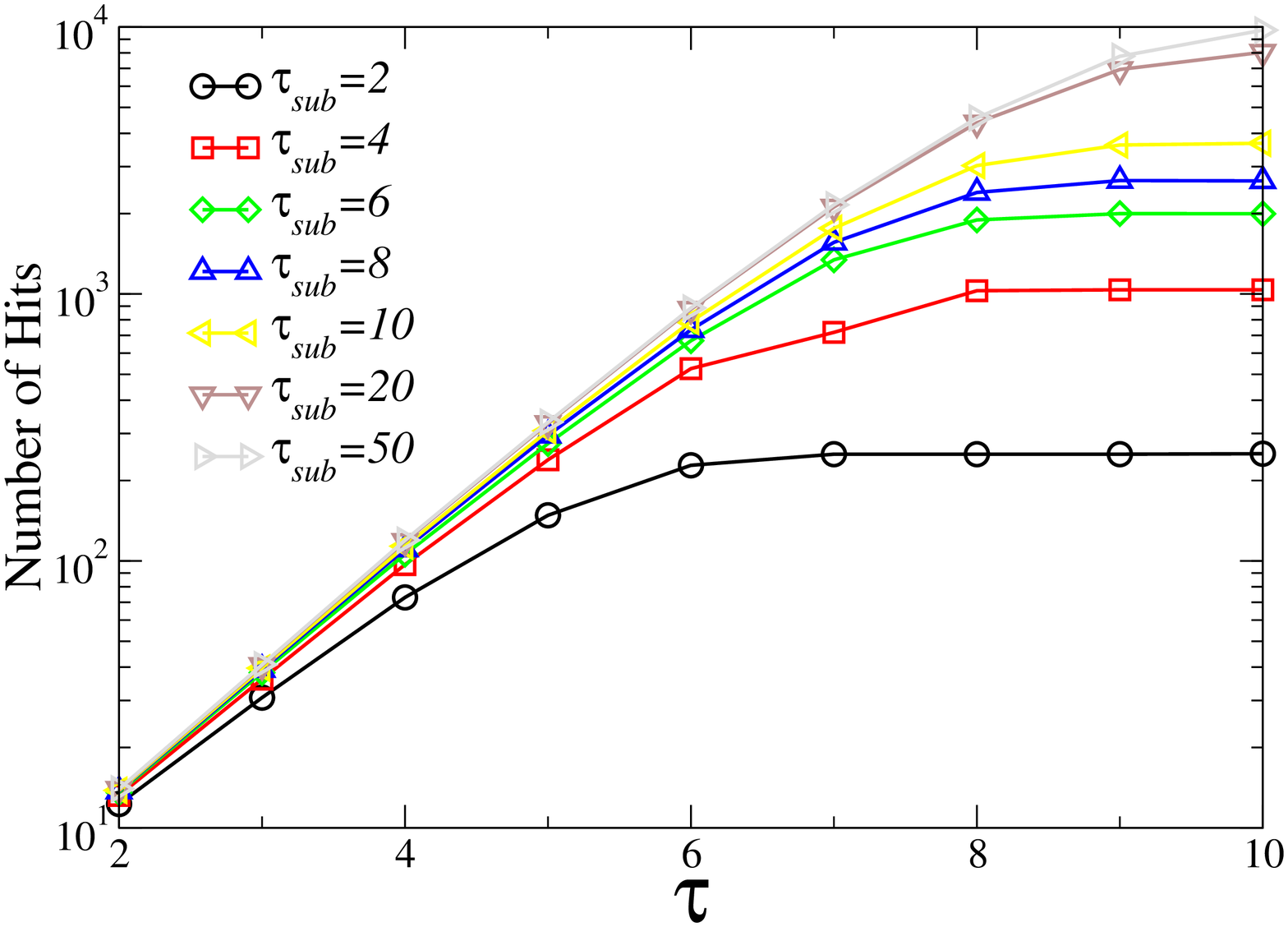}
& \hspace{-2mm}
\includegraphics[keepaspectratio=true,angle=0,width=60mm]
{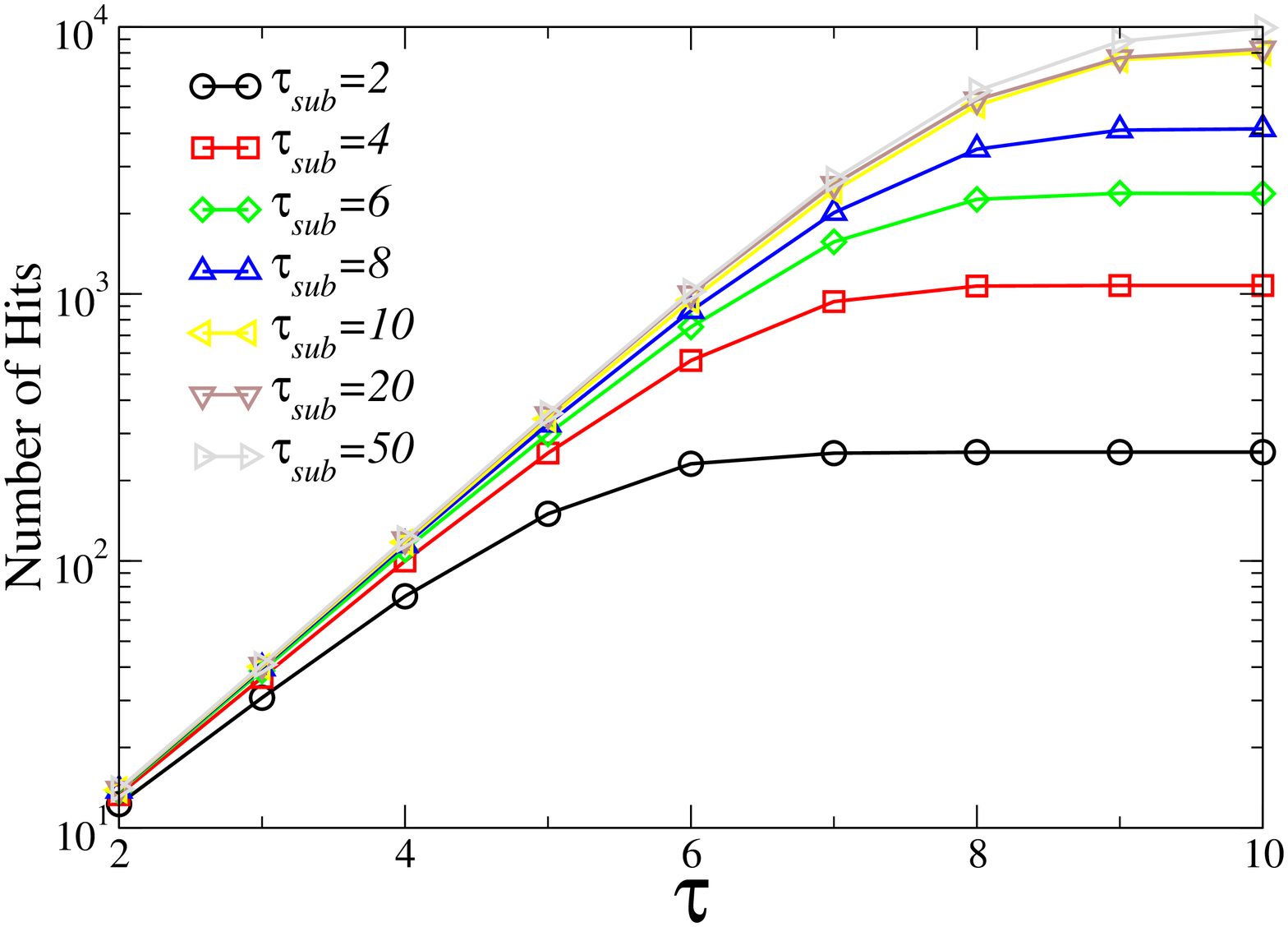}
\vspace{-5mm}
\\
\small{(g) $m=3$, no cutoff} & \small{(h) $m=3$, $k_c=50$} &
\small{(i) $m=3$, $k_c=10$}
\end{tabular}
\end{center}
\caption{RW results for DAPA model.} \label{fig_rw_da}
\end{figure*}

\subsection{Results}

We simulated the three search algorithms FL, NF, and RW on the
topologies generated by the four methods PA, CM, HAPA, and DAPA, and
provide results all the combinations with various hard cutoffs.
Through the PA, CM, HAPA, and DAPA methods, we generated topologies
with 10000 nodes. We used cutoff values of 10 and 40 (or 50 in some
cases), in addition to the natural cutoff, i.e., no hard cutoff.
When generating DAPA topologies, we used $\tau_{sub}$ values of 2,
4, 6, 8, 10, 20, and 50 with expectation that larger $\tau_{sub}$
should yield better search efficiency. Minimum degree values (or
$m$) in our topologies were 1, 2, or 3. We varied the $\tau$ values
of search queries in FL up to the point we reach the system size and
for NF/RW up to 10. To compare search efficiencies of RW and NF
fairly in our simulations, we equated $\tau$ of RW searches to the
number of messages incurred by the NF searches in the same scenario.
Thus, for the search efficiency graphs of RW [e.g.,
Fig.~\ref{fig_pk-pa}] when $\tau$ is equal to a particular value
such as 4, this means that the number of hits data-point
corresponding to that $\tau=4$ value is obtained by simulating a RW
search with $\tau$ equal to the number of messages that were caused
by an NF search using a $\tau$ value of 4. A similar normalization
was done in \cite{GMS05}.

\subsubsection{Search Efficiency}

As the $\tau$ varies, Fig.~\ref{fig_fl-pa} shows the number of hits
achieved by FL on various topologies generated by the PA method.
Similarly, Fig.~\ref{fig_fl-mr} shows search efficiency of FL on the
topologies generated by the CM. In both of these figures, as
expected, when there is no hard cutoff in the topology, the FL
algorithm can achieve higher search efficiency by capturing more of
the peers in the network for a specific $\tau$ value. Also, the
affect of imposing a hard cutoff reduces when minimum degree in the
topology is higher. One interesting feature of CM is when the
minimum number of links is one (i.e., $m=1$) the number of hits
cannot reach system size even for very large $\tau$ values because
the network is not a connected one for $m=1$. Fig.~\ref{fig_fl-hapa}
shows a similar search efficiency behavior for FL on HAPA
topologies, with even more apparent effect of hard cutoff. For small
values of cutoff, PA and HAPA give similar performances in FL,
whereas for higher values of cutoff HAPA has better hits results due
to the star-like topology. The FL in DAPA is less efficient than in
PA, although for higher values of $\tau_{sub}$ it gets closer to PA
and efficiency of FL increases as can be seen in
Fig.~\ref{fig_fl-dapa}.

\emph{A minimum of three links for all peers eliminates negative
effects of hard cutoffs:} An interesting observation is that
negative effect of hard cutoffs on the FL performance on the PA and
HAPA topologies can be easily reduced to negligible values by
increasing the number of stubs $m$ (or connectedness). The number of
stubs as small as 3 leaves virtually no difference between the
search performances of overlay topologies with or without hard
cutoffs. This result provides the guideline that \emph{to achieve a
better FL performance a requirement of having at least three links
to the rest of the network will be adequate to assure that no one
else in the network will need to maintain unbearably large number of
links}. However, the necessity of complete or partial \emph{global}
information about the overall when constructing a PA topology is a
major discouragement of using the PA and HAPA methods for overlay
topologies of unstructured P2P networks.

\emph{There exists an interplay between connectedness and the degree
distribution exponent for a fixed cutoff:} As the DAPA method is a
purely local method, it is more interesting to observe search
performance on the DAPA topologies. Figs.~\ref{fig_fl-dapa}(a),
\ref{fig_fl-dapa}(b), and \ref{fig_fl-dapa}(c) show the FL
performance on DAPA topologies generated with minimum degrees (or
the number of stubs) of 1, 2, and 3 respectively. In each of these
figures, search performance is shown for different $\tau_{sub}$
values 2, 4, 6, 8, 10, 20, and 50. Interestingly, when there is weak
connectedness (i.e. $m=1$), Fig.~\ref{fig_fl-dapa}(a) shows that
imposing hard cutoffs improves the search performance. This is due
to the fact that hard cutoffs increase the connectedness of the
topology by moving the links that were normally go to a hub in a
topology without a hard cutoff. However, when the number of stubs is
larger, in Figs.~\ref{fig_fl-dapa}(a-c), we observe an interplay
between the degree distribution exponent and connectedness for a
fixed cutoff. We observe that improvement caused by hard cutoffs
depend on the value of the hard cutoff, suggesting that reducing
hard cutoff value hurts the search performance after a while. That
is, \emph{potential improvements by having smaller hard cutoffs
diminishes as the performance starts to become dominated by the
degree distribution exponent rather than the connectedness}. Another
observation to be made is that impact of local information plays a
major role in the search performance, as can be seen from
Figs.~\ref{fig_fl-dapa}(a-c).

\emph{Hard cutoffs may improve search efficiency in NF and RW:} More
interestingly, for NF and RW, improvements due to having hard
cutoffs are apparent in all three topology generation methods,
including the PA topologies, regardless of the number of stubs $m$.
The only exception to this behavior is the CM, as shown in
Figs.~\ref{fig_nf}(b) and \ref{fig_rw}(b) for NW and RW
respectively. This means that practical search algorithms like NF
and multiple RWs are affected better by having hard cutoffs on the
overlay topology. For NF, this is evidenced by
Figs.~\ref{fig_nf}(a,d) and \ref{fig_nf}(c,f) for the PA and HAPA
topologies respectively. As it can be seen, having a little more
local connectivity to the network by having a minimum of 2-3 links
in every peer, the search performance increases rapidly for the same
$\tau$ values (i.e. by comparing Fig.~\ref{fig_nf}(a) and
Fig.~\ref{fig_nf}(d)). For RW, a very similar behavior is exhibited
in Figs.~\ref{fig_rw}(a,d) and \ref{fig_rw}(c,f), with only
difference that effect of hard cutoffs is more apparent due to the
fact that NF does better averaging of search possibilities. The
observed behavior of RW illustrates how bad the effect of hard
cutoffs can be on the search performance. It is intuitive that
multiple RWs would perform more similar to NF in terms of
performance.

\emph{More global information is more important when target
connectedness is high:} Fig.~\ref{fig_nf_da} shows the performance
of NF on various DAPA topologies with different parameters.
Figs.~\ref{fig_nf_da}(a-c) shows the search performance on linear
scale when $m=1$, while Figs.~\ref{fig_nf_da}(d-i) show it on
semi-logarithmic scale when $m=2$ and $m=3$. We observe, again, that
as the hard cutoff is getting smaller, the search efficiency
improves regardless of the connectedness $m$. Also, having a little
better connectedness (e.g. $m=3$) improves the search performance
greatly. An interesting observation is that, when constructing the
overlay topology, having more information (i.e. larger $\tau_{sub}$)
about the global topology (thus more scale-freeness in the overall
topology) yields more important improvements on the search
performance for topologies with more connectedness, i.e. larger $m$.
This means that, for the purpose of constructing topologies with
better search performance, when the target connectedness value is
high one needs to be more patient and obtain as much information as
possible before finalizing its links to the rest of the peers.

\emph{DAPA and HAPA models perform almost as optimal as the CM:} An
interesting characteristic to observe is how close the performances
of DAPA and HAPA are to the best possible correspondent CM for the
NF and RW search algorithms. Unlike the other topology construction
mechanisms studied in this paper, CM achieves a perfect
scale-freeness for a given target hard cutoff value, with the cost
of global information. Specifically, topologies generated by the CM
do not have big jumps at the hard cutoff values [e.g.,
Fig.~\ref{fig_pk-pa}(b)] in their degree distributions, in such a
way that the links are configured in the perfect manner to assure
that no node has links more than the target hard cutoff and the
degrees of nodes follow exactly a power-law. This can be seen by
comparing Fig.~\ref{fig_pk-mr} with its counterparts
Figs.~\ref{fig_pk-pa}, \ref{fig_pk-hapa}, and \ref{fig_pk-dapa}. As
can be seen from Figs. \ref{fig_nf}(e) and \ref{fig_nf}(f), with
connectedness $m=2$ or $m=3$, HAPA performs slightly worse than CM
when using NF. Similarly, DAPA performance for moderate $\tau_{sub}$
values (e.g. 6) is very close to the optimal possible by the CM. For
small or no connectedness $m=1$, the behavior is the same, in that
DAPA and HAPA performance is close to the CM performance as shown in
Figs.~\ref{fig_nf_da}(a)-(c), \ref{fig_nf}(c), and \ref{fig_nf}(b)
for DAPA, HAPA, and CM respectively.

\subsubsection{Messaging Complexity}

We also looked at the complexity of messaging overhead for the
search algorithm and topology combinations\footnote{Due to space
constraints we have not included the results for our messaging
complexity study. The results are available upon request from
authors.}. We specifically looked at the average number of messages
incurred by a search request. As the FL algorithm is an extreme and
not scalable in terms of messaging complexity, we did not study its
performance. In all cases, NF performs better than RW consistently,
though the difference between the two algorithms diminishes as
$\tau$ increases for weak connectedness, i.e. $m=1$. But, for
stronger connectedness, i.e. $m>1$, the difference between NF and RW
is more apparent. More importantly, although the effect of hard
cutoffs is negative in terms of messaging complexity, we observed
that this negative effect is very minimal and negligible, given that
improvements on the search performance were observed for smaller
hard cutoffs.

\section{Summary and Discussions}
\label{sec:summary}

We studied effects of the hard cutoffs peers impose on the number of
entries they store on the search efficiency. Specifically, we showed
that the exponent of the degree distribution reduces as hard cutoffs
imposed by peers become smaller. We introduced new scale-free
topology generation mechanisms (e.g., HAPA and DAPA) that use
completely or partially local information unlike traditional
scale-free topology generation mechanisms (i.e., PA and CM) using
global topology information. We showed that topologies generated by
our mechanisms allow better search efficiency in practical search
algorithms like normalized FL and RW. Our study also revealed that
interplay between the degree distribution exponent with a fixed hard
cutoff and connectedness is likely to occur when using our
mechanisms. We also showed that this interplay can be exploited by
enforcing simple join rules to peers such as requiring each peer to
have a minimum of 2-3 links to the rest of the unstructured P2P
network.

Future work will include study of join/leave scenarios for the
overlay topologies while attempting to maintain the scale-freeness
of the overall topology. The challenge is to achieve minimal
messaging overhead for join and leave operations of peers while
keeping the scale-freeness in a topology with a hard cutoff.

\renewcommand{\baselinestretch}{1}

\appendix
The pseudo-codes of the algorithms we used for this work are
presented here. In all these codes it is assumed that the programmer
had routines to create and maintain a network data structure at hand
such as $ADD\_EDGE(i,j)$ creating an undirected edge between nodes
$i$ and $j$. The other routines commonly used by these algorithms
are $RANDOM(i,j)$ creating a random integer $x$ such that $i\leq x
\leq j$, $fRANDOM()$ creating a real-valued random number in
$[0,1]$; and the variables are $k_i$ (stores the degree of the node
$i$), $m$ (the number of stubs or initial links), $Adj[i]$ (all the
nodes connected to node $i$), $k_{total}$ (the total number of
degrees in the network) and $k_c$ (the hard cutoff value).

\subsection{PA} \label{app_pa}

The preferential attachment is a simple model to generate a
scale-free network as in Alg.~\ref{alg_pa}. The algorithm assumes
that the user has already created a network with $m+1$ fully
connected nodes and adds one node at a time starting with the
$m+2^{nd}$ node and fills its stubs by attempting to connect it to
the existing nodes until the total number of nodes in the network
reaches $N$. The repeat-until loop (lines 3-10) repeated $m$ times
is responsible for making sure that the new node $i$ is connected to
one of the old nodes (line 7). To link the new node to the old one
some conditions have to be satisfied (line 6), i.e., they should not
be already connected, a random real number generated during the
execution should be less than the ratio of the degree of the old
node and the total degree (preference in the attachment), and the
old node should have less than $k_c$ connections.

\begin{algorithm}
\caption{PA algorithm}
\label{alg_pa}
\begin{algorithmic}[1]
\FOR {$i=m+2$ to $N$}
\FOR {$j=1$ to $m$}
\REPEAT
\STATE $try \leftarrow \mathbf{true}$
\STATE $node \leftarrow RANDOM(1,i-1)$; $rnd \leftarrow fRANDOM()$
\IF {$node \notin Adj[i] $ AND $rnd < k_{node}/k_{total}$ AND $k_{node}<k_c$}
\STATE $ADD\_EDGE(i,node)$
\STATE $try \leftarrow \mathbf{false}$
\ENDIF
\UNTIL {$try=\mathbf{false}$}
\ENDFOR
\ENDFOR
\end{algorithmic}
\end{algorithm}

\subsection{CM} \label{app_cm}

The main characteristic of the CM is that it uses a predefined
degree sequence. We assume that the user has already generated a
degree sequence consisting of random integers drawn from a specific
distribution such as a power-law distribution. The CM algorithm
(Alg.~\ref{alg_cm}) uses this user-supplied degree sequence
$\{k^{pre}_i\}$ in a for loop and connects the node $i$ to randomly
selected another node (line 3) until all the connections are made.
After that the algorithm deletes the self-loops and multiple
connections (line 9-10). The implementations of these functions
depend on the data structures used.

\begin{algorithm}
\caption{CM algorithm}
\label{alg_cm}

\begin{algorithmic}[1]
\FOR {$i=1$ to $N$}
\WHILE {$k^{pre}_i > 0$}
\STATE $node \leftarrow RANDOM(1,N)$
\STATE $ADD\_EDGE(i, node)$
\STATE $k^{pre}_i \leftarrow k^{pre}_i-1$
\STATE $k^{pre}_{node} \leftarrow k^{pre}_{node}-1$
\ENDWHILE
\ENDFOR
\STATE $DELETE\_SELF\_LOOPS()$
\STATE $DELETE\_MULTIPLE\_LINKS()$
\end{algorithmic}
\end{algorithm}

\subsection{HAPA} \label{app_hapa}

The HAPA algorithm (Alg.~\ref{alg_hapa}) also assumes, like PA
algorithm (Alg.~\ref{alg_pa}), that the user created a network with
$m+1$ fully connected nodes. Then the new node tries to attempt to a
randomly selected node in this initial network by using the
preferential attachment rule and hard cutoff condition (lines 4-7).
Independently from the success of this attempt the new node hops by
using the existing links (lines 10-14) until it fills all of its
stubs (or the number of connections reaches $m$). The new node
avoids connecting itself with an additional condition in line 11.
This process is repeated for all new nodes until the number of nodes
in the network reaches $N$. For this algorithm we define an
additional function $RANDOM\_LINK(i)$ returning a randomly chosen
neighbor of the node $i$.

\begin{algorithm}
\caption{HAPA algorithm}
\label{alg_hapa}

\begin{algorithmic}[1]

\FOR {$i=m+2$ to $N$}
\STATE $j \leftarrow 0$
\STATE $node \leftarrow RANDOM(1,i-1)$; $rnd \leftarrow fRANDOM()$
\IF {$i \notin Adj[node]$ AND $rnd < k_{node}/k_{total}$ AND $k_{node} < k_c$}
\STATE $ADD\_EDGE(i, node)$
\STATE $j \leftarrow j+1$
\ENDIF
\STATE $node \leftarrow i$
\WHILE {$j < m$}
\STATE $node \leftarrow RANDOM\_LINK(node)$; $rnd \leftarrow fRANDOM()$
\IF {$i\neq node$ AND $i \notin Adj[node]$ AND $rnd < k_{node}/k_{total}$ AND $k_{node} < k_c$}
\STATE $ADD\_EDGE(i, node)$
\STATE $j \leftarrow j+1$
\ENDIF
\ENDWHILE
\ENDFOR
\end{algorithmic}

\end{algorithm}

\subsection{DAPA} \label{app_dapa}

The DAPA algorithm (Alg.~\ref{alg_dapa}) is different than
previously defined algorithms in the sense that it maintains two
different networks; a substrate network $G_S$ and an overlay or P2P
network $G_O$ on top of $G_S$. We assumed that the user has already
created the substrate network $G_S$ with number of nodes $N_S$, and
with a GRN topology or a two-dimensional regular mesh topology as
explained in Section~\ref{sec_dapa}. Throughout our simulations we
used a GRN with size $N_S = 2 \times 10^4$ and average degree
$\overline{k}=10$ as a substrate to generate an overlay network with
size $N_O=10^4$. We also assume that the user created an overlay
network $G_O$ with some nodes (2 in our simulations) randomly chosen
from the substrate network. A new function $ADD(G,i)$ which adds the
node $i$ to the network or list $G$ is also defined.

The algorithm has a big while loop checking at every time step the
size of the overlay network, or the number of nodes in it, $|G_O|$
(line 1). Then a random $node$ is chosen from the substrate network
and if that node is not already in the overlay network the algorithm
enters a big $if$ block. In this $if$ block first a bread-first
search algorithm is run on the substrate network to find the
distances of the nodes to $node$ and the values are stored in
$\{d_i\}$. After all the distances to node are determined the nodes
which belong to $G_O$, having distances to $node$ less than
$\tau_{sub}$ and degree less than the hard cutoff are added to the
list $L_{PH}$ (the list of peers in the horizon of $node$) as in
lines 6-10. If the number of peers in the horizon ($|L_{PH}|$) in
the list is less than $m$ then all the peers are connected to $node$
(lines 11-15). Otherwise the preferential attachment procedure is
used, i.e. a random node $peer$ is selected from the list $L_{PH}$
and conditions are checked (lines 18-31). This process is continued
until all the stubs are filled. Every selected $node$ becomes a peer
in $G_O$ once they are connected to any other peer in their horizon.
This algorithm does not guarantee that the minimum degree will be
$m$ since in some cases the peer candidates may not find enough
number of peers in their horizon especially for low $\tau_{sub}$
values. The algorithm continues until the number of peers in $G_O$
reaches $N_O$ with the condition that a node which became a peer
cannot selected again to connect other peers.

\begin{algorithm}
\caption{DAPA algorithm}
\label{alg_dapa}

\begin{algorithmic}[1]

\WHILE {$|G_O|<N_O$}
\STATE $node \leftarrow RANDOM(1,N_S)$

\IF {$node \notin G_O$}

\STATE $BFS(G_S,node)$

\STATE $L_{PH} \leftarrow NULL$

\FOR {$i=1$ to $N_S$}
\IF {$i \in G_O$ AND $d_i > 0$ AND $d_i \leq \tau_{sub}$ AND $k_i < k_c$}
\STATE $ADD(L_{PH},i)$
\ENDIF
\ENDFOR

\IF {$|L_{PH}| \leq m$}
\FOR {$i=1$ to $|L_{PH}|$}
\STATE $ADD\_EDGE(node, L_{PH}[i])$
\ENDFOR
\STATE $ADD(G_O, node)$

\ELSE

\STATE $j \leftarrow 0$
\WHILE {$j<m$}
\REPEAT
\STATE $try \leftarrow \mathbf{true}$
\STATE $peer \leftarrow RANDOM(1,|L_{PH}|)$
\STATE $rnd \leftarrow fRANDOM()$
\IF {$node \notin Adj[peer] $ AND $rnd < k_{peer}/k_{total}$ AND $k_{peer}<k_c$}
\STATE $ADD\_EDGE(node,peer)$
\STATE $try \leftarrow \mathbf{false}$
\ENDIF
\UNTIL {$try=\mathbf{false}$}
\STATE $j \leftarrow j+1$
\ENDWHILE

\STATE $ADD(G_O, node)$

\ENDIF
\ENDIF
\ENDWHILE

\end{algorithmic}
\end{algorithm}

\begin{table}
    \caption{Table of Symbols and Acronyms}
    \label{tab:symbols}
    \begin{center}
\vspace{-5mm}
    \begin{tabular}{cc}
    \begin{tabular}{|c|c|}
    \hline
        \textbf{Acronym} & \textbf{Meaning} \\
    \hline
        PA & Preferential Attachment \\
        CM & Configuration Model \\
        HAPA & Hop-and-Attempt Preferential Attachment \\
        DAPA & Discover-and-Attempt Preferential Attachment \\
        FL & Flooding \\
        NF & Normalized Flooding \\
        RW & Random Walk \\
    \hline
    \end{tabular}
    &
        \begin{tabular}{|c|c|}
    \hline
        \textbf{Symbol} & \textbf{Meaning} \\
    \hline
        $N$ & Number of Nodes \\
        $k$ & Node Degree \\
        $P(k)$ & Probability that node degree is $k$ \\
        $\gamma$ & Power-law exponent, Exponent of the degree distribution \\
        $k_c$ & Hard cutoff on the node degree \\
        $m$ & Number of Stubs, Connectedness \\
        $\tau$ & Time-to-live (TTL)\\
        $\tau_{sub}$ & Local time-to-live in DAPA\\
    \hline
    \end{tabular}
    \\
    \end{tabular}
\vspace{-5mm}
    \end{center}
\end{table}

\section*{Acknowledgment}
This work was supported by the U.S. Department of Energy with
contract number W-7405-ENG-36 and by the National Science Foundation
with award number 0627039.

%H.G. acknowledges support from the U.S. Department of Energy with
%contract number W-7405-ENG-36. M.Y was supported by NSF award number
%0627039.

\renewcommand{\baselinestretch}{1}

\bibliographystyle{IEEE}
\bibliography{hard-cutoff}

\end{document}